\documentclass[12pt]{article}
\pdfoutput=1

\usepackage{putex}
\usepackage{amsmath,amssymb,amsfonts}
\usepackage{psfrag}
\usepackage{footmisc}
\usepackage{url}
\usepackage{mathtools}
\usepackage{color}

\usepackage{tikz}
    \usepackage{amssymb,amsfonts,amsmath}
    \usepackage{tkz-euclide}
        \usetikzlibrary{arrows,calc,patterns}
\usepackage{pgfplots}

\definecolor{darkblue}{rgb}{0.1,0.1,.7}
\definecolor{purple}{rgb}{0.6,0,0.6}
\definecolor{orange}{rgb}{0.9,0.6,0}
\usepackage[colorlinks, linkcolor=darkblue, citecolor=darkblue, urlcolor=darkblue, linktocpage]{hyperref} 
\usepackage[square, comma, compress,numbers]{natbib}
\usepackage[]{amsmath}
\usepackage[]{graphicx}
\usepackage[]{latexsym}
\usepackage[utf8]{inputenc}
\usepackage{geometry}
\usepackage{amscd}
\usepackage[all,cmtip]{xy}
\usepackage{mathrsfs}
\usepackage[margin=10pt,font=small,labelfont=bf]{caption}
\usepackage{dsdshorthand}
\usepackage{changepage}
\usepackage{setspace}
\usepackage{tikz}
\usepackage{subcaption}

\usepackage[titles]{tocloft}
\usetikzlibrary{arrows,positioning}

\def\SL2{\widetilde{SL}(2,\mathbb R)}

\def\mC{\mathcal C}

\newcommand\mR{\mathbb{R}}
\newcommand\mZ{\mathbb{Z}}

\newcommand\lpl{\ell_\text{Pl}}

\numberwithin{equation}{section}

\newcommand{\str}{\text{Str}\,}

\newcommand {\bes} {\begin {equation*}}
\newcommand {\ees} {\end {equation*}}
\newcommand {\beq} {\begin {equation}}
\newcommand {\eeq} {\end {equation}}
\newcommand {\bea} {\begin {eqnarray}}
\newcommand {\ea} {\end {eqnarray}}
\newcommand {\eea} {\end {eqnarray}}

\newcommand{\Sch}{\text{Sch}}

\numberwithin{equation}{section}

\def\<{\langle}
\def\>{\rangle}

\tikzset{
    >=stealth',
    punkt/.style={
           rectangle,
           rounded corners,
           draw=black, very thick,
           text width=15em,
           minimum height=2em,
           text centered},
    pil/.style={
           ->,
           thick,
           shorten <=2pt,
           shorten >=2pt,}
}

 \def\ie{\begin{equation}\begin{aligned}}
\def\fe{\end{aligned}\end{equation}}

\begin{document}


\preprint{PUPT-2621}

\institution{PU}{${}^1$ Joseph Henry Laboratories, Princeton University, Princeton, NJ 08544, USA}
\institution{IAS}{${}^2$ School of Natural Sciences, Institute for Advanced Study, Princeton, NJ 08540, USA}
\institution{SU}{${}^3$ Stanford Institute for Theoretical Physics, Stanford University, Stanford, CA 94305}
\institution{UCSB}{${}^4$ Physics Department, University of California, Santa Barbara, CA 93106, USA}

\title{
The statistical mechanics of near-BPS\\ black holes
}

\authors{ Matthew Heydeman${}^{1,2}$, Luca V. Iliesiu${}^{1,3}$, Gustavo J. Turiaci${}^4$ and Wenli Zhao${}^1$}

\abstract{
Due to the failure of thermodynamics for low temperature near-extremal black holes, it has long been conjectured that a ``thermodynamic mass gap'' exists between an extremal black hole and the lightest near-extremal state.  For non-supersymmetric near-extremal black holes  in Einstein gravity with an AdS$_2$ throat, no such gap was found. Rather, at that energy scale, the spectrum exhibits a continuum of states, up to non-perturbative corrections.  
In this paper, we compute the partition function of near-BPS black holes in supergravity where the emergent, broken, symmetry is $PSU(1,1|2)$. To reliably compute this partition function, we show that the gravitational path integral can be reduced to that of a $\cN=4$ supersymmetric extension of the Schwarzian theory, which we define and exactly quantize. In contrast to the non-supersymmetric case,  we find that black holes in supergravity have a mass gap and a large extremal black hole degeneracy consistent with the Bekenstein-Hawking area. Our results verify a plethora of string theory conjectures, concerning the scale of the mass gap and the counting of extremal micro-states.  
}
\date{}

\maketitle
\tableofcontents

\newpage
\section{Introduction and outline}
\label{sec:intro}

The goal of this paper is to understand the energy levels of near-extremal charged black holes in D dimensions from the perspective of the Euclidean gravitational path integral. The description of such black holes simplifies due to their  $AdS_2 \times X_{D-2}$ near-horizon geometry, where $X_{D-2}$ is a compact space, specifying the geometry of the horizon.

The spectrum of near-extremal black holes has been a source of confusion over the years.\footnote{By near-extremal we mean large charge black holes with a small but non-zero temperature ($T\ll r_0^{-1}$, where $r_0$ is the horizon radius at extremality). } The first manifestation of this was pointed out by Preskill et al \cite{Preskill:1991tb} (see also \cite{Maldacena:1998uz} and \cite{Page:2000dk}). The thermodynamics derived from a semiclassical evaluation of the gravitational path integral gives a temperature-dependent mass above extremality $E = M-M_0$ as $E \sim 2\pi^2 \Phi_r T^2 $, where $M_0$ is the extremal mass. A similar analysis gives the semiclassical entropy $S=S_0 + 4\pi^2 \Phi_r T$, where $S_0=A_0/(4G_N)$ is proportional to the extremal area $A_0$. This behavior is universal, but the precise value of the parameter $\Phi_r$ depends on the details of the model. Because of this scaling,  it was noticed in \cite{Preskill:1991tb} that the thermodynamic description of near-extremal black holes breaks down at low enough temperatures, $T\lesssim 1/\Phi_r$. At such temperatures, the naive semiclassical analysis suggests that emitting even a single Hawking quanta is sufficient to change the black hole's temperature by a large amount.\footnote{An equivalent issue was pointed out in \cite{Maldacena:1998uz} due to the dependence of $\Phi_r$ with Newton constant $G_N$ in particular models.} String theory microstate counting examples \cite{Maldacena:1996ds} indicate the resolution of this issue is the presence of a mass gap of order $E_{gap}\sim 1/\Phi_r$ in the black hole spectrum.\footnote{In this paper, we will focus on scales of order $\sim 1/\Phi_r$ also captured by the gravity description. Of course, at a completely different regime of even lower temperatures $T \sim e^{-S_0}$, we expect to find a discrete spectrum with spacing of order $e^{-S_0}$. This scale requires a UV completion of the gravitational theory.} However, its origin from the gravitational description has, so far, been elusive.

A second manifestation of the gap problem is the question of the extremal black hole degeneracy.  If the gravity description somehow breaks down at the energy scale $E \sim 1/\Phi_r$ then it is possible that the black hole entropy $S_0$, obtained by a semiclassical computation, does not correctly capture the degeneracy of the extremal black hole; rather, $e^{S_0}$ could instead capture the number of extremal and near-extremal states within a $\sim 1/\Phi_r$ energy interval. Nevertheless, while some authors claimed the extremal entropy vanishes \cite{Hawking:1994ii}, string theory examples that preserve sufficient supersymmetry showed that the degeneracy at extremality (assumed to be captured by an index) matches with the Bekenstein-Hawking area $S_0$ \cite{Strominger:1996sh}. 

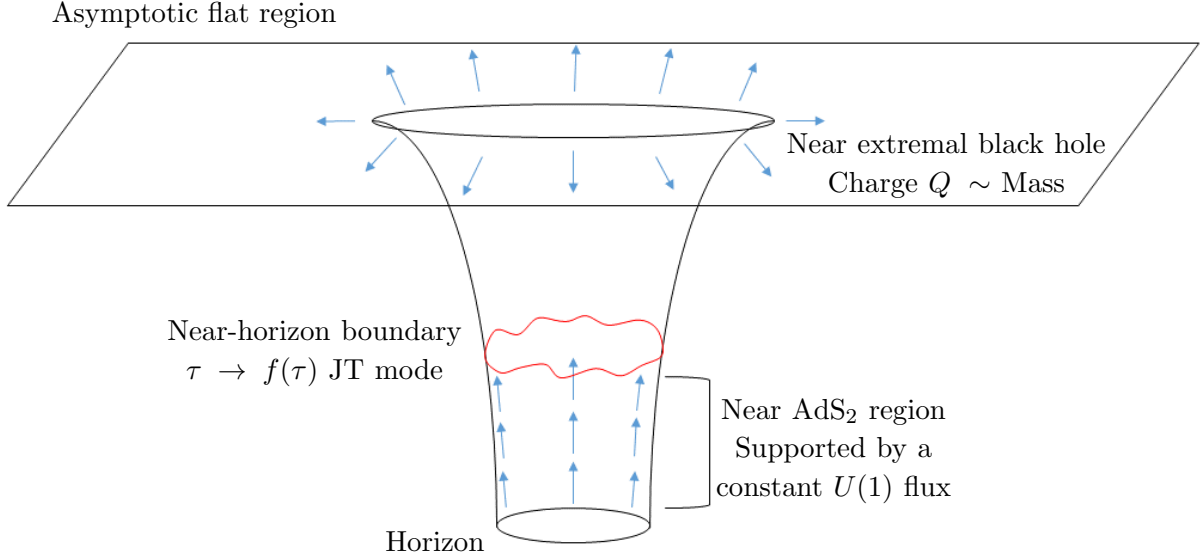
\begin{figure}[t!]
\begin{center}
 \begin{tikzpicture}[scale=1]
 \pgftext{\includegraphics[scale=0.6]{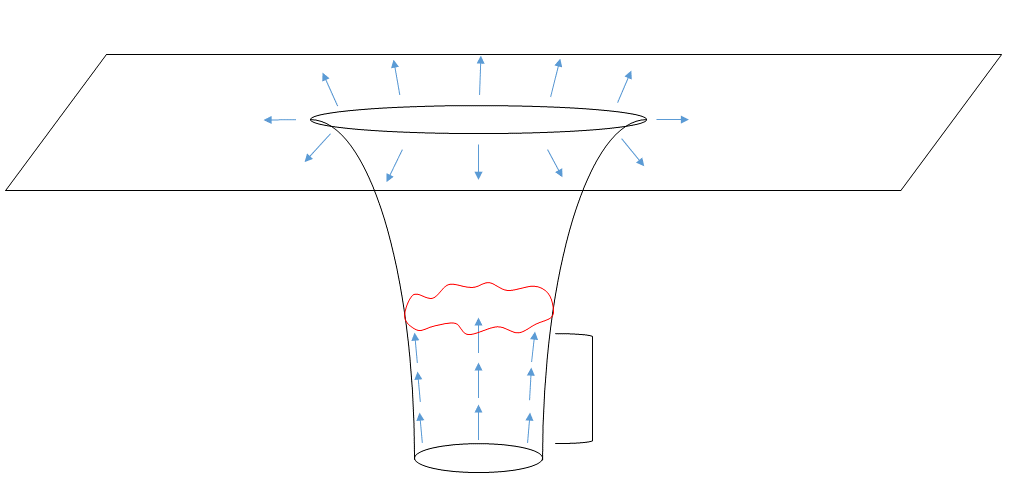}} at (0,0);
  \draw (-5.5,3.5) node  {\small Asymptotic flat region};
    \draw (-3.9,-1) node[text width=5cm,align=center]  {\small Near-horizon boundary \\  $\tau \rightarrow f(\tau)$ JT mode};
  \draw (-2.3,-3.5) node[text width=10cm,align=center]  {\small Horizon};
   \draw (3,-2.3) node[text width=5cm,align=center]  {\small Near AdS$_2$ region\\ Supported by a constant $U(1)$ flux};
    \draw (4.5,1.5) node[text width=5cm,align=center]  {\small Near extremal black hole\\ Charge $Q \sim $ Mass};
  \end{tikzpicture}
  \caption{\label{fig:4DnearBH} The near extremal 4D black hole. A similar picture applies to near-extremal rotating BTZ black hole in AdS$_3$.}
    \end{center}
  \end{figure}

In \cite{Iliesiu:2020qvm} these questions have been revisited using lessons from SYK \cite{kitaevTalks, Maldacena:2016hyu} and Jackiw-Teitelboim (JT) gravity \cite{Teitelboim:1983ux, Jackiw:1984je, Almheiri:2014cka, Jensen:2016pah, Maldacena:2016upp, Engelsoy:2016xyb,Iliesiu:2019lfc, Kapec:2019ecr}. The procedure used to compute the higher dimensional Euclidean path integral for non-supersymmetric black holes is the following. First, we separate the full near extremal geometry into the near-horizon $AdS_2 \times X_{D-2}$ throat (where the interesting physics takes place) and the far away region, which we take to be 
flat. The second step is to recast the $D$-dimensional theory on the throat as a 2D theory on $AdS_2$. It was argued in \cite{Iliesiu:2020qvm} that the temperature-dependence of the Euclidean path integral near-extremality is solely captured by a JT gravity theory (composed of fluctuations around the $AdS_2$ metric and a dilaton which captures the size of $X_{D-2}$), a dimensional reduction of $D$-dimensional gauge fields to 2D, and a gauge field originating from isometries of the horizon $X_{D-2}$. This reduction was considered classically in \cite{Sachdev:2015efa, Almheiri:2016fws, Anninos:2017cnw, Turiaci:2017zwd, Nayak:2018qej, Moitra:2018jqs, Hadar:2018izi, Castro:2018ffi, Larsen:2018cts,  Moitra:2019bub, Sachdev:2019bjn, Hong:2019tsx, Castro:2019crn, Charles:2019tiu,Larsen:2020lhg} but the new feature of \cite{Iliesiu:2020qvm} is the explanation that quantum effects at low temperatures are also captured by this simple theory. It can be shown that JT gravity exactly reduces to a boundary mode, the Schwarzian theory, which lives in the boundary of the throat (shown in figure \ref{fig:4DnearBH}), and describes the breaking of the emergent $SL(2,\mathbb{R})$ symmetry in the throat. In this theory, one can compute the path integral exactly and extract the near-extremal spectrum.\footnote{Integrating out the massive KK modes and other fields have the only possible effect of introducing temperature-independent logarithmic corrections to $S_0$ previously computed in \cite{Banerjee:2010qc, Banerjee:2011jp, Sen:2011ba, Sen:2012cj, Iliesiu:2020qvm} and thus such modes will not affect the computation of the density of states.}

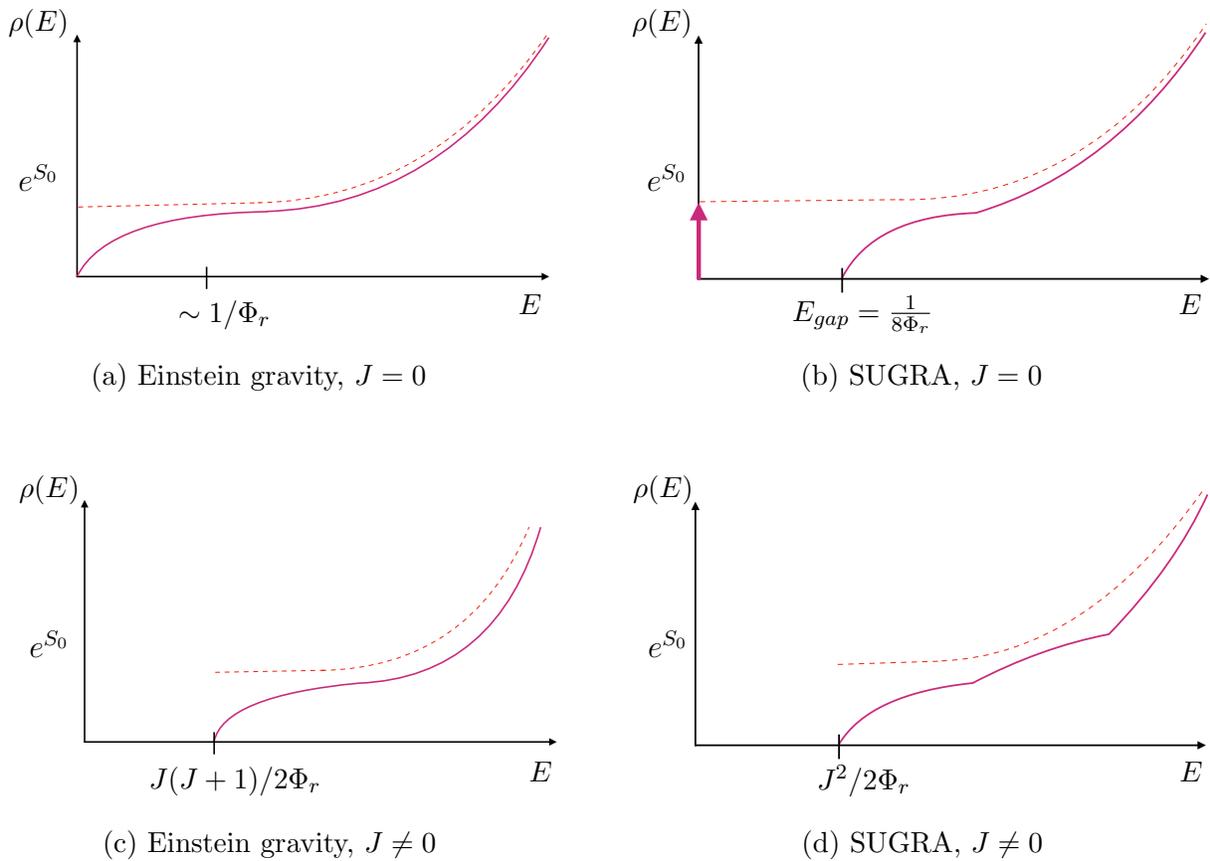
\begin{figure}[t!]
\begin{center}
 \begin{tikzpicture}[scale=0.65]
 \pgftext{\includegraphics[scale=0.46]{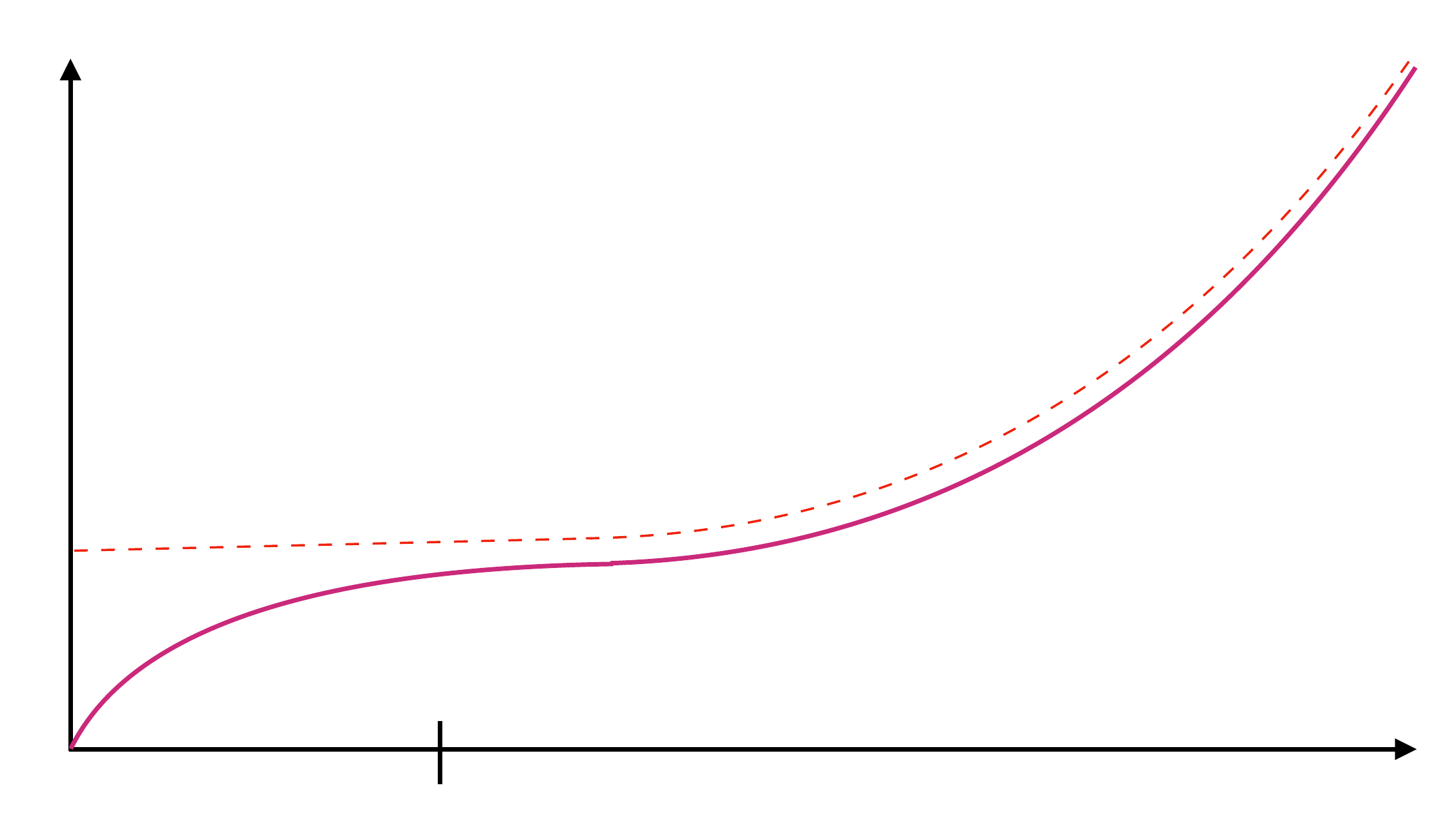}} at (0,0);
 \draw (-5.5,-0.4) node  {$e^{S_0}$};
  \draw (-5.5,2.75) node  {\small $\rho(E)$};
  \draw (-1.7,-3.2) node  {\small $\sim 1/\Phi_r$};
   \draw (4.55,-3) node  {\small $E$};
      \draw (-1,-4.5) node  {\small (a) Einstein gravity, $J=0$};
  \end{tikzpicture}
  \hspace{0.5cm}
   \begin{tikzpicture}[scale=0.65]
 \pgftext{\includegraphics[scale=0.51]{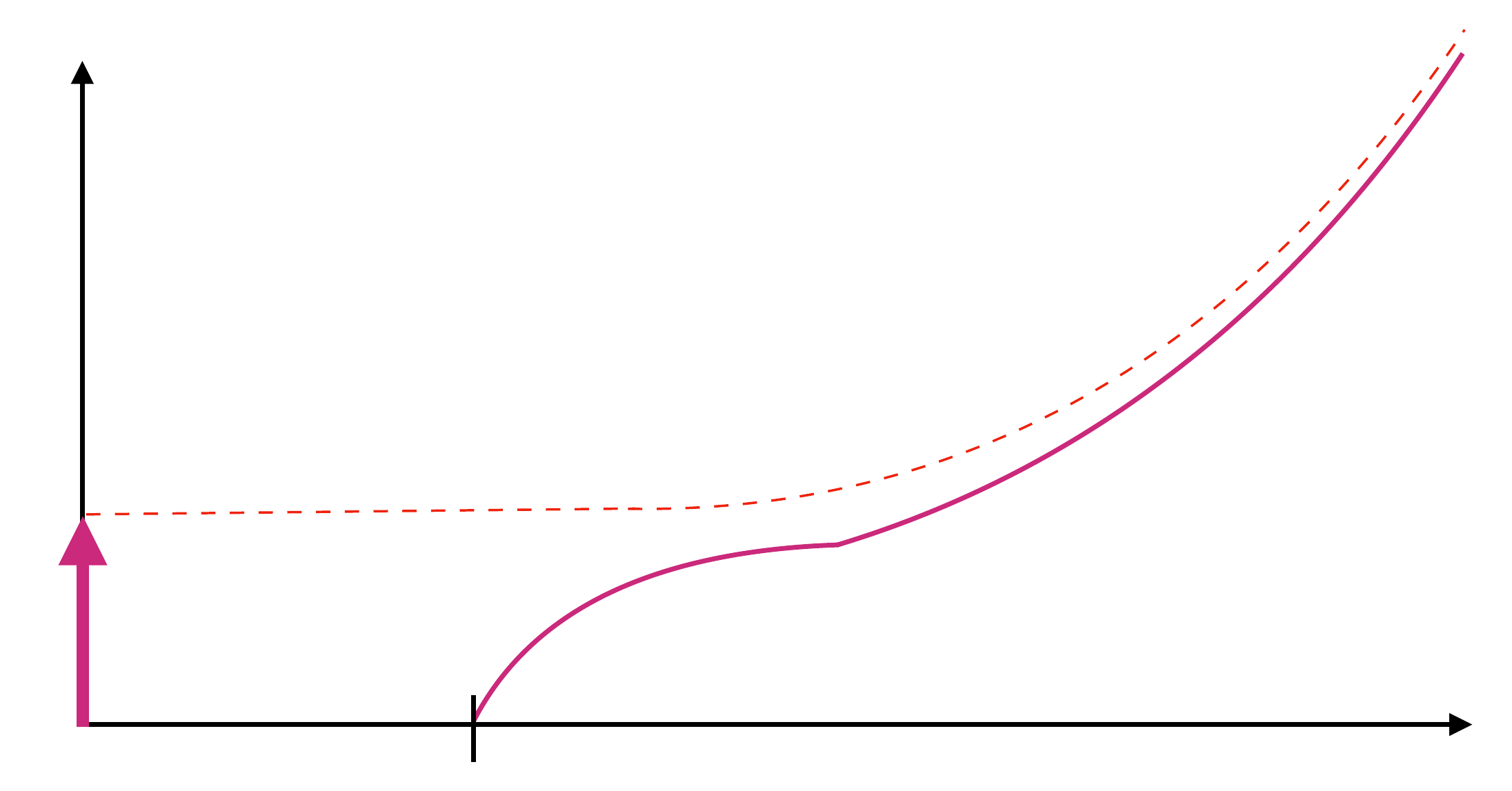}} at (0,0);
 \draw (-5.7,-0.4) node  {$e^{S_0}$};
   \draw (-5.8,2.75) node  {\small $\rho(E)$};
  \draw (-1.7,-3.2) node  {\small $E_{gap}= \frac{1}{8 \Phi_r}$};
   \draw (5,-3) node  {\small $E$};
     \draw (-0.5,-4.5) node  {\small (b) SUGRA, $J=0$};
  \end{tikzpicture}
 \\
 \vspace{0.8cm}
   \begin{tikzpicture}[scale=0.65]
 \pgftext{\includegraphics[scale=0.46]{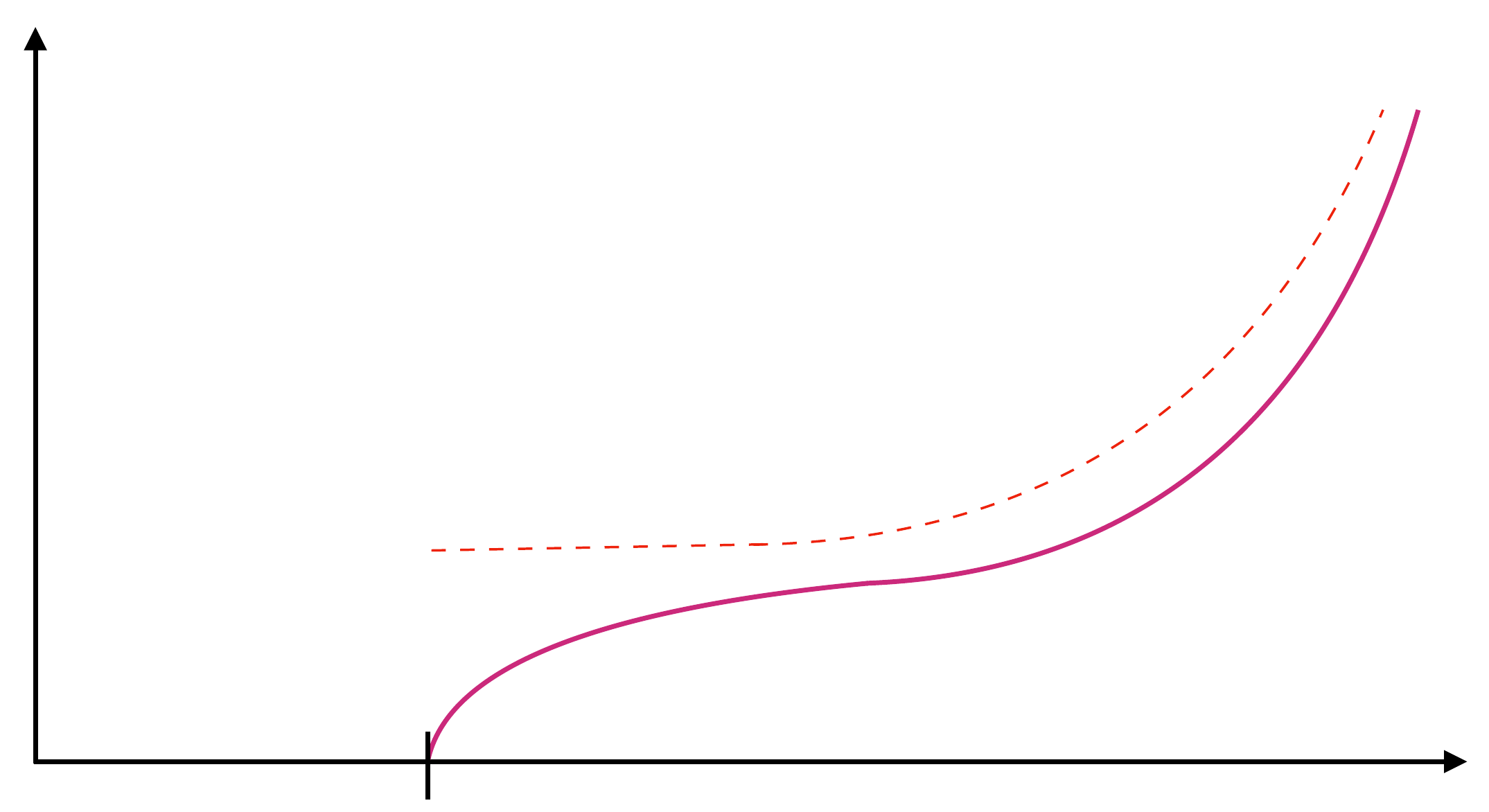}} at (0,0);
 \draw (-5.5,-0.4) node  {$e^{S_0}$};
  \draw (-5.5,2.75) node  {\small $\rho(E)$};
  \draw (-1.7,-3.2) node  {\small $J(J+1)/2\Phi_r$};
   \draw (4.55,-3) node  {\small $E$};
      \draw (-1,-4.5) node  {\small (c) Einstein gravity, $J\neq0$};
  \end{tikzpicture}
  \hspace{0.5cm}
   \begin{tikzpicture}[scale=0.65]
 \pgftext{\includegraphics[scale=0.51]{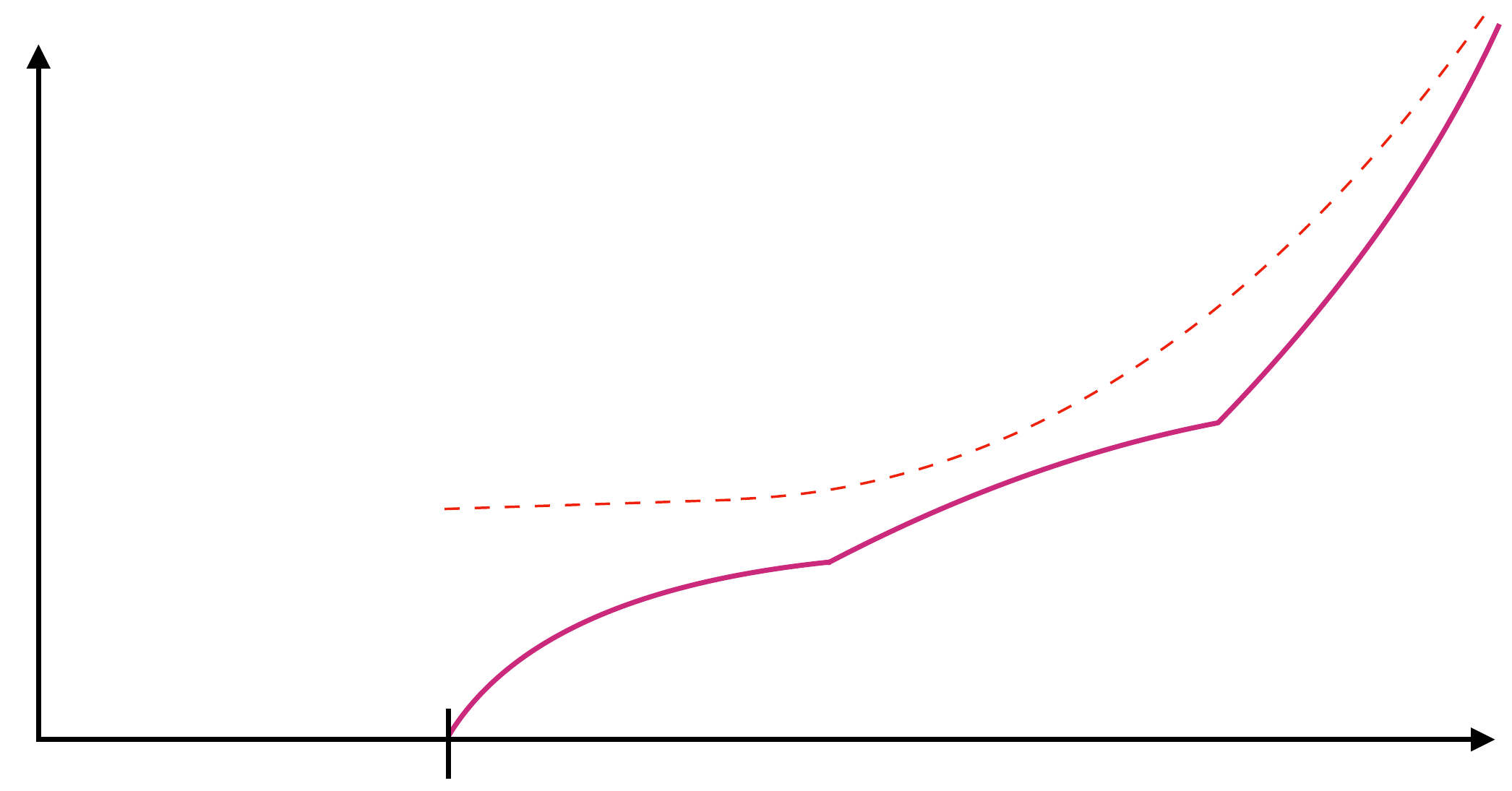}} at (0,0);
 \draw (-5.7,-0.4) node  {$e^{S_0}$};
   \draw (-5.8,2.75) node  {\small $\rho(E)$};
  \draw (-1.7,-3.2) node  {\small $J^2/2\Phi_r$};
   \draw (5,-3) node  {\small $E$};
     \draw (-0.5,-4.5) node  {\small (d) SUGRA, $J\neq0$};
  \end{tikzpicture}
  \caption{\label{fig:plot-of-rho} Schematic shape of the black hole spectrum at fixed $SU(2)$ charge $J$ as a function of energy above extremality $E$. In 4D $J$ is angular momentum while in $AdS_3$ it is the $SU(2)$ charge (one of the two rotations along extra $S^3$ in string theory) that breaks supersymmetry. We show the semiclassical answer (red dashed) and the solution including quantum effects (purple). (a) Answer for Einstein gravity. We see there is no gap at scale $E\sim 1/\Phi_r$ and the extremal entropy goes to zero. (b) Answer for supergravity (either $\cN=2$ in 4D or $\cN=(4,4)$ in 3D). We find a gap at the scale $E_{gap}=\frac{1}{8\Phi_r}$ and a number $e^{S_0}$ of extremal states, consistent with string theory expectations. (c) Einstein gravity spectrum for $J\neq 0$. (d) Supergravity spectrum for $J\neq 0$, the jumps indicate contributions from different supermultiplets $\mathbf{J}, \mathbf{J+1/2}$ and $\mathbf{J+1}$.}
    \end{center}
  \end{figure}

With this new perspective, the authors of \cite{Iliesiu:2020qvm} addressed the puzzling questions about the thermodynamics of non-supersymmetric near-extremal Reissner-Nordstr\"om black holes. Regarding the first confusion pointed out above, the scale $T\sim 1/\Phi_r$ identified by \cite{Preskill:1991tb} was found to be the scale at which quantum effects in the gravitational path integral become large, and the Schwarzian description becomes strongly coupled. This effect signals that the breaking of the $SL(2, \mathbb R)$ conformal symmetry in the throat becomes important \cite{Almheiri:2016fws}. However, in contrast with string theory proposals \cite{Maldacena:1996ds}, it was found that for non-supersymmetric black holes, there is no gap at the scale $E\sim 1/\Phi_r$ and the gravitational path integral predicts a density of states that goes smoothly to zero as $E\to0$. For convenience, this density of states is reproduced here in figure \ref{fig:plot-of-rho} in the left column. Regarding the second confusion about the ground state degeneracy of the extremal black hole, \cite{Iliesiu:2020qvm} also predicts the entropy of the extremal black hole to be much smaller than the naive prediction given by the area of the extremal horizon. A similar conclusion was obtained for near-extremal rotating black holes in 3D gravity \cite{Ghosh:2019rcj,Maxfield:2020ale}.

The results of \cite{Iliesiu:2020qvm} thus show a very different spectrum of near-extremal black holes compared to results obtained from microscopic counting: string theory predicts that the ground state degeneracy of extremal black holes should agree with Bekenstein-Hawking extremal area \cite{Strominger:1996sh} and that the value of the mass gap  $E_{gap}\sim 1/\Phi_r$ can be extrapolated from the weak coupling regime  \cite{Maldacena:1996ds}. The purpose of this paper is to resolve this puzzle. The solution is that most examples in string theory involve supergravity and the extremal black hole preserves some supersymmetries. In the cases studied in this paper, the temperature dependence of the Euclidean path integral is captured by a supersymmetric generalization of JT gravity, which describes the breaking of an emergent superconformal symmetry.

 In this paper, we will discuss the case of near-BPS near-extremal charged black holes in two setups. The first is in four dimensions ($D=4$), where such objects are described classically by the Reissner-Nordstr\"om solution with a $AdS_2 \times S^2$ throat.   We will consider such black holes in 4D ungauged $\cN=2$ supergravity in asymptotically flat space \cite{Freedman:1976aw}.  The situation is depicted in Figure~\ref{fig:4DnearBH}. In our conventions $M_0=Q/\sqrt{G_N}$, proportional to the charge of the black hole, and $\Phi_r= \sqrt{G_N}\hspace{0.1cm} Q^3$. Such a 4D theory has the right ingredients for SUSY preserving extremal black holes \cite{Gibbons:1982fy,Tod:1983pm,Ferrara:1995ih}. The BPS nature of such gravitational solutions allows one to identify them with corresponding string theory constructions, for example \cite{Maldacena:1997de, Ooguri:2004zv,Simons:2004nm,Denef:2007vg}. The second setup we will consider is a near-extremal rotating black hole in $(4,4)$ supergravity in $AdS_3$, with a near-horizon geometry $AdS_2 \times S^1$. This system is relevant for comparison with the D1/D5 system \cite{Strominger:1996sh}. Even though some questions about the spectrum can be addressed using solely the $(4,4)$ super-Virasoro algebra in this particular case thanks to the presence of an $AdS_3$ throat, we will be able to show in section \ref{sec:ads3} that the spectrum matches with the one derived from JT gravity. The Virasoro analysis obviously does not generalize to other cases, for example the Reissner-Nordstr\"om black hole mentioned above, but the JT analysis does and is universal as we explain below.

The main feature in 4D $\cN=2$ supergravity and $(4,4)$ supergravity in $AdS_3$, compared to Einstein gravity, is the fact that the emerging symmetry in the throat becomes the superconformal group $PSU(1,1|2) \supset SL(2,\mathbb{R})\times SU(2)$. In 4D, this includes the $AdS_2$ conformal symmetry and the $S^2$ isometries as bosonic subgroups \cite{Kallosh:1997qw, Boonstra:1998yu,Claus:1998yw,Michelson:1999kn}. In $AdS_3$ the $SU(2)$ arises from a 3D gauge field. In the examples from string theory there is a second $SU(2)$ gauge field coming from the isometries of an extra $S^3$ factor. This second $SU(2)$ charge can be turned on without breaking supersymmetry, and corresponds to the black hole of \cite{Breckenridge:1996is} which has rotation along the $S^3$. We find in both cases the relevant 2D mode in the throat controlling the temperature dependence of the partition function is given by $\cN=4$ super-JT gravity, which describes the breaking of $PSU(1,1|2)$ and becomes strongly coupled at low temperatures. We will first define this theory and solve it exactly to extract the temperature dependence of the partition function, and from it the black hole spectrum. In order to do this, we will show that $\cN=4$ super-JT gravity is equivalent to a $\cN=4$ super-Schwarzian theory, which we introduce in this paper. We solve this theory using either path integral localization or canonical quantization.\footnote{Previous attempts of defining the $\mathcal{N}=4$ super-Schwarzian are \cite{Aoyama:2018lfc, Aoyama:2018voj, Galajinsky:2020hsy} however, explicit formulae for all the components of the  $\mathcal{N}=4$ super-Schwarzian were not presented. Furthermore, to our knowledge, no previous quantization attempt has been made. }
  
To contrast the results found in this paper, we show in Figure~\ref{fig:plot-of-rho} the spectrum of near-extremal black holes derived from Einstein gravity in 4D coupled to a Maxwell field in the left panel. As previously mentioned, we see a small extremal entropy and a lack of a gap. In the right column, we show the main result of this paper, the density of states for near-extremal black holes in 4D $\cN=2$ supergravity and $(4,4)$ supergravity in asymptotically $AdS_3$. Through a computation of the Euclidean path integral, independent of the UV completion of the theory, our analysis verifies and predicts the following results:
\begin{itemize}
\item We find that extremal BPS states exhibit an exact degeneracy. This degeneracy is given by the Bekenstein-Hawking horizon area, which is consistent with extremal microstate countings in string theory \cite{Strominger:1996sh}. This is not true for extremal non-BPS states with $J\neq 0$.

\item We observe the presence of a mass gap $E_{gap}=1/(8\Phi_r)$. In the context of $(4,4)$ supergravity in asymptotically $AdS_3$ and the D1/D5 system, we verify the mass gap estimated at weak coupling from long strings \cite{Maldacena:1996ds}. Here, this prediction is also expanded to black holes in 4D $\cN=2$ supergravity. 

\item We find that the extremal states are solely bosonic, implying the Witten index from a microscopic model coincides with the black hole degeneracy (For previous arguments see for example \cite{Sen:2009vz, Mandal:2010cj, Benini:2015eyy}). This implies that the typical counting of microstates in string theory using an index, such as \cite{Strominger:1996sh}, is correct.

\item A previous attempt to obtain the gap from a gravitational perspective was studied by Maldacena and Strominger \cite{Maldacena:1997ih}. Their argument is only correct for black holes in AdS$_3$, and not for Reissner-Nordstr\"om for example. Our analysis that takes into account the breaking of $PSU(1,1|2)$ can in principle be applied to any system with this pattern of symmetry breaking and applies to situations without an $AdS_3$ factor in the throat.

\item In combination with the prior non-supersymmetric results from \cite{Iliesiu:2020qvm}, we conclude that these features found in string theory examples heavily rely on supersymmetry and are special to supergravity. 
\end{itemize}

Without jumping into all the details, we will give a summary of the technical results derived below. The $\cN=4$ super-Schwarzian theory describing the spectrum of supersymmetric black holes, is given by the following action
\beq
\label{eq:intro-N=4-Schw-action}
I_{\cN=4} =-\Phi_r \int d\tau \left[ \Sch(f,\tau) + {\rm Tr} \left( g^{-1} \partial_\tau g\right)^2 + ({\rm fermions}) \right],
\eeq
where the Schwarzian derivative is defined as 
\beq
\Sch(f,\tau) \equiv \frac{\partial_\tau^3 f}{\partial_\tau f} - \frac{3}2 \left(\frac{\partial_\tau^2 f}{\partial_\tau f} \right)^2.
\eeq
In the action \eqref{eq:intro-N=4-Schw-action},  $f(\tau)$ is a reparametrization mode, $g(\tau)$ is a $SU(2)$ element, and we ignore the terms involving fermionic fields $\eta(\tau)$ and $\bar{\eta}(\tau)$ in the doublet of $SU(2)$. The field $g(\tau)$ describes fluctuations in the angular momentum $J$ of the black hole in the 4D setup comes from the isometries of $S^2$. This theory has $\cN=4$ Poincar\'e supersymmetry but \textbf{breaks} the emergent superconformal $PSU(1,1|2)$ space-time symmetry.\footnote{Nonetheless, $PSU(1,1|2)$ is an important global symmetry for the action \eqref{eq:intro-N=4-Schw-action}. We will clarify this in section \ref{sec:space-time-and-global-symm}.} According to the unbroken $\cN=4$ Poincar\'e supersymmetry, supermultiplets organize into $\mathbf{J}=(J)\oplus2(J-\frac{1}{2})\oplus(J-1)$ for $E\neq0$ and states $(J)$ for $E=0$. Solving this theory exactly gives the density of supermultiplet states labeled by their highest angular momentum $J$ as
\beq
\rho(J,E) = e^{S_0} \delta_{J,0}\delta(E)+ 
\frac{e^{S_0}J}{4\pi^2 \Phi_r E^2}\sinh \left(2 \pi \sqrt{2\Phi_r(E-E_0(J))} \right)\hspace{0.1cM} \Theta\Big(E-E_0(J)\Big)\,, 
\eeq
where we define $E_0(J)\equiv J^2/(2\Phi_r)$. The details of this formula can be found in section \ref{sec:N=4-super-Schw}. For example, states with zero angular momentum come from the contributions with $J=0$, $J=1/2$ since $\mathbf{1/2}=(1/2)\oplus 2 (0)$ and $J=1$ since $\mathbf{1}=(1)\oplus 2(1/2)\oplus (0)$. This is the origin of the plot in the right panel of figure \ref{fig:plot-of-rho}. In the continuous part states with angular momentum $J$ start at an energy $E_0(J)$. This is not necessarily surprising since the same feature appears in the non-supersymmetric case \cite{Iliesiu:2020qvm}. From the gravity perspective this is the correction to the extremality bound of non-BPS extremal charged slightly-rotating black holes. The surprising feature in the supersymmetric theory is that there are no states with zero angular momentum in the range $0<E<1/(8\Phi_r)$.

The results of \cite{Iliesiu:2020qvm} and the present work on the $\cN=4$ super-Schwarzian theory can be seen as a low energy effective theory for black holes without and with supersymmetry, respectively. We are obtaining this effective theory from a detailed analysis of the gravitational path integral in specific examples, but we expect it to be universal and depend only on the pattern of symmetry breaking at low temperatures. We leave for future work the extensions of these ideas to cases such as near-BPS black holes in gauged supergravity in asymptotically $AdS_4$ or $AdS_5$, relevant within the AdS/CFT correspondence. The symmetries suggest the near-extremal spectrum is described by the $\cN=2$ super-Schwarzian theory introduced in \cite{Fu:2016vas},\footnote{To our knowledge, such a super-Schwarzian cannot describe the spectrum of near-BPS black holes in flatspace because the number of supercharges that are preserved by the extremal solution is a multiple of $4$ and 1D $\cN=2$ super-Poincar\'e only has two supercharges.} describing the breaking of $PSU(1,1|1) \supset SL(2,\mathbb{R})\times U(1)$.\footnote{We describe properties of the spectrum of the $\cN=2$ super-Schwarzian in Appendix \ref{app:N2}. Progress on relating the dynamics of near-BPS black holes in  $AdS_4$ to this super-Schwarzian theory was reported in \cite{Forste:2020xwx}. Such a theory is, of course, also relevant when studying near-extremal black holes in $(2, 2)$ supergravity in AdS$_3$.} Such a super-Schwarzian theory sometimes exhibits a gap (depending on the value of the fundamental $U(1)$ R-symmetry charge and whether or not a certain mixed anomaly is present in the theory) but, in contrast to the $\cN=4$ super-Schwarzian, does not generically have purely bosonic ground states.

Finally, our investigations are relevant to the relation between 2D gravity and SYK models. While SYK models with $\cN=1$ and $\cN=2$ supersymmetry were constructed in \cite{Fu:2016vas, Murugan:2017eto}, finding models with $\cN=4$ has been elusive. We hope understanding $\cN=4$ super-JT and super-Schwarzian can help find an SYK model with $\cN=4$ supersymmetry, which is described at low temperatures by the $\cN=4$ super-Schwarzian theory we described. We leave this for future work, but discuss difficulties in constructing such theories towards the end of the paper.

The rest of the paper is organized as follows. In section \ref{sec:N=4superJT} we introduce $\cN=4$ supersymmetric Jackiw-Teitelboim gravity, a 2D dilaton-gravity theory. We will use its BF description to argue that the disk partition function is computed by an $\cN=4$ version of the Schwarzian theory living on the boundary. In section \ref{sec:N=4-super-Schw}, we will study this boundary mode in its own right and compute its partition function and density of states as a function of energy and charge. We will explain in all details the features briefly described above. In section \ref{sec:higherD}, we will perform the dimensional reduction in the throat of a 4D near extremal black holes in $\cN=2$ ungauged supergravity and argue the resulting theory is $\cN=4$ super-JT gravity. Using the results of section \ref{sec:N=4-super-Schw}, we will present the final result for the spectrum of near-extremal black holes in supergravity. In section \ref{sec:ads3} we analyze near-BPS near-extremal black holes in $(4,4)$ supergravity in asymptotically $AdS_3$ spaces, reaching similar conclusions regarding the black hole spectrum. This example helps to make a precise connection with the long string origin of the black hole gap in \cite{Maldacena:1996ds}. In section \ref{sec:conclusion} we conclude and discuss future directions.

\section{ $\cN = 4$ Jackiw-Teitelboim gravity and the $\cN = 4$ super-Schwarzian}
\label{sec:N=4superJT}

In order to be able to study the more complicated case of near-extremal black holes in 4D $\cN=2$ supergravity, we first need to introduce 2D $\cN = 4$ super-JT gravity and its relation to the $\cN=4$ super-Schwarzian. As we shall explain, the latter can be studied using a conventional reformulation in terms of a BF theory. The difficulty will be in relating boundary conditions in this BF theory to SUSY preserving boundary conditions, typically imposed in the second formalism of gravity. The analysis of the $\cN=4$ super-Schwarzian proves even more complicated. While there have been past papers defining the $\cN=4$ super-Schwarzian derivative (in particular, we follow the definitions of \cite{Matsuda:1988qf} and \cite{Matsuda:1989kp}), extracting all the components of the super-symmetric action has, to our knowledge, not been previously done.

\subsection{Formulation as a $\mathfrak{psu}(1,1|2)$ BF theory}
\label{sec:PSU(1,1|2)-BF-thy}

 To obtain the bulk action of $\cN = 4$ super-JT gravity we start by considering a BF theory with a $\mathfrak{psu}(1,1|2)$ superalgebra, 
\be 
\label{eq:psu(1,1|2)-superalgebra}
[L_m, L_{n}] &=(m-n)L_{m-n}\,, \qquad 
[T_i, T_j] = i \epsilon_{ijk} T_k\,,\nn \\
[L_m, G_p{}^\a] &=  \left(\frac{m}2 - \a\right) G_p{}^{\a+m}\,, \qquad [L_m, \bar G^p{}_{\a}] = \left(\frac{m}2 - \a\right)\bar  G^p{}_{\a+m}\,,\nn \\
[T_i,G_p{}^\a] &= -\frac{1}2 \left(\sigma^i\right)_p{}^q G_{q}{}^\a\,, \qquad [T_i,\bar G^p{}_\a] = \frac{1}2 \left(\sigma^i\right)^*{}^p{}_q G^{q}{}_\a\,,\nn \\
\{G_p{}^\a, \bar G^q{}_\b\}&=  2 \delta^q{}_{p} L_{\a+\b}-2(\a-\b)(\sigma^i)^q{}_p T_i\,,
\ee
with all other commutators/anti-commutators between the generators vanishing. Here $L_0$ and $L_{\pm 1}$ are bosonic generators forming an $\mathfrak{sl}(2, \mR)$ sub-algebra, $T_i$, $i = 1,\,\dots \,3$, are bosonic generators forming an $\mathfrak{su}(2)$ sub-algebra, and $G_p{}^\a$ and $\bar G^q{}_\b$ are the fermionic generators with $p,q = 1, 2$ and $\a, \b = -\frac{1}2, \, \frac{1}2$. 

The action of $\cN = 4$ super-JT gravity  is given by, 
\be
I_{BF} =- i \int \str \phi F\,, ~~~~ F = dA - A\wedge A\,,
\ee
where $A$ is a $\mathfrak{psu}(1,1|2)$ gauge field with a field strength $F$ and $\phi$ is a zero-form field transforming in the adjoint of $\mathfrak{psu}(1,1|2)$. Above,  $\str \phi F = \< \phi,\,F\>$ is a quadratic bilinear form invariant under adjoint transformations.  Such a form can be obtained from the quadratic Casimir of $\mathfrak{psu}(1,1|2)$.\footnote{The quadratic Casimir of the superalgebra can be written as, \be
C_2 = L_0^2 - \frac{1}2 \{L_1, L_{-1}\} - T_i T^i +\frac{1}4  (i \sigma^2)^\b{}_\a [G_p{}^\a ,\bar G^{\,p}{}_\b],.\nn
\ee
Rewriting $C_2 = g_{AB} X^A X^B$ with $X = \{L_0, L_\pm,\,T_i,\,G_p{}^\a,\, \bar G^{\,p}{}_\a\}$, we can define the invariant quadratic form by using the metric, $
 \eta_{AB}^{\mathfrak{psu(1,1|2)}} = (-1)^{[X_A]} (g_{AB}^{-1})
$, 
where $[X_A] = 0,1$ for bosonic, and respectively, fermionic generators. In such a case, we can define the supertrace as, $
\text{Str} B F = \< B, \, F\> = \eta_{AB} B^A F^B$.
} We define the gauge field in terms of the supermultiplet of the frame $e^a$ and spin connection $\omega$, also consisting of the SU(2) gauge field $B^i$ and the gravitinos $\psi_{p}{}^{ \a}$ as:\footnote{We choose such conventions for the $\mathfrak{sl}(2, \mR)$ components of the gauge field in order to agree with \cite{Saad:2019lba}. }
\be 
\label{eq:gauge-field-ansatz}
A(x) &= \sqrt{\frac{\Lambda}{2}} \left[e^1(x) L_0 + \frac{e^2(x)}2 \left(L_{1} - L_{-1}\right) \right]- \frac{\omega(x)}2 \left(L_1+L_{-1}\right) + B^i(x) T_i + \nonumber \\ &+   \left(\frac{\Lambda}{2}\right)^{\frac{1}{4}}\left(\bar{\psi}^{p}{}_{ \a}(x) G_{p}{}^{ \a} +  \psi_{p}{}^{ \a}(x) \bar{G}^p{}_{ \a}\right)\,.
\ee
The zero-form field $\phi(x)$ is fixed in terms of the supermultiplet of the $SL(2, \mR)$ Lagrange multipliers $\phi^{1, 2}$ and $\phi^0$, 
\be
\label{eq:Lagr-multiplier}
\phi(x) &= 2\phi^1(x) L_0 + {\phi^2(x)} \left(L_1 - L_{-1} \right)- {\phi^0(x)} \left(L_1+L_{-1}\right) + b^i (x) T_i + \nn \\ &+ \left(\frac{\Lambda}{2}\right)^{-\frac{1}{4}}\left(\bar{\lambda}^{p}{}_{ \a }(x) G_{p}{}^{ \a}+   \lambda_{p}{}^{ \a }(x) \bar G^{p}{}_{ \a}\right)\,.
\ee
Here, $\lambda$ and $\bar \lambda$, and $\psi$ and $\bar \psi$ should be understood as independent components of $A$ and are not related to the Hermitian conjugates of each other. In such a case, the action can be written as,
\be
\label{eq:JT-N=4-action}
I_{JT}^{\cN=4}& = {i} \int \bigg( 
\underbrace{\phi^0}_{\sim\text{ Dilaton}}\bigg[\underbrace{d\omega + \frac{\Lambda}4 \epsilon_{ab} e^a \wedge e^b}_{\frac{d^2 x}2 \sqrt{g} (R+\Lambda)} -\underbrace{\sqrt{2\Lambda}\bar{\psi}^p{}_\alpha \wedge \psi_p {}^\alpha }_{\substack{\text{Gravitino $\psi_p{}^\a$ contribution}\\\text{multiplying the dilaton}}}\bigg]  \nn \\ &-  \sqrt{\frac{\Lambda}{2}}\underbrace{\phi^a}_{\substack{\text{Super-torsion}\\ \text{Lagr.~multiplier}}}\underbrace{\left[de^a - \epsilon_{ab}\,\omega\wedge e^b +2\left(\bar \gamma_a\right)^{\alpha}{}_{\beta}\bar{\psi}^{p}{}_\alpha\wedge \psi_p{}^{\beta} \right]}_{\text{super-torsion component }\tau^a} \\ 
&-  \tr_{SU(2)} \underbrace{\left[b \left(dB + B \wedge B \right)\right]}_{SU(2)\text{ BF theory}} \,\,+ \,\, \sqrt{\frac{\Lambda}{2}}\underbrace{b^q{}_p \bar{\psi}^p{} \wedge \bar{\gamma}_3\psi_q{} }_{SU(2)\text{ BF super-partner}}\nn 
+ \, \underbrace{2\lambda\mathcal{D} \bar{\psi} +2\mathcal{D}^*\psi \bar{\lambda}}_{\substack{\text{Gravitino $\psi^p{}_\a$ and }\\\text{dilatino $\lambda_p{}^\a$ contribution}}}\bigg)\,, 
\ee
where we have outlined the important terms in the action which will ease the comparison with the near-horizon action which we shall obtain in section \ref{ssec:4dSugra}.\footnote{Above, our normalization of the $SU(2)$ trace is given such that $\tr_{SU(2)}(T^i T^j) = -\frac{1}2 \delta^{ij}$.  The covariant derivative is given by $\mathcal{D}=\bar{\gamma}_3d+ \sqrt{\frac{\Lambda}{2}}\bar{\gamma}_ae^a+\frac{1}{2}\omega+\frac{\bar{\gamma}_3}{2}B^i (\sigma^i)$, and $\cD^*$ is the conjugate of $\cD$. We choose $\bar{\gamma}_1=\sigma_1,\,\, \bar{\gamma}_2=-\sigma_3$, and $\bar \gamma_3 = \bar \gamma_1 \bar \gamma_2 = i\sigma_2$. }

The equations of motion for $\phi^{1,2}$ act as Lagrange multipliers and imposes that the super-torsion vanishes. After integrating out $\phi^{1,2}$, one can replace $e^1$, $e^2$ and $\omega$ in terms of the metric $g_{\mu \nu}$ to obtain the supergravity action \eqref{eq:JT-N=4-action} in the second order formalism. The JT gravity dilaton is given by $-i \phi^0 \equiv \Phi$. When comparing the gravitational theory obtained from the dimensional reduction of the near-horizon region of near-BPS black holes to the $\cN=4$ super-JT action we will use this latter form. For simplicity, in the remainder of this section we will fix the cosmological constant, $\Lambda = 2$. Later, when discussing the effective action for the near-horizon region of the near-BP black holes we will revert to a general cosmological constant, determined by the radius of the black hole.

We can complete the dictionary between the BF theory \eqref{eq:JT-N=4-action} and the second-order super-JT gravity by noting that  infinitesimal $PSU(1,1|2)$ gauge transformations in the former, map to infinitesimal diffeomorphisms in the latter. In particular, the infinitesimal supersymmetric transformations on all the fields in \eqref{eq:JT-N=4-action} can be obtained by considering the infintesimal gauge transformation whose gauge parameter takes the form $\Lambda = \epsilon^{p\a} G_{p \a} + \bar \epsilon^{p \a} \bar G_{p \a}$.

With this mapping between the BF theory \eqref{eq:JT-N=4-action} and $\cN=4$ super-JT gravity in mind, we can now analyze the supersymmetric boundary conditions applied to the theory  \eqref{eq:JT-N=4-action} which will be necessary in section \ref{sec:higherD} to understand the gluing of the near-horizon region of near-BPS black holes to asymptotic flatspace. As we will see, these boundary conditions will  reduce the gravitational path integral to that of the $\cN=4$ super-Schwarzian.

\subsection{The super-JT  boundary conditions}
\label{sec:super-JT-bdy-cond}

We begin, just like in the non-supersymmetric case, by imposing that the boundary metric is fixed and proper boundary length is large, $L = \beta/\varepsilon$, with $\varepsilon$ small. Working in Fefferman-Graham gauge, where the metric can be written as
\be 
\label{eq:FG-gauge-metric}
ds^2 = dr^2 + \left(\frac{1}4 e^{2r} - \tilde \cL(\tau) + \dots \right) d\tau^2 \,,
\ee
we consider the boundary to be at fixed, but large, $r$. To satisfy the boundary conditions we identify $\tau \sim \tau+\beta$ and cut-off the geometry at $e^{-r} = \varepsilon/2$. We fix the leading component of the boundary metric and allow the sub-leading component, $\tilde \cL(\tau)$, to vary. The $\dots$ represent terms that are further sub-leading in $r$ which we do not need to fix.

Similarly, we require that the dilaton takes an asymptotically large and constant value at the boundary, $\Phi|_{\partial \cM} \equiv -i\phi_0|_{\partial \cM} = \Phi_r/\varepsilon$. Next, we discuss the boundary conditions for the super-partners of the frame and spin-connection. If working in the  Fefferman-Graham gauge \eqref{eq:FG-gauge-metric}, then, in order to preserve supersymmetry, we impose that the leading component of the gravitino is fixed and vanishes $\psi^{p \alpha} = O(e^{-r/2})$ and $\bar \psi^{q \alpha} = O(e^{-r/2})$. Similarly, to again preserve supersymmetry, we impose that the leading contribution of the dilatino vanishes at asymptotically large values $\lambda  = O(e^{-r/2})$. Finally, we describe the boundary conditions for the $SU(2)$ gauge field and its Lagrange multiplier. From the perspective of the higher dimensional  black holes which we will study in section \ref{sec:higherD}, the $SU(2)$ gauge field appears from the isometry of $S^2$ along which we perform the dimensional reduction. We would therefore, like to fix Dirichlet boundary conditions at the boundary of the asymptotically flat region. As we will describe in detail in section \ref{sec:higherD}, imposing these boundary conditions in the asymptotically flat region translates to fixing a linear combination of the zero-form field $b$ and the $SU(2)$ gauge field $B$ at the boundary of the near-horizon region.

We would now like to translate the boundary conditions discussed above in the second-order formalism to boundary conditions in the BF theory \eqref{eq:JT-N=4-action}. We will follow the steps outlined in \cite{Grumiller:2017qao} (explained also recently in \cite{Saad:2019lba}) in non-supersymmetric JT gravity and will obtain results similar to \cite{Cardenas:2018krd}, which studied boundary conditions in JT supergravity with an $OSp(2, \cN)$ isometry group. Fixing the Fefferman-Graham gauge for the metric \eqref{eq:FG-gauge-metric} yields the value of the frame $e^{1, \, 2}$ and spin connection $\omega$ \cite{Saad:2019lba}
\be 
\label{eq:FG-1st-order-formalism}
e^1 = dr \,, \qquad e^2= \left(\frac{1}2 e^r - \tilde \cL(\tau) e^{-r}\right)d\tau \,, \qquad \omega  = -\left(\frac{1}2 e^r + \tilde \cL(\tau) e^{-r}\right)d\tau \,.
\ee 
Fixing the decaying piece of the gravitino we can gauge fix all other components of the  $PSU(1,1|2)$ gauge field along the boundary to be
\be
\label{eq:PSU(1,1|2)-gauge-field-form}
A_\tau(\tau) &= L_1 \frac{e^r}2 + L_{-1} \tilde \cL(\tau) e^{-r} + B^i_\tau(\tau) T_i \nn \\ &+ e^{-\frac{r}2} \frac{\bar\psi^p(\tau)}2  G_{p, \, -\frac{1}2} + e^{-\frac{r}2} \frac{\psi^p(\tau)}2 \bar G_{p,\, -\frac{1}2} + O(e^{-2r})\,,
\ee
where we have used the shorthand notation $\psi^{p}(\tau) \equiv \psi^{p,\, \frac{1}2} $ and $\bar \psi^{p}(\tau) \equiv \bar \psi^{p,\, \frac{1}2} $ on the boundary $\partial \cM$. 
We can now impose the equations of motion of $A_\tau$ in the proximity of the boundary $D_\tau \phi = 0$, or rather due to only fixing the leading orders of $A_\tau$ and $\phi$, $D_\tau \phi = O(e^{-r})$, where $D$ denotes the $PSU(1,1|2)$ covariant derivative. This implies that the zero-form field $\phi$ should be constrained to take the form 
\be
\label{eq:b.c.-for-dilaton}
\phi(\tau)  &= -{i\Phi_r} L_1 e^r +  {2 i\Phi_r'} L_0 +(\bar \psi \lambda + \psi \bar \lambda -  2i \tilde\cL \Phi_r - 2i\Phi_r'')L_{-1} e^{-r} + b^i(x) T_i \nn \\&+ \bar \lambda^p G_{p,\frac{1}2} e^{r/2} - \bigg(2(\bar \lambda^p)' - i\Phi_r \bar \lambda^p  -  B_\tau^i (\sigma^i{}^*)^p{}_q\, \bar \lambda^q\bigg)  G_{p,-\frac{1}2}e^{-r/2} \nn \\ &+ \lambda^p \bar G_{p,\frac{1}2}e^{r/2} - \bigg(2 (\lambda^p)' - i \Phi_r  \lambda^p  -  B_\tau^i (\sigma^i\,)^p{}_q\,  \lambda^q\bigg)  \bar G_{p,-\frac{1}2} e^{-r/2} + O(e^{-2r})\,,
\ee
where, again, we have used the short-hand notation $\lambda^{p}(\tau) = \lambda^{p, \frac{1}2}$ and $\bar \lambda^{p}(\tau) =\bar \lambda^{p, \frac{1}2}$ on the boundary $\partial \cM$ and where $\Phi'_r \equiv \partial_\tau \Phi_r(\tau)$ denotes the derivative with respect to the boundary time $\tau$.\footnote{We will motivate why the parameter $\Phi_r$ in \eqref{eq:b.c.-for-dilaton} can be identified with the renormalized dilaton shortly.}

If we want to impose the boundary conditions for the dilaton ($\Phi_r = \text{const}$), the dilatinos ($\lambda^{p, \,\frac{1}2} = 0$ and $\bar \lambda^{p,\,\frac{1}2} =0$) and the zero-form field $b^i(x)$ we can then relate the gauge field in \eqref{eq:PSU(1,1|2)-gauge-field-form} to the zero-form field $\phi(\tau)$ as $-2i \Phi_r \hspace{0.1cm}A_\tau(\tau) = \phi(\tau)$, where $\Phi_r$ is the renormalized value of the dilaton. Thus, in the first-order formalism and in the gauge in which $A_\tau$ takes the form \eqref{eq:PSU(1,1|2)-gauge-field-form}, the boundary condition that we want to impose is $\delta(2i\Phi_r A_\tau(\tau) + \phi(\tau))|_{\partial \cM}=0$. The boundary term necessary for such a boundary condition to have a well defined variational principle is \cite{Cardenas:2018krd}:
\be 
\label{eq:bdy-term-JT-grav-first-order}
I_{BF,\,\text{bdy.}} = \frac{i}2  \int_{\partial \cM} \text{Str}\, \phi A = { \Phi_r} \int_{\partial \cM} d\tau\, \text{Str}\,  A_\tau^2 \, . 
\ee
Integrating out the field $\Phi$ in the bulk we find that the bulk term yields no contribution. 
Replacing $A_\tau$ in \eqref{eq:bdy-term-JT-grav-first-order} we find that the action can then be rewritten as
\be 
I_{BF,\,\text{bdy.}} = -{\Phi_r} \int_{\partial \cM} d\tau \left[\tilde \cL(\tau) + \frac{1}2 \left((B^1_\tau)^2+(B^2_\tau)^2+ (B^3_\tau)^2\right)\right]
\ee
Thus, it is convenient to define 
\be
\label{eq:def-cL(u)-and-bdy-Lagr}
\cL(\tau)  = \tilde \cL(\tau) + \frac{1}2 \left((B^1_\tau)^2+(B^2_\tau)^2+ (B^3_\tau)^2\right)\,, \,\, \text{ from which } \,\, I_{BF,\,\text{bdy.}} = -{\Phi_r} \int_{\partial \cM} d\tau \, \cL(\tau)\,.
\ee
We will not determine $\cL(\tau)$ explicitly. Rather we will see how $\cL(\tau)$ (and all other variables in \eqref{eq:PSU(1,1|2)-gauge-field-form}) transform under the gauge transformations that preserve the asymptotic form of \eqref{eq:PSU(1,1|2)-gauge-field-form}. In general in a BF theory with gauge group $G$,  boundary gauge  transformations  are parametrized by functions $g$ in $\text{loop}(G)/G$. However, since we are preserving the asymptotic form \eqref{eq:PSU(1,1|2)-gauge-field-form} which comes from working in the Fefferman-Graham gauge \eqref{eq:FG-1st-order-formalism} the space of gauge transformations is instead parametrized by $\text{Diff}(S^{1|4})/PSU(1, 1|2)$. Therefore, the way in which $\cL(\tau)$ transforms under this special class of gauge transformations will yield how the boundary Lagrangian transforms under $\text{Diff}(S^{1|4})/PSU(1, 1|2)$. From here, we will show in section \ref{sec:N=4-transf-law} that the boundary Lagrangian can be identified with the $\cN=4$ super-Schwarzian derivative. Thus, to do this identification, we note that $\cL(\tau)$, $B_\tau(\tau)$, and $\psi^p(\tau)$ transform as 
\bea 
\label{eq:gauge-transf-preserving-asymp}
\delta_\L \cL &=&\xi \cL'+2 \cL \xi'+ \xi'''- B^i_\tau (t^i)'+\frac{1}{2}\left(3\bar{\psi} \epsilon'+\bar \psi'\epsilon -3\bar{\epsilon}' \psi -\bar{\epsilon} \psi'\right), \nn \\ 
\delta_\L \psi^p &=& \xi ( \psi^p)'+\frac{3}{2}\psi^p \xi'-\epsilon^p \cL -2(\epsilon^p)''-\frac{1}{2}t^i(\sigma^i){}^p{}_q\, \psi^q + (B^i_\tau)' (\sigma^i){}^p_q \epsilon^q +2 B^i_\tau (\sigma^i{})^p{}_q\,(\epsilon^q)',\nn\\ 
\delta_\L B^i_\tau&=& \left( \xi B^i_\tau\right)'-(t^i)'+ \frac{1}2 \bar \psi \sigma^i\epsilon+ \frac{1}2 \bar{\epsilon} \sigma^i \psi + i \epsilon^{ijk} \,t^j\,B^k_\tau\,,
\eea
under the gauge transformation which preserves the form of $A_\tau$ in \eqref{eq:PSU(1,1|2)-gauge-field-form}:
\be
\Lambda(\tau) &= \frac{\xi}2 L_1 - \xi' L_{0} + \left(\frac{1}2 \bar \psi   \epsilon +\frac{1}2 \psi \bar \epsilon - \frac{1}2  \xi (B^a)^2 +\cL \xi + \xi'' \right) L_{-1} - (t^i -\xi B^i_\tau) T_i \nn \\ &+ \frac{1}2  \bar \epsilon^p G_{p,\frac{1}2} - \left((\bar \epsilon^p)' - \frac{1}2 \xi \bar \psi^p  - \frac{1}2 B_\tau^i (\sigma^i{}^*){}^p_q\, \bar \epsilon^q\right)  G_{p,-\frac{1}2} + \frac{1}2 \epsilon^p \bar G_{p,\frac{1}2} \nn \\ &- \left( \epsilon^p - \frac{1}2 \xi  \psi^p  - \frac{1}2 B_\tau^i (\sigma^i)^p_q\,  \epsilon^q\right)  \bar G_{p,-\frac{1}2}
\ee
It is not a coincidence that $\Lambda(\tau)$ takes the same form (up to the redefinition \eqref{eq:def-cL(u)-and-bdy-Lagr}) as $\phi(\tau)$ in \eqref{eq:b.c.-for-dilaton}. This follows from requiring that the leading result in both $D_\tau \Lambda$ (since we require that the gauge transformation $\Lambda$ not change the asymptotic form of $A_\tau$ \eqref{eq:PSU(1,1|2)-gauge-field-form})  and $D_\tau \phi$ (since we impose the equation of motion for $A$ by also using \eqref{eq:PSU(1,1|2)-gauge-field-form}) both vanish. With these transformations in mind, we now proceed by introducing the necessary technology to define the $\cN=4$ Schwarzian derivative. Following that, we will finally show that this Schwarzian derivative can be identified with the boundary Lagrangian  $\cL(\tau)$ from \eqref{eq:def-cL(u)-and-bdy-Lagr}.

Finally, we can extend this analysis to the case when the $SU(2)$ chemical potential, $\a$, is turned on. In order to do this we can generalize the previous boundary condition from $ A_\tau - \frac{i \phi}{2\Phi_r} = 0$ to $ A_\tau - \frac{i\phi}{2\Phi_r} = \frac{2\pi i}{\beta} \alpha T^3 $, where $\alpha\sim \alpha +1$. This does not modify the boundary conditions of gravity and the fermions but now the $SU(2)$ gauge field boundary condition is $B_\tau - \Phi^{-1}_r b = \frac{2\pi i}{\beta} \alpha T^3$, supporting the identification of $\alpha$ with a chemical potential. In section \ref{sec:subtleties} we explain how from the 4D near-extremal black hole perspective this boundary condition is equivalent to fixing the holonomy of the gauge field in the asymptotically flat region, which is related to fixing the boundary 4D angular velocity.

\subsection{The $\cN=4$ supersymmetric boundary mode} 

So far we have seen that $\cN=4$ super-JT gravity can be reduced to a boundary theory. In this section, we will see this theory can be recasted as a $\cN=4$ super-reparametrization mode with a super-Schwarzian action. We will first review in \ref{sec:N=4-superDiffeos} the definition of super-diffeomorphisms. In \ref{sec:super-Schw-action} we will define the super-Schwarzian derivative that will be the action and in \ref{sec:N=4-transf-law} we will put everything together to write down the final boundary action. 

\subsubsection{Super-diffeomorphisms}
\label{sec:N=4-superDiffeos}

To match with the $\cN=4$ super-JT theory defined previously we will be interested in $SU(2)$ extended $\cN=4$ supersymmetry.\footnote{Other choices with also $\cN=4$ are the $O(4)$ extended algebra studied for example in \cite{Schoutens:1988ig}, we will not be interested in those for the purposes of this paper.} We will begin by giving a super-space description of $\cN=4$ super-diffeomorphisms. 

Consider an $\cN=4$ super-line parametrized by a bosonic variable $\tau$ and fermionic variables $\theta^p$ and $\bar{\theta}_q$, where $p,q=1,2$. The coordinates $\theta^p$ and $\bar{\theta}_q$ transform respectively in the fundamental and antifundamental  representations of a local $SU(2)$ symmetry. We will omit the indices and simply write $\theta$ and $\bar{\theta}$ when it is clear by context. We define the four super-derivatives as 
\beq
D_p = \frac{\partial}{\partial \theta^p} + \frac{1}{2} \bar{\theta}_p \partial_\tau ,~~~~~\bar{D}^q = \frac{\partial}{\partial \bar{\theta}_q} + \frac{1}{2} \theta^q \partial_\tau.
\eeq
The $SU(2)$ indices will be raised and lowered by the antisymmetric tensors $\varepsilon_{pq}$ and $\varepsilon^{pq}$ with $\varepsilon_{12}=\varepsilon^{21}=1$. We will denote by $\sigma^i$ with $i=1,2,3$ the Pauli matrices and indices will be contracted from `left-bottom' to `right-up', e.g. $\bar{\theta} \sigma^i \theta = \bar{\theta}_p (\sigma^i)^p_q \theta^q$. We will sometimes group the coordinates of the $\cN=4$ super-line as $Z=(\tau, \theta^p, \bar{\theta}_p)$. 

We will study general reparametrizations of the super-line, which will become the degrees of freedom in the path integral that defines the Schwarzian theory. These have the following form 
\beq\label{reparametrizations}
\tau \to \tau'(\tau, \theta, \bar{\theta}) ,~~~~\theta^p \to \theta'{}^p(\tau, \theta, \bar{\theta}), ~~~~\bar{\theta}_q \to \bar{\theta}'_q(\tau, \theta, \bar{\theta}),
\eeq
and satisfy a set of constrains given by 
\bea\label{repa_constraints}
D_p \bar{\theta}'_q =0&,&~~~\bar{D}^p \theta'{}^b=0,\\
D_p \tau' - \frac{1}{2} (D_p \theta'{}^q)\bar{\theta}'_q=0&,&~~~~\bar{D}^p \tau' - \frac{1}{2} (\bar{D}^p \bar{\theta}'_q)\theta'{}^q=0
\eea
analyzed in \cite{Matsuda:1988qf} and \cite{Matsuda:1989kp}. They guarantee that the superderivatives transform homogeneously and preserve the global $SU(2)$ symmetry. We will refer to the space of solutions of these constrains as $\cN=4$ super-diffeomorphisms and denote it as ${\rm Diff}(S^{1|4})$, indicating one bosonic and four fermionic directions. 

Next we will look at some examples. The simplest super-reparametrizations are purely bosonic. They are given in terms of an arbitrary function $f(\tau)$ and an arbitrary $SU(2)$ matrix $g(\tau)$. The solutions of the constraints have the form 
\bea\label{eq:repbos}
\tau &\to& f(\tau) + \frac{1}{8} f''(\tau) (\bar{\theta}\theta)^2,\\
\theta^p &\to& g\Big( \tau -\frac{1}{2}\bar{\theta}\theta\Big)^p_q  ~\theta^q~ \sqrt{f'\Big(\tau-\frac{1}{2}\bar{\theta}\theta\Big)} , \\
\bar{\theta}_p &\to& \bar{\theta}_q~ \bar{g}\Big(\tau + \frac{1}{2}\bar{\theta}\theta\Big)^q_p ~\sqrt{f'\Big(\tau+\frac{1}{2}\bar{\theta}\theta\Big)}.
\eea
Another simple example are the `chiral' and `anti-chiral' fermionic transformations. The chiral ones are parametrized in terms of two fermionic functions $\eta^p(\tau)$ in the fundamental of $SU(2)$ and the reparametrization is given by
\bea\label{eq:repfer}
\tau &\to&\tau + \frac{1}{2} \bar{\theta} \eta\Big(\tau+\frac{1}{2}\bar{\theta}\theta\Big) + \frac{1}{4} \partial_\tau (\bar{\theta} \eta(\tau))^2,\\
\theta^p &\to& \theta^p + \eta^p\Big(\tau-\frac{1}{2}\bar{\theta}\theta\Big)  , \\
\bar{\theta}_p &\to& \bar{\theta}_p\Big( 1 + \bar{\theta} \eta'\Big(\tau-\frac{1}{2}\bar{\theta}\theta\Big)  \Big).
\eea
The anti-chiral are parametrized by $\bar{\eta}_p(\tau)$ in the antifundamental, and the reparametrization is given by
\bea\label{eq:repfer2}
\tau &\to&\tau + \frac{1}{2} \theta \bar{\eta}\Big(\tau-\frac{1}{2}\bar{\theta}\theta\Big) + \frac{1}{4} \partial_\tau (\theta \bar{\eta}(\tau))^2,\\
\theta^p &\to&\theta^p\Big( 1 +\theta \bar{\eta}'\Big(\tau+\frac{1}{2}\bar{\theta}\theta\Big)  \Big) , \\
\bar{\theta}_p &\to&  \bar{\theta}_p +\bar{\eta}_p\Big(\tau+\frac{1}{2}\bar{\theta}\theta\Big).
\eea
The most general element of ${\rm Diff}(S^{1|4})$ is obtained by a general fermionic transformation followed by a bosonic one.\footnote{ In writing the Schwarzian derivative, the order of the composition is important. In particular, to compute the $\cN=4$ super-Schwarzian we compose the bosonic transformation with a fermionic one in this order. The order of compositions will however be unimportant when localizing the path integral.} We parametrize ${\rm Diff}(S^{1|4})$ in terms of the degrees of freedom
$
f(\tau),~~g(\tau)\in SU(2),~~\eta^p(\tau),~~\bar{\eta}_p(\tau),
$
where $p=1,2$. It is hard to determine the finite form that has all parameters turned on since such answer contains a large number of terms. Therefore, we will not write it down explicitly although its straightforward to get from the results presented so far. Instead, it is useful to analyze the most general infinitesimal transformation. Up to $\mathcal{O}(\eta)$ is given by 
\bea
\tau &\to&f(\tau-\frac{1}{2}\bar{\eta}(\tau-\frac{1}{2}\bar{\theta}\theta)\theta+\frac{1}{2}\bar{\theta}\eta(\tau+\frac{1}{2}\bar{\theta}\theta)) + \frac{1}{8}f''(\tau)\left((\bar{\theta}+\bar{\eta})(\theta+\eta)\right)^2\\
\theta^p&\to& F^p{}_q\left(\tau-\frac{1}{2}\bar{\theta}\theta\right)\left(\theta^q+\eta^q\left(\tau-\frac{1}{2}\bar{\theta}\theta\right)\right)+\partial_{\tau}\left(F^p{}_q\left(\tau\right)\theta^q \theta \bar{\eta}\left(\tau\right)\right)\\ 
\bar{\theta}_p &\to& \left(\bar{\theta}_q+\bar{\eta}_q\left(\tau+\frac{1}{2}\bar{\theta}\theta\right)\right)F^q{}_p\left(\tau+\frac{1}{2}\bar{\theta}\theta\right)+ \partial_{\tau}\left(\eta\left(\tau\right)\bar{\theta}\bar{\theta}_qF^q{}_p\left(\tau\right)\right)
\eea
where we use $F^p{}_q(\tau)=\sqrt{f'(\tau)}g^p{}_q(\tau) $.

We introduce now an important sub-group of the super-reparametrizations, the global superconformal group $PSU(1,1|2)$. It is generated by six bosonic variables (three from $SL(2)$ and three from $SU(2)$) and eight fermionic. The general case can be found in \cite{Matsuda:1989kp}. Instead we will write down the bosonic generators 
\bea\label{infintesimalparameterb}
\tau &\to& \frac{a\tau+b}{c\tau+d} - \frac{c}{4(c\tau+d)^3}(\bar{\theta}\theta)^2  ,\\
\theta^p &\to&[e^{i \vec{t} \cdot \vec{\sigma}}]^p_q\theta^q \frac{1}{c(\tau - \frac{1}{2} \bar{\theta}\theta)+d} ,\\
\bar{\theta}_p &\to& \bar{\theta}_q [e^{-i \vec{t} \cdot \vec{\sigma}}]^q_p \frac{1}{c(\tau + \frac{1}{2} \bar{\theta}\theta)+d}\label{infintesimalparameterb-last} ,
\eea
with $a,b,c,d\in \mathbb{R}$ such that $a d- bc=1$ and $c>0$. This is precisely of the form \eqref{eq:repbos} where $f(\tau)$ is a global conformal transformation $PSL(2,\mathbb{R})$ while $(t^1, t^2, t^3)$ parametrize a global $SU(2)$ transformation.  The fermionic generators can be parametrized by two constant spinor doublets $\eta$ and $\tilde{\eta}$ and act as 
\bea
\label{infintesimalparameterf-first}
\tau &\to& \tau + \frac{1}{2} (\bar{\theta}-\bar{\tilde{\eta}}) \eta - \frac{1}{2} \bar{\eta} (\theta-\tilde{\eta})  ,\\
\theta^p &\to& \theta^p+ \eta^p  - \tilde{\eta}^p,\\
\bar{\theta}_p &\to&\bar{\theta}_p + \bar{\eta}_p   - \bar{\tilde{\eta}}_p,
\label{infintesimalparameterf}
\eea
 As anticipated there are eight fermionic generators. As explained above, the most general case is a composition of bosonic and fermionic. 

\subsubsection{Super-Schwarzian action}
\label{sec:super-Schw-action}

The Schwarzian derivative associated to reparametrizations of the $\cN=4$ super-circle was defined by Matsuda and Uematsu in \cite{Matsuda:1989kp} (see also \cite{Matsuda:1988qf}). The Schwarzian derivative is given in terms of superspace variables as 
\beq\label{Sderivative}
S^i (Z; Z') = - 2 D \sigma^i \bar{D} \log \left(\frac{1}{2} (D_p \theta'{}^q)(\bar{D}^p \bar{\theta}'_q)\right).
\eeq
It satisfies the following chain rule 
\beq
\label{chainrule}
S^i (Z; Z') = \frac{1}{ 2} (\sigma^i)^p_q(\sigma^j)^{q'}_{p'} (D_p \theta''{}^{p'})(\bar{D}^q \bar{\theta}''_{q'}) S^j (Z'';Z')+S^i (Z;Z'').
\eeq
Another defining property is the fact that $S^i (Z;Z')=0$ whenever the super-reparametrization is an element of the global $PSU(1,1|2)$ as in \eqref{infintesimalparameterb}. This will prove consequential in section \ref{sec:space-time-and-global-symm} when studying the global symmetries of the $\cN=4$ super-Schwarzian action which we shall define shortly.  

The derivative $S^i$ is a superfield with several components. We can extract the bosonic piece we want to associate to the Schwarzian action. In the notation of \cite{Matsuda:1989kp} it is 
\beq
S^i (Z; Z') \supset - \bar{\theta} \sigma^i \theta \hspace{0.1cm}S_b(\tau, \theta, \bar{\theta}; \tau',\theta',\bar{\theta}')
\eeq
One motivation of this choice is to look at purely bosonic transformations defined in equation \eqref{eq:repbos}. For simplicity let's briefly consider $Z'=(\tau',\theta',\bar{\theta}')$ where $\tau'$, $\theta'$ and $\bar \theta'$ take the special form \eqref{eq:repbos}, in terms of an arbitrary $f(\tau)$ and set $g(\tau)=1$. Then the definition above gives the super-Schwarzian 
\beq
\label{eq:just-in-terms-of-Schw}
S^i (Z,Z') = -{\rm Sch}(f,\tau) \bar{\theta} \sigma^i \theta.
\eeq
Another motivation is that when we interpret $S^i$ as a superconformal generator, that component generates bosonic translations along the circle and we want to identify this as the action of the Schwarzian theory.  

The super-Schwarzian satisfies the constrains $D_p D_q S^i = \bar{D}^p \bar{D}^q S^i=0$. Therefore, as a superfield it can be expanded in the following components 
\bea\label{sschwarzian}
S^i &=& 2S_{T}^i + \bar{\theta} \sigma^i S_f + \bar{S}_f \sigma^i \theta - \bar{\theta} \sigma^i \theta S_b  + i \epsilon^{ijk} \bar{\theta} \sigma^j \theta \partial_\tau S_{T}^k \nonumber\\
&&- \frac{1}{2}(\bar{\theta}\theta) \bar{\theta}\sigma^i \partial_\tau S_f + \frac{1}{2} (\bar{\theta} \theta) \partial_\tau \bar{S}_f \sigma^i \theta + \frac{1}{4} (\bar{\theta} \theta)^2 \partial^2_\tau S_T^i 
\ea
Then the super-Schwarzian action is 
\beq
\label{Saction}
I_{\cN=4} = - \Phi_r\int d\tau S_b[f(\tau),g(\tau),\eta(\tau)]
\eeq
where $\Phi_r$ can be viewed as a coupling constant whose role we shall soon discuss and the factor of $12$ above is chosen such that it simplifies the factor in \eqref{eq:just-in-terms-of-Schw}.  

We can rewrite the action in super-field notation by defining $S = \bar{\theta} \sigma^i \theta S^i,$ where we sum over $i=1,2,3$. Then, using the expansion of $S^i$ gives,\footnote{This is a simpler expression of $S^i$ for which more components are present. We can construct $S$ from an $S^i$ as long as $S^i$ satisfies  $D_p D_q S^i = \bar{D}^p \bar{D}^q S^i=0$. To derive this we use that 
\bea
&&(\bar{\theta} \sigma^i \theta)(\bar{\theta} \sigma^i \theta)=-3(\bar{\theta} \theta)^2,~~~~~~\varepsilon^{ijk}(\bar{\theta} \sigma^i \theta)(\bar{\theta} \sigma^j \theta)=0,\nn\\
&& (\bar{\theta} \sigma^i \theta) (\bar{\theta} \sigma^i G) = -3 (\bar{\theta} \theta) (\bar{\theta} G),~~~~(\bar{\theta} \sigma^i \theta) (\bar{G} \sigma^i \theta) = 3 (\bar{\theta} \theta) (\bar{G} \theta) \nn\,.
\eea
}
\beq
S =2 \bar{\theta} \sigma^i S_T^i \theta  - 3 (\bar{\theta} \theta) \bar{\theta} S_f + 3 (\bar{\theta} \theta) \bar{S}_f \theta + 3 S_b (\bar{\theta}\theta)^2\,.
\eeq
Then the action \eqref{Saction} can be rewritten as $I_{\cN=4}\sim \Phi_r \int d\tau d^4\theta S$.
Note that in terms of the $ S[f(\tau),g(\tau),\eta(\tau)]$ there is no obvious chain rule analogous to \eqref{chainrule}. For this reason it will oftentimes be easier to work with $S^i$ instead of the super-field $S$. To make things concrete, it is informative to write the Schwarzian action when focusing on purely bosonic components, when setting $\eta(\tau) = 0$ in \eqref{Saction}:
\be
\label{eq:bosonic-compoents-of-the-action}
I_{\cN=4,\text{ bosonic}}[f(\tau), g(\tau), \eta(\tau) = 0] =- \int_0^\b d\tau \Phi_r\left[\Sch( f, \tau ) + \Tr(g^{-1} \partial_\tau g)^2\right] \,.
\ee
Since $\eta(\tau)=0$ is a solution to the equations of motion we will soon use the above action to extract the classical saddle point when quantizing the super-Schwarzian action at the level of the path integral.

\subsubsection{Transformation law and a match with the JT boundary term } 
\label{sec:N=4-transf-law}

In this section, we shall derive explicitly the infinitesimal transformation rules of the $\cN=4$ super-Schwarzian (\ref{sschwarzian}). We will expand $f(\tau)\approx\tau+\xi(\tau)$, $g \approx 1 + i t^{i}(\tau) \sigma^i$, $\eta \approx \epsilon(\tau)$ and $\bar{\eta}\approx\bar{\epsilon}(\tau)$ and work to linear order in $\xi$, $t^i$, $\epsilon$ and $\bar{\epsilon}$. A convenient way to encode the infinitesimal reparametrizations of the super line (\ref{reparametrizations}) that automatically satisfies the constraints (\ref{repa_constraints}) is to use a super-field as in \cite{Matsuda:1988qf,Matsuda:1989kp}: 
\bea\label{repara_superfield}
E(Z)=\xi(\tau+\frac{1}{2}\bar{\theta}\theta)+\xi(\tau-\frac{1}{2}\bar{\theta}\theta)+\bar{\theta} \epsilon(\tau-\frac{1}{2}\bar{\theta}\theta)-\bar{\epsilon}(\tau+\frac{1}{2}\bar{\theta}\theta)\theta+\frac{1}{2}\bar{\theta}\sigma^i\theta t^i(\tau),
\eea
Under such a reparametrization (\ref{repara_superfield}), the super-Schwarzian (\ref{sschwarzian}) transforms in the same way as the super-energy momentum tensor, which is given in \cite{Matsuda:1989kp}: 
\bea\label{SStransformation}
\delta_E S^i= \partial_{\tau}\left(E(Z)S^i\right)+D E(Z)\bar{D} S^i+ \bar{D}E(Z)D S^i - i \epsilon^{ijk}\left(D\sigma^j\bar{D}E\right) S^k-2D\sigma^i\bar{D}\partial_{\tau}E(z),
\eea
 Now substitute (\ref{sschwarzian}) and (\ref{repara_superfield}) into (\ref{SStransformation}), and collect in components: 
\bea
\delta_E S^i= 2\delta_E S_T^i + \bar{\theta} \sigma^i \delta_ES_f + \delta_E\bar{S}_f \sigma^i \theta - \bar{\theta} \sigma^i \theta \delta_E S_b  + i \epsilon^{ijk} \bar{\theta} \sigma^j \theta \partial_\tau \delta_E S_T^k+\dots,
\eea
we obtain the infinitesimal transformation of $S_T^i, S_f, \bar{S}_f, S_b$. Note that the terms in $\dots$ are purely determined by lower components, and thus it is enough to focus on the terms up to $\mathcal{O}(\bar{\theta}\theta).$ As a result, the transformations of $S_T^i, S_f, \bar{S}_f, S_b$ are given by:
\bea 
\label{eq:Sb-transformation}
\delta_E S_b&=&\xi S_b'+2S_b \xi'+ \xi'''- S_T^i (t^i)'+\frac{1}{2}\left(3\bar{S}_f\epsilon'+\bar{S}_f'\epsilon-3\bar{\epsilon}'S_f-\bar{\epsilon} S_f'\right), \nonumber\\ 
\delta_E S_f^p&=& \xi (S_f^{p})'+\frac{3}{2}S_f^p \xi'-\epsilon^p S_b-2(\epsilon^p)''-\frac{1}{2}t^i (\sigma^i){}^p_qS_f^q+(S_T^i)' (\sigma^i{})^p_q\epsilon^q+2 S_T^i (\sigma^i{})^p_q(\epsilon^q)',\nonumber\\ 
\delta_E S_T^i&=& \left( \xi S_T^i\right)'-(t^i)'+\frac{1}2 \bar{S}_f\sigma^i\epsilon+ \frac{1}2\bar{\epsilon}\sigma^iS_f+i \epsilon^{ijk}t^j S_T^k.
\eea

We note that they exactly agree with the infinitesimal transformation deduced from the boundary action of the BF theory \eqref{eq:gauge-transf-preserving-asymp} with the field identification:
\be 
\cL &\leftrightarrow S_b\,,\nn \\ 
\psi^p &\leftrightarrow S_f^p\,,\nn \\ 
B^i_\tau &\leftrightarrow S_T^i\,.
\ee
Since the infinitesimal transformations and ${PSU}(1,1|2)$ invariance suffices to determine the global form of the action as the $\cN=4$ super-Schwarzian, it then follows that the boundary action \eqref{eq:def-cL(u)-and-bdy-Lagr} from BF theory agrees with our definition of the Schwarzian action \eqref{Saction}. Therefore, the path integral in the $\cN=4$ super-JT gravity with the boundary conditions discussed in section \ref{sec:super-JT-bdy-cond} can be reduced to that for the $\cN=4$ super-Schwarzian action defined by \eqref{Saction}.  

\subsection{Spacetime and global symmetries}
\label{sec:space-time-and-global-symm}
Before analyzing the quantization of the super-Schwarzian theory, it is useful to study the space-time and global symmetries present in this action. 

A useful way to discuss symmetries of a quantum field theory is to cast it in terms of \emph{internal symmetries} and \emph{spacetime symmetries}.  The continuous internal symmetries of (\ref{Saction}) form PSU$(1,1|2)$,\footnote{It is sometimes confusing whether the group is SU$(1,1|2)$ or PSU$(1,1|2)$. We note that SU$(1,1|2)$ contains an extra $U(1)$ factor generated by identity, due to the cancellation of SL$(2)$ part and $SU(2)$ part of the super-trace in the algebra \cite{BrittoPacumio:1999ax}. We do not have such an $U(1)$ factor in \ref{infintesimalparameterb}, and, therefore, our symmetry group is PSU$(1,1|2).$ } generated by (\ref{infintesimalparameterb})--(\ref{infintesimalparameterf}), directly acting on $\left(f(\tau), g(\tau), \bar{\eta}_p(\tau), \eta^p(\tau)\right).$ They are zero modes of the $\cN=4$ super-Schwarzian derivative, and the action (\ref{Saction}) is invariant due to the chain rule (\ref{chainrule}). Specifically,
\be
\label{eq:invariance-under-PSU(1,1|2)}
S^i(Z, h \circ Z'(Z)) = S^i(Z,  Z'(Z)) \,,
\ee
where $h$ is a composition of the bosonic and fermionic transformations in (\ref{infintesimalparameterb})--(\ref{infintesimalparameterf}) which we apply to the super-reparametrization $Z'(Z)$.
These transformations are the supersymmetric generalization of the $SL(2,\,\mR):f(\tau) \to \frac{a f(\tau)+b}{c f(\tau) +d}$, and we will have to quotient out such transformations as we proceed to compute the partition function \cite{Jensen:2016pah,Maldacena:2016upp,Engelsoy:2016xyb}, in order to obtain a well-defined partition function. Furthermore, aside from the $PSU(1, 1|2)$ that has an $SU(2)$ subgroup which acts on the left of the field $g(\tau)$ there is an additional $SU(2)$ symmetry which acts on the right of the field $g(\tau)$. This additional symmetry does not act on $f(\tau)$ or on the fermionic fields $\bar \eta$ and $\eta$. Thus, to summarize the continuous internal symmetry is, up to discrete factors, $PSU(1,1|2) \times SU(2)$.\footnote{There is also an outer $SU(2)$ inherited from the $PSU(1,1|2)$ algebra. It acts on $(\eta^1\left(\tau\right), -\bar{\eta}_2\left(\tau\right))$ as a doublet. In the $4d$ setup we later consider, this is the $R$ symmetry of the $\mathcal{N}=2$ supergravity that is broken by stringy effects. }

The $\cN=4$ super-Schwarzian theory also admits \emph{spacetime} symmetries.  To see this consider the transformation \eqref{eq:Sb-transformation}; for the transformation $\xi(\tau) = \xi$ corresponding to $\tau \to \tau + \xi$ (time translations), $t^i(\tau) =t^i$ corresponding to  constant infinitesimal rotation of $\theta^p$ ($R$-symmetry rotations), and $\epsilon(\tau) = \epsilon$ corresponding to $\theta \to \theta+\epsilon$ (super-translations), the action $S_b$ is invariant up to a total derivative. Together, all such infinitesimal transformations generate the $\cN = 4$ super-Poincar\'e spacetime symmetry. They satisfy the algebra
\bea
\{\bar{Q}_p, Q^q\}=2\delta^q_p H, ~~~\text{with }~~\bar{Q}_p=i\frac{\partial}{\partial \theta^p}+\bar{\theta}_p\partial_{\tau},\,\,\, Q^p=-i\frac{\partial}{\partial \bar{\theta}_p}-\theta^p \partial_{\tau},\,\,\,H=\partial_\tau
\eea
while all other commutators vanish. 
It is straightforward to identify these $\cN=4$ supercharges from the Noether procedure using the transformation \eqref{eq:Sb-transformation}: $S_b$ is the charge generating time translations (and is thus the Hamiltonian of the theory), $S_f$ is the generator of super-translations, and $S_T^i$ is the generator of R-symmetry rotations.  Note that the Hamiltonian here is \emph{not} one of the generators of PSU$(1,1|2),$ since it is in fact the super-Schwarzian itself, and thus proportional to the quadratic Casimir of the PSU$(1,1|2).$ This is quite analogous to the non-supersymmetric case \cite{kitaevTalks, Mertens:2017mtv, Lin:2019qwu}. 

We would like to stress that even though $PSU(1,1|2)$ plays a big role in constructing the Schwarzian action, the theory is not invariant under spacetime superconformal $PSU(1,1|2)$ symmetries. The spectrum is not organized according to $PSU(1,1|2)$ representations. Only the $\cN=4$ super-Poincar\'e sub-group is a spacetime symmetry. 

It is also useful to briefly discuss some of the discrete internal and spacetime symmetries of the theory. The $\cN=4$ super-Schwarzian has a time reversal symmetry $\cT$ which acts on the fermionic fields as $\cT \eta(\tau) = i \eta(-\tau)$ and $\cT\bar \eta(\tau) = i \bar \eta(-\tau)$ and on the bosonic fields as $\cT f(\tau) = f(-\tau)$ and $\cT g(\tau) = g(-\tau)$; in such a case, $\cT^2 = 1$ and the symmetry is $\mathbb Z_2^\cT$.  There is also a $\mathbb Z_2^F$ fermionic symmetry $(-1)^F$ which solely acts on the fermionic fields.\footnote{The same time-reversal properties are also true in the $\cN=1$ theory \cite{Stanford:2019vob}. } Thus, the symmetry is $\mZ_2^F \times \mZ_2^\cT$. We can now address discrete factors for the internal symmetry group of the theory. In (\ref{infintesimalparameterb})--(\ref{infintesimalparameterf}) we note that the transformation given by the center of $SL(2, \mR)$ with $a=d=-1$ and $b=c=0$ (acting as $\eta \to -\eta$ and $\bar \eta\to -\bar \eta$) is redundant with the composition of the two center transformations for the two $SU(2)$, the first of which acts as $g \to -g$, $\eta \to -\eta$, $\bar \eta\to -\bar \eta$, and the second of which acts solely on $g \to -g$. Furthermore, this transformation is again redundant with the $(-1)^F$ symmetry mentioned previously. Thus, the bosonic subgroup of the symmetry group for the theory is given by 
\be
\frac{SL(2, \mR) \times SU(2) \times SU(2) \times \mZ_2^F}{\mZ_2} \times \mZ_2^\cT\,.
\ee 
Consequently, we note that even-dimensional representations of $SU(2)$ (half-integer spins) need to be fermionic and odd-dimensional representations of $SU(2)$ (integer spins) need to be bosonic. Since we will be able to decompose our partition function as a sum over $SU(2)$ characters, this fact will play an important role in easily obtaining the supersymmetric index from the theory by using the result for the partition function. 

\section{Quantizing the $\cN=4$ Schwarzian theory}
\label{sec:N=4-super-Schw}

In this section, we will study the $\cN=4$ super-Schwarzian theory in more detail. In particular we will compute the exact partition function and density of states. We will do the calculation in two ways, first using the localization method of Stanford and Witten \cite{Stanford:2017thb} and second using the 2D CFT approach of \cite{Mertens:2017mtv}, and find agreement. Then we will analyze the spectrum that we derive and point out some salient features like the large zero temperature degeneracy and the presence of a gap.

\subsection{The action}

The $\cN=4$ super-line can be parametrized in superspace by $(\tau, \theta^p, \bar{\theta}_p)$, $p=1,2$, where $\theta$ and $\bar{\theta}$ are Grassman variables transforming as fundamental and anti-fundamental of an $SU(2)$ symmetry.  $\cN=4$ super-reparametrizations are parametrized by a bosonic field $f(\tau)\in {\rm Diff}(S^1)$, a local transformation $g(\tau) \in SU(2)$ (or more precisely the loop group) and fermionic fields $\eta^p(\tau)$ and $\bar{\eta}_p(\tau)$. In terms of a super-reparametrization these fields can be roughly written as 
\bea
\tau &\to& f(\tau) + \ldots,\\
\theta^p &\to& g^p_{~q}(\tau) \theta^q \sqrt{f'(\tau)} + \eta^p(\tau) + \ldots, \\
\bar{\theta}_p &\to& \bar{\theta}_q \hspace{0.1cm}\bar{g}^q_{~p}(\tau) \sqrt{f'(\tau)} + \bar{\eta}_p(\tau) + \ldots.
\ea
The dots correspond to terms that are fixed by the super-reparametrization constrains and can be found in the previous section. We also defined a Schwarzian action $I_{\cN=4} =-\Phi_r \int d\tau S_b [ f, g, \eta, \bar{\eta}]$ invariant under $PSU(1,1|2)$ transformations acting on the fields $(f,g,\eta,\bar{\eta})$. The bosonic component of this action is 
\beq
I_{\cN=4} =- \Phi_r \int_0^\b d\tau \left[\Sch(f,\tau) + \Tr(g^{-1} \partial_\tau g)^2 + ({\rm fermions})\right]
\eeq
which gives the usual Schwarzian action and a particle moving on $SU(2)$. The extra terms involve the fermions $\eta$ and $\bar{\eta}$. 

In this section, we will compute the Euclidean path integral giving the partition function,\footnote{We leave the measure implicit in this formula. We take the measure to be the Pfaffian of the symplectic form over the integration space ${\rm Diff}(S^{1|4})/PSU(1,1|2)$, studied in \cite{Aoyama:2018lfc}.}
\beq
Z (\beta,\alpha) = \int \frac{\mathcal{D} f \mathcal{D} g \mathcal{D} \eta \mathcal{D} \bar{\eta}}{PSU(1,1|2)} ~\exp\left( \Phi_r \int d\tau S_b[f,g,\eta,\bar{\eta}] \right)\,, 
\eeq
where $\Phi_r$ is a dimensionful coupling constant of the theory. The inverse temperature $\beta$ and chemical potentials $\alpha$ appears in the path integral through the boundary conditions of the fields:
\beq
f(\tau+\beta)=f(\tau),~~~~g(\tau+\beta) = e^{2 \pi i \alpha \sigma^3} g(\tau),~~~  \eta(\tau+ \beta) =- e^{2\pi i \alpha \sigma^3} \eta(\tau),
\eeq
and similarly for $\bar{\eta} $. In the rest of this section we will evaluate this path integral as a function of $\beta$, $\alpha$ and the coupling $\Phi_r$, and rewrite it as a trace over a Hilbert space with a possibly continuous spectrum. 
 
 \subsection{The partition function}
 
\subsubsection{Method 1: Fermionic localization}
\label{sec:fermionic-localization}

In this section, we will solve the theory following \cite{Stanford:2017thb}. The integration space is a coadjoint orbit and the super-Schwarzian action generates a $U(1)$ symmetry. Even though we will not work out the measure and symplectic form in detail, we will assume it is chosen such that we can apply the Duistermaat-Heckman theorem. Therefore, we will compute the classical saddles, the one-loop determinants, and put everything together into the final answer \eqref{eq:localization-part-function}.

\paragraph{The $\cN=4$ saddle-point:} As previously mentioned, the bosonic part of the $\cN=4$ Schwarzian action is given by
\be
\label{eq:bosonic-compoents-of-the-action-2}
I_{\cN=4,\text{ bosonic}} =- \int_0^\b d\tau \Phi_r\left[\Sch(f,\tau) + \Tr(g^{-1} \partial_\tau g)^2\right] \,.
\ee
The equations of motion for $f(\tau)$ and $g(\tau)$ imply that:
\be
\partial_\tau \Sch(f,\tau) = 0\,, \qquad \partial_\tau \Tr (g^{-1} \partial_\tau g)^2 =  0
\ee
The solution for the Schwarzian is well-known and is given by $f(\tau) = \tan(\pi \tau/\b)$. The solution for the $SU(2)$ adjoint field takes the form $g = \exp(i t_i \e^i \tau)$, where $\e^i$ is a constant that needs is set by the boundary conditions for the field $g(\tau)$. To make the computation easier we note that all solutions can be transformed to the diagonal form ($g = \exp(i \sigma_3 \e^3 \tau)$) using an $SU(2)$ transformation. If  require that the field $g$ be periodic, than we have that  $\e^{3} = 2\pi n/\beta$, with $n \in \mZ$. More generally, the $SU(2)$ symmetry could have a fugacity   which would imply that the field $g$ is no longer periodic; rather, it has $g(\beta) = z\, g(0)$ where $z \in SU(2)$ is the fugacity. Once again, since the partition function only depends on the conjugacy class of $z$, we can perform an $SU(2)$ transformation to diagonalize $z =  \exp(2\pi i \a \sigma_3 )$. The solution for $h$ is then given by $g = \exp\left(2\pi i \sigma_3 (n+\a) \frac{\tau}\beta\right)$. 

In such a case, the on-shell value of the action $I_{\cN=4,\text{bosonic}}$ is given by: 
\be 
\label{eq:on-shell-action-N=4}
I_{\cN=4,\text{bosonic}}^{\,\text{on-shell}} = -\frac{2\pi^2 \Phi_r} {\b} \left(1 -4  (n+\a) ^2 \right)\,,
\ee
Now that we have the  on-shell action we can proceed by computing the one-loop determinant which is sufficient for fully computing the partition function.

\paragraph{The one-loop determinant:} To compute the one-loop determinant, we have to account for all quadratic fluctuations in the theory. The quadratic fluctuations of the Schwarzian field have been analyzed in great detail \cite{Maldacena:2016upp, Stanford:2017thb}, and its contributions to the one-loop determinant is given, up to an overall proportionality constant, by
\be 
\label{eq:Schwarzian-one-loop-determinant}
\det_{\text{Schw., one-loop}} = \left(\frac{\Phi_r}{\beta}\right)^\frac{3}2\,.
\ee
The quadratic fluctuation around the saddle-point of the $SU(2)$-group element can be parametrized as $g(\tau) = \exp\left( \sigma_3 \left[2\pi i(n+\a) \frac{\tau}\beta + \epsilon^3(\tau) \right]\right)\exp\left( \epsilon^2(\tau) \sigma_2\right) \exp\left( \epsilon^1(\tau) \sigma_1\right)$, and yields a contribution to the action \cite{Picken:190160}
\be 
I_{SU(2), \text{ quad}} &=  \frac{8\pi^2 \Phi_r} {\b}  (n+\a) ^2\nn \\ &-2 \Phi_r \int_0^\b d\tau \left((\epsilon^1(\tau)')^2 + (\epsilon^2(\tau)')^2 - (\epsilon^3(\tau)')^2 + \frac{8\pi (n+\a)}{\b} \epsilon^2(\tau) \epsilon^1(\tau)'  \right)\,,
\ee
and the one-loop determinant obtained from integrating out these modes for each value of $n$ is given by \cite{Picken:190160}:
\be
\label{eq:SO3-one-loop-determinant}
 \det_{SU(2), \text{ one-loop}}  = \frac{\Phi_r^{3/2} (n+\a) }{ \beta^{3/2} \sin(2\pi \a)} \,.
\ee
Finally, we discuss the quadratic contribution of the fermionic fields. By using the saddle-point solution for $f(\tau)$ and $g(\tau)$ and by quadratically expanding the super-Schwarzian action:
\be
 I_{\text{ferm., quad.}} =  \Phi_r\int_0^{\beta} d\tau\, &\bigg(\eta^p \left[\frac{2\pi^2}{\beta^2}\left(1+2(n+\a)^2 \right) \partial_\tau - \partial_\tau^3\right]\bar \eta_p + \nn \\&+ \partial_\tau \eta^p \left[\frac{12\pi^2(n+\a)^2}{\b^2} + \frac{8i  \pi (n+\a)\partial_\tau}{\beta} - 3\partial_\tau^2 \right] \bar \eta_p\bigg) 
 \ee
Expanding  the fermionic fields in Fourier modes, 
\bea \eta^1(\tau) &=&e^{i\frac{2\pi (n+\a)\tau}{\b} }\sum_{m_1 \in \dots, -\frac{1}2, \, \frac{1}2, \dots} \sqrt{\frac{\b}{2\pi}} \,\eta^1_{m_1} e^{-i\frac{2\pi m_1\tau}{\b} }\\
\eta^2(\tau) &=&e^{-i\frac{2\pi (n+\a)\tau}{\b} }\sum_{m_2 \in \dots, -\frac{1}2, \, \frac{1}2, \dots} \sqrt{\frac{\b}{2\pi}}\, \eta^2_{m_2} e^{i\frac{2\pi m_2\tau}{\b}}\,,
\ea
where we impose anti-periodic boundary conditions for the fermionic fields when $\alpha= 0$ and we impose boundary conditions consistent with the introduction of the fugacity $z$ when $\alpha\neq 0$. We can then rewrite the action as, 
\be
 I_{\text{ferm., quad.}} = \frac{2\pi^2 i \Phi_r}{\beta} \left[\sum_{p=1, 2}\,\sum_{m_p  \in \dots, -\frac{1}2, \, \frac{1}2, \dots} (m_p - n - 	\a) (4m_p^2-1) \eta^p_{m_p}  \bar \eta^p_{-m_p}\right]\,.
\ee
We are interested in computing the dependence of the one-loop determinant on $n$, $\alpha$, $\beta$ and $\Phi_r$. The $\beta$ and $\Phi_r$ dependence is captured by the existence of the four-fermionic zero modes with $m_p = \pm 1/2$. As in \cite{Stanford:2017thb}, to compute the rest of the one-loop determinant, we will regularize this result by dividing the result by the one-loop determinant with $n=0$ and $\alpha = 0$. We thus find that the regularized one-loop determinant is given, again up to a proportionality constant, by 
\be
\label{eq:ferm-one-loop}
\det_{\text{ferm., one-loop}} = \frac{\beta^4}{\Phi_r^4} \prod_{p=1, 2}\,\,\prod_{m_p \in \dots, -\frac{5}2, -\frac{3}2, \frac{3}2, \frac{5}2, \dots} \frac{m-n -\a}{m} = \frac{\beta^4}{\Phi_r^4} \frac{\cos(\pi\a)^2}{(1 - 4(n+\a)^2)^2 }\,. 
\ee

\paragraph{Final answer:} Thus, accounting for the saddle-point value of the action \eqref{eq:on-shell-action-N=4} together with the one-loop determinants \eqref{eq:Schwarzian-one-loop-determinant}, \eqref{eq:SO3-one-loop-determinant}, and \eqref{eq:ferm-one-loop}, we find that the partition function of the $\cN=4$ Schwarzian theory is given, up to an overall proportionality constant, by: 
\be 
\label{eq:localization-part-function}
Z_{\cN=4\,\,\,\text{Schw.}}  &= \sum_{n \in \mathbb Z} \det_{\text{Schw., one-loop}} \,\,\det_{SU(2), \text{ one-loop}}\,    \det_{\text{ferm., one-loop}} e^{-I_{\cN=4,\text{bosonic}}^{\,\text{on-shell}} } \nn \\ &=\sum_{n \in \mathbb Z} \frac{ \beta \cot(\pi \a) (\a+n)}{\Phi_r (1- 4(n+\a)^2)^2} e^{\frac{2\pi^2 \Phi_r} {\b} \left(1 -4  (n+\a) ^2 \right)}
\ee 
We will thus continue by matching this result using the completely distinct method of canonical quantization, after which we will come back to a detailed analysis of the spectrum associated to \eqref{eq:localization-part-function} in section \ref{sec:density-of-states}. 

\subsubsection{Method 2: Canonical quantization} \label{sec:meth2cq}
In this section, we will compute the partition function of the $\cN=4$ super-Schwarzian theory using the canonical quantization approach of \cite{Mertens:2017mtv} (for a very recent discussion explaining the connection to the localization approach see also \cite{Alekseev:2020jja}). 

We will illustrate briefly the idea first. The localization formula we used above can be applied to integrals that generally have the following form 
\beq
Z = \int dqdp ~e^{-H(p,q)},
\eeq
where the integral is over a symplectic space (a classical phase space) with coordinates $(q,p)$ and $H(q,p)$ generates via the Poisson brackets a $U(1)$ symmetry. In the case of the bosonic Schwarzian theory the integration manifold is ${\rm Diff}(S^1)/SL(2,\mathbb{R})$ which a coadjoint orbit of the Virasoro group (and, therefore, symplectic), and $H(p,q)$ is the Schwarzian action. Instead of using localization, we can obtain this integral using the following identity 
\beq\label{eq:SKJDKD}
\lim_{\hbar\to 0} {\rm Tr}\left( e^{-H(p,q)}\right) = \int dqdp ~e^{-H(p,q)},
\eeq
where the left-hand side is $\hbar\to 0$ limit of the trace evaluated over the Hilbert space obtained by quantizing the phase space. In the case of the Schwarzian, the quantization of ${\rm Diff}(S^1)/SL(2,\mathbb{R})$ is the identity representation of the Virasoro algebra with central charge $c \sim 1/\hbar$. The left hand side of \eqref{eq:SKJDKD} can by very easily computed at finite $c$ as a Virasoro vacuum character by counting descendants, and a very simple calculation gives the Schwarzian path integral \cite{Mertens:2017mtv}. In the bosonic case, the main advantage of this method is the possibility to compute correlation functions which are not available using localization. In this case, we will use it as a double check on our previous result.

For the case of the $\cN=4$ super-Schwarzian the integration space is ${\rm Diff}(S^{1|4})/PSU(1,1|2)$ which is a coadjoint orbit of super-Virasoro and, therefore, also symplectic. We will assume that the quantization of this phase space, the Hilbert space in \eqref{eq:SKJDKD}, is equivalent to the identity representation of the small $\cN=4$ Virasoro algebra with central charge $c\sim 1/\hbar$. The $\cN=4$ super-Schwarzian partition function is then the semiclassical limit of the vacuum character.  

Lets begin then by recalling the super-Virasoro algebra involved in this problem. The bosonic generators are $L_n$ and $T_n^i$ where $n$ is an integer and $i=1,2,3$ label the generators of a Kac-Moody $SU(2)$ at level $k$. Their algebra is 
\bea
[ L_m,L_n ]&=& (m-n)L_{m+n}+\frac{k}{2} m(m^2-1)\delta_{n+m,0} \label{eq:smalln4gen1}\\
\text{[} T^i_m,T^j_n \text{]} &=& i \epsilon^{ijk}T^k_{m+n} + \frac{k}{2} m \delta_{m+n,0} \delta_{i,j} \label{eq:smalln4gen2}\\
\text{[} L_m , T_n^i\text{]}  & = & -n T_{m+n}^i.\label{eq:smalln4gen3}
\ea
The central charge of the bosonic Virasoro sector is $c=6k$, which is fixed by a Jacobi identity. The fermionic generators are $G_r^p$ and $\bar{G}_s^p$, $p=1,2$. They transform in the fundamental and antifundamental of the $SU(2)$. The Fourier mode parameter $r,s$ are integer in the Ramond sector or half-integer in the Neveu-Schwarz sector. The rest of the algebra, involving the fermionic generators, can be found for example in \cite{Eguchi:1987sm}, and is given by
\bea
&&\{ G_r^p, \bar{G}_s^q\} = 2 \delta^{pq} L_{r+s}-2(r-s)\sigma^i_{pq} T^{i}_{r+s}+\frac{k}{2}(4r^2-1)\delta_{r+s,0},\nn \\
&& [T_m^i,G_r^p]=-\frac{1}{2}\sigma^i_{pq}G^q_{m+r},~~[T^i_m,\bar{G}_r^p]=\frac{1}{2}\sigma^i_{pq}{}^\star \bar{G}_{m+r}^q,~~ \{ G_r^p, G_s^q\}=\{\bar{G}_r^p,\bar{G}_s^q\}=0\nn  \\
&&[L_m,G_r^p]=\left(\frac{m}{2}-r\right)G_{m+r}^p,~~[L_m,\bar{G}_r^p]=\left(\frac{m}{2}-r\right)\bar{G}_{m+r}^p,\label{eq:smalln4gen4}
\ea
where $\sigma^i_{pq}$ are the Pauli matrices. In that reference, Eguchi and Taormina also construct the unitary representations of the algebra.  

For the application we have in mind in this paper, the Schwarzian path integral, we will only need the massless representations in the NS sector, due to the fact that we want the Schwarzian fermions to be antiperiodic as explained in \cite{Mertens:2017mtv}.\footnote{If we wanted the Schwarzian theory Witten index we would use the characters in the Ramond sector.} General representations are labeled by $h$, the eigenvalue of $L_0$, and $\ell$, the spin of the $SU(2)$ representation, and the massless sector has $(h=\ell, \ell)$ with half-integer $\ell=0, \frac{1}{2}, \ldots, \frac{k}{2}$. For the Schwarzian path integral we will need the $\ell=0$ representation. The characters are defined by
\beq
\chi_{\ell} (k;q,z) \equiv {\rm Tr}_{NS}\left[ (-1)^F q^{L_0-\frac{c}{24}} z^{T^3_0}\right],
\eeq
over a representation $\ell$ of the algebra. We need to insert $(-1)^F$ such that the fermions along the quantization direction are periodic and survive the semiclassical limit (see discussion in \cite{Mertens:2017mtv}). 

These characters were computed by Eguchi and Taormina \cite{Eguchi:1987wf} by simply counting states. They are given by the following expression\footnote{The origin of the first factor in the right hand side is explained in section 5 of \cite{Kraus:2006nb}.}
\beq
\chi_{\ell} (k;q,z=e^{2\pi i y}) = e^{2\pi i k \frac{y^2}{\tau}} q^{-\frac{k}{4}}  q^{\ell+\frac{1}{4}} \frac{i \theta_3(q,-z) \theta_3(q,-z^{-1})}{\eta(q)^3 \theta_1(q,z^2)} \left[ \mu(z,q) - \mu(z^{-1},q) \right],
\eeq
where $\theta_3(q,z)$ is the Jacobi theta function and we defined the function 
\beq
\mu(z,q) \equiv \sum_{n\in\mathbb{Z}} \frac{q^{(k+1)n^2+(2\ell+1)n}z^{2(k+1)n + 2 \ell + 1}}{(1- z q^{n+\frac{1}{2}})(1- z q^{n+\frac{1}{2}})}.
\eeq

Now we have all the ingredients to extract the Schwarzian partition function from the $\hbar \sim 1/k \to 0$ ($c\to\infty$) limit applied to the above expression for the identity $\ell=0$ representation. As explained in \cite{Mertens:2017mtv} we need to consider the following scaling  
\beq
z=e^{2 \pi i \alpha \tau},~q=e^{2\pi i \tau}~~~~~{\rm with}~~~~\tau = \frac{i}{k} \frac{4\pi \Phi_r}{\beta}.
\eeq
We then take $\tau \cdot k$ fixed in the limit and this constant is related to the ratio $\Phi_r/\beta$ in the Schwarzian theory. This choice of $z$ and $q$ is written directly in terms of $\alpha$ and $\beta$ which will become the chemical potential and inverse temperature in the Schwarzian limit. 

When taking the Schwarzian limit we will only keep track of the dependence on $\alpha$ and $\beta$ since any prefactor can be absorbed in a redefinition of the zero-point entropy and energy. We will not go over all the details but some useful intermediate steps are 
\beq
\mu(z,q)-\mu(z^{-1},q) \sim \frac{8 e^{\frac{2\pi^2 \Phi_r}{\beta}4\alpha^2}}{\pi^2 |\tau|^2} \sum_{n\in \mathbb{Z}} \frac{(\alpha+n)}{(1-4(\alpha+n)^2)^2} e^{-\frac{2\pi^2 \Phi_r}{\beta} 4(n+\alpha)^2}.
\eeq
Using eq (3.15) of \cite{Ahn:2003tt} gives the following limit
\beq
i \left( \frac{\theta_4(q,z)}{\eta(q)^3}\right)^2 \frac{\eta(q)^3}{\theta_4(q,z^2q^{\frac{1}{2}})} \sim \frac{\tau}{\tan \pi \alpha},
\eeq
which is related in a simple way to the Jacobi theta functions appearing in the character. Including the rest of the terms the semiclassical $k\to\infty$ limit of the vacuum character is 
\beq
\chi_{\ell=0} (k\to\infty;q,z) \sim  \sum_{n\in \mathbb{Z}} \frac{\beta \cot(\pi \alpha)(\alpha+n)}{\Phi_r(1-4(n+\alpha)^2)^2} e^{\frac{2 \pi^2 \Phi_r}{\beta}(1-4(n+\alpha)^2)}.
\eeq
This precisely reproduces the partition function computed by localization, given in equation \eqref{eq:localization-part-function} (An analogous match was checked in \cite{Mertens:2017mtv} for the case of $\cN=1$ and $\cN=2$ super-Schwarzian).

We can mention some interesting features of this expression. First of all the factor of $\cot \pi \alpha$ is important for the formula to make sense. When $\alpha \to 0$ or $1$ it is crucial to include this factor for the final answer to be finite. The same happens when $\alpha \to 1/2$ since otherwise the sum would be divergent. 

From the 2D CFT perspective the identity representation is invariant under the generators of the global $PSU(1,1|2)$ algebra. In terms of the Virasoro algebra those generators are  
\beq
\text{Bosonic:}~~~~L_{-1},L_0,L_1,~~~~T_0^1,T_0^2,T_0^3,~~~~\text{Fermionic:}~~~~G_{\pm \frac{1}{2}}^a,~~~\bar{G}^a_{\pm \frac{1}{2}},
\eeq
which satisfy the same superalgebra as \eqref{eq:psu(1,1|2)-superalgebra}. 
It is important to take the fermionic generators in the NS sector. These produce the pre-factor of $\beta^1$ in the character. In the localization calculation this factor basically counts the number of bosonic and fermionic zero modes $Z \sim \beta^{(\#{\rm fermion})/2 - (\#{\rm bosons})/2}$. In the case of the small $\cN=4$ algebra there are $8$ fermionic zero modes and $6$ bosonic zero modes, giving a factor of $\beta$. 

\subsection{$\cN=4$ supermultiplets}

Before extracting the spectrum from the exact partition function we first explain what properties we expect it to have. The super-Schwarzian theory we are studying captures the explicit breaking of the superconformal symmetry group $PSU(1,1|2)$. Still, as we have seen in section \ref{sec:space-time-and-global-symm}, translations, super-translations and rigid $SU(2)$ rotations are symmetries. We can write the fermionic generators as $Q_p$ and $\bar{Q}^p$ with $p=1,2$. Then a part of the algebra that we will use here is 
\beq
\{ Q_p, \bar{Q}^q\} = 2 \delta_a^b H,~~~\{Q_p,Q_q\}=\{ \bar{Q}_p, \bar{Q}_q\} =0
\eeq
These generators can be written in terms of the Schwarzian fields but we will not need it for the manipulations here. Imagine we first diagonalize $H$ and look at some states with energy $E$. Then as long as $E\neq 0$ the operators $Q \sim \hat{a}$ act as a $SU(2)$ doublet of lowering fermionic operators and $\bar{Q}\sim \hat{a}^\dagger$ as a $SU(2)$ doublet of rising fermionic operators. To construct a representation we can begin with a state $|J\rangle$ which transforms as a spin $J$ representation of $SU(2)$, constructed such that $Q_p|J\rangle=0$. The supermultiplet will have states acting with a single charge $\bar{Q}^q|J\rangle$, which can be expanded into $(1/2)\otimes J = (J-1/2) \oplus (J+1/2)$; and acting with two charges $\bar{Q}_1 \bar{Q}_2 |J\rangle$ of spin $J$. Therefore, the supermultiplet with $E\neq 0$, starting with $J\neq 0$ is made of $(J-1/2)\oplus 2 (J) \oplus (J+1/2)$. When we construct a supermultiplet starting with a singlet $|0\rangle$, the $\bar{Q}^q|0\rangle$ transforms as a doublet and $\bar{Q}_1\bar{Q}_2|0\rangle$ as another singlet, giving $2(0)\oplus(1/2)$. Labeling the supermultiplet by the state with highest $SU(2)$ spin, the $E\neq 0 $ part of the spectrum should organize as 
\bea
\mathbf{J}&=&(J) \oplus 2 (J-1/2) \oplus (J-1),~~~~J\geq 1\\
\mathbf{1/2}&=&(1/2) \oplus 2 (0).
\ea
Finally we might also have states with $E=0$. Starting with a spin-$J$ representation $|J\rangle$, having $H|J\rangle =0$ implies that all supercharges annihilate the state and, therefore, that's the whole supermultiplet.

Taking these considerations into account, we can expect the partition function of the $\cN=4$ super-Schwarzian theory to be expanded in the following way
\bea
Z(\beta,\alpha) &=& \sum_J \chi_J(\alpha) \rho_{\rm ext}(J) + \int dE ~e^{-\beta E} \left( \chi_{1/2}(\alpha) +2\chi_{0}(\alpha)\right) \rho_{\rm cont}(1/2,E) \nonumber\\
&&+\sum_{J\geq 1} \int dE ~e^{-\beta E} \left( \chi_{J}(\alpha) +2\chi_{J-\frac{1}{2}}(\alpha)+ \chi_{J-1}(\alpha)\right) \rho_{\rm cont}(J,E),\label{sqwewq}
\ea
where the sums are over half-integer $J$ and $\chi_J(\alpha)\equiv \sum_{m=-J}^J e^{4\pi i \alpha m} = \frac{\sin (2J+1)2\pi \alpha}{\sin 2\pi \alpha}$ is the character of a spin-$J$ representation of $SU(2)$. In the first line, the first term corresponds to states with $E=0$ while the second term to the $E\neq 0$ multiplet $\mathbf{1/2}$. The second line corresponds to the sum over all other $E\neq 0$ supermultiplets. Therefore, $\rho(J,E)$ is the density of supermultiplets with energy $E\neq 0$ and highest spin $J$, while $\rho_{\rm ext}(J)$ is the density of $E=0$ states of spin $J$. 

We will see in the next section that the spectrum of the $\cN=4$ super-Schwarzian derived from the exact partition function we computed above has precisely this form (although with only singlet $J=0$ zero energy states).

\subsection{Exact density of states}
\label{sec:density-of-states}
The final answer for the exact $\mathcal{N}=4$ super-Schwarzian theory partition function is given by the following function of inverse temperature $\beta$ and $SU(2)$ chemical potential $\alpha$ as
\beq\label{exactPFN4d}
Z(\beta,\alpha) =e^{S_0} \sum_{n\in \mathbb{Z}} \frac{\beta}{\Phi_r} \frac{2 \cot(\pi \alpha)(\alpha+n)}{\pi^3(1-4(n+\alpha)^2)^2} e^{\frac{2 \pi^2 \Phi_r}{\beta}(1-4(n+\alpha)^2)}\,.
\eeq
We have fixed the overall normalization in a way that will be convenient later. We will write this answer as a trace over a Hilbert space (albeit with continuous spectrum) realizing it has precisely the form \eqref{sqwewq}.

To understand the physics of this partition function we want to extract the density of states as a function of energy at fixed $SU(2)$ charge, which we will refer to as angular momentum (anticipating the application to near extremal black holes in 4D). To do that we begin by performing an inverse Laplace transform and define the fixed-chemical-potential density of states 
\beq
Z(\beta,\alpha) = \int dE e^{-\beta E} D(\alpha, E).
\eeq
Applying this to our result \eqref{exactPFN4d} gives
\beq
D(\alpha,E) = D_{E=0}(\alpha) \delta(E) + D_{\rm cont}(\alpha,E),
\eeq
where we separate the BPS and continuous part of the spectrum,\footnote{The sum in $D_{E=0}(\alpha)$ is at face value divergent. To regulate it we used the following prescription $\lim_{N\to\infty} \sum_{n=-N}^N\frac{4(\alpha+n)}{\pi \tan \pi \alpha (1-4(\alpha+n)^2)} = 1$. We can verify that this is the correct prescription by checking that after integrating over energies, this gives back the original partition function.}
\bea
\label{toto}D_{E=0}(\alpha) &=&e^{S_0} \sum_{n\in\mathbb{Z}} \frac{4(\alpha+n)}{\pi \tan \pi \alpha (1-4(\alpha+n)^2)} = e^{S_0}\label{extD} \\\label{contD}
D_{\rm cont}(\alpha,E)&=&e^{S_0}\sum_{n\in \mathbb{Z}} \frac{4(\alpha+n)}{\pi \tan \pi \alpha} \frac{I_2\left(2\pi \sqrt{2 \Phi_r E(1-4(\alpha+n)^2)}\right)}{E(1-4(\alpha+n)^2)}
\ea
We see the first line corresponding only to states with zero energy is independent of $\alpha$. This means it only gets contributions from zero charge (angular momentum) states. We chose the normalization of the partition function such that this gives $\exp \left( S_0 \right)$ and can be interpreted as the degeneracy of ground states.

To find the density of states we use the following identity to rewrite \eqref{contD} as 
\beq\label{ble}
D_{\rm cont}(\alpha,E) =e^{S_0} \sum_{m=1}^\infty \frac{m \sin 2\pi m \alpha}{ \tan \pi \alpha} \frac{\sinh \left(2 \pi \sqrt{2\Phi_rE-\frac{1}{4}m^2}\right)}{2 \pi^2 \Phi_r E^2} \Theta\Big(E-\frac{m^2}{8\Phi_r}\Big).
\eeq
We defined the Heaviside function $\Theta(x)$ such that $\Theta(x>0)=1$ and $\Theta(x<0)=0$. The dependence with the chemical potential can be expanded in $SU(2)$ characters in the following simple way 
\bea\label{charactedID4}
\frac{2 \sin 2\pi m \alpha}{ \tan \pi \alpha} &=&  \chi_{J}(\alpha) +2\chi_{J-\frac{1}{2}}(\alpha)+ \chi_{J-1}(\alpha),~~~J\equiv m/2,~{\rm with}~m>1\\
\frac{2 \sin 2\pi \alpha}{ \tan \pi \alpha} &=&  \chi_{1/2}(\alpha) +2\chi_{0}(\alpha),~~\hspace{2.3cm}J\equiv m/2,~{\rm with}~m=1
\eea
where in the right hand side we defined the angular momentum $J$ in terms of the integer $m$. In principle we can use this formula to extract the density of states for each $SU(2)$ representation. Instead we will notice this is precisely the combination in equation \eqref{sqwewq}
where $J$ now labels the supermultiplet $\mathbf{J}$. The second line with $m=1$ involves the special case $\mathbf{1/2}$. Comparing \eqref{sqwewq} with \eqref{ble} we can extract the density of supermultiplets $\rho_{\rm cont}(J,E)$ for $E\neq0$ and using \eqref{extD} we can write the density of $E=0$ states $\rho_{\rm ext}(J)$. The final answer is given by
\bea
\rho_{\rm ext}(J) &=& e^{S_0} \delta_{J,0}.\label{sksks}\\
\rho_{\rm cont}(J,E) &=& 
\frac{e^{S_0}J}{4\pi^2 \Phi_r E^2}\sinh \left(2 \pi \sqrt{2\Phi_r(E-E_0(J))} \right)\hspace{0.1cM} \Theta\Big(E-E_0(J)\Big), \hspace{0.1cm}\text{for }J\geq\frac{1}{2},\label{sksks2}
\ea
 where the gap for each supermultiplet labeled by $J$ is given by $E_0(J) \equiv J^2/(2\Phi_r)$.
\begin{figure}
    \centering
     \begin{tikzpicture}[scale=0.65]
 \pgftext{\includegraphics[scale=0.5]{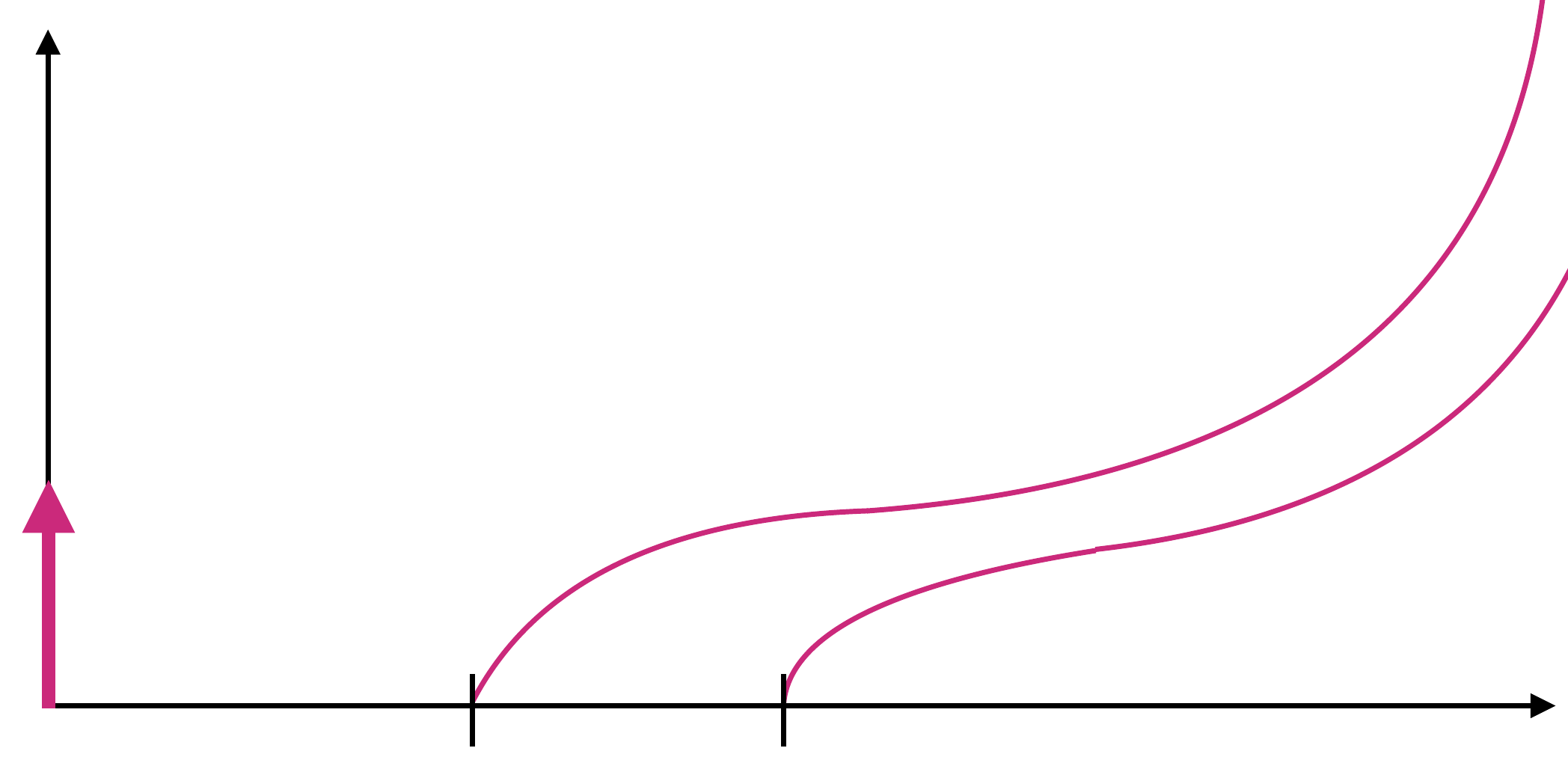}} at (0,0);
 \draw (-5.7,-0.4) node  {$e^{S_0}$};
   \draw (-5.8,2.75) node  {\small $\rho(E)$};
   \draw (-2,-3) node  {\small $E_{\rm gap}$};
   \draw (-0,-3) node  {\small $E_0(1)$};
   \draw (-0,-1.5) node  {\small $\mathbf{1}$};
   \draw (-2.3,-1.5) node  {\small $\mathbf{1/2}$};
   \draw (-4.6,-1.5) node  {\small $\mathbf{0}$};
   \draw (5,-3) node  {\small $E$};
  \end{tikzpicture}
  \hspace{0.6cm}
   \begin{tikzpicture}[scale=0.65]
 \pgftext{\includegraphics[scale=0.47]{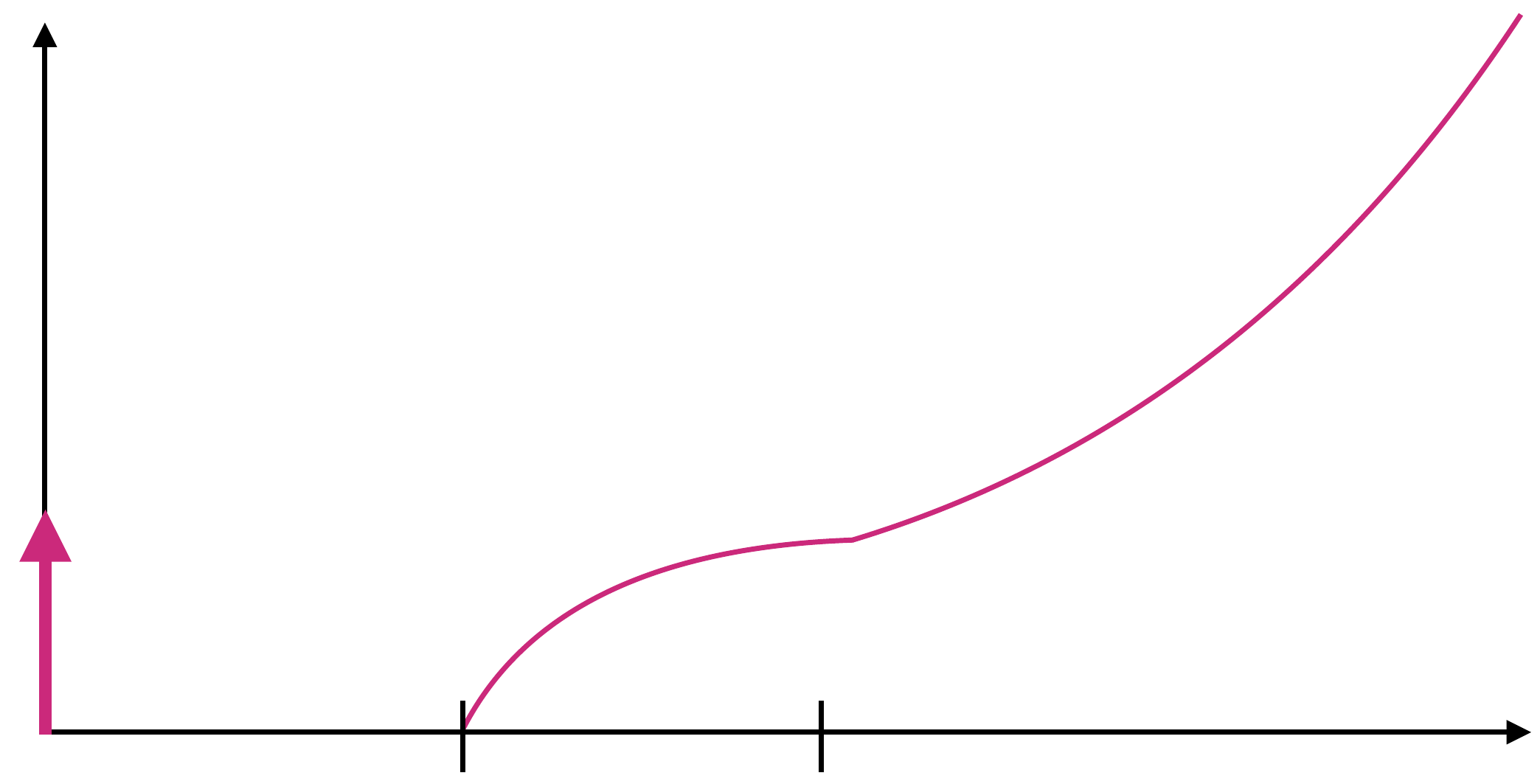}} at (0,0);
 \draw (-5.4,-0.4) node  {$e^{S_0}$};
   \draw (-5.4,2.75) node  {\small $\rho(E)$};
    \draw (-2,-3) node  {\small $E_{\rm gap}$};
   \draw (0.2,-3) node  {\small $E_0(1)$};
   \draw (5,-3) node  {\small $E$};
  \end{tikzpicture}
    \caption{\textbf{Left:} Density of supermultiplets labeled by the highest spin $\mathbf{J}$. We show $\mathbf{0}$, which is simply a delta function at $E=0$; $\mathbf{1/2}$ which is continuous but starts at $E_{\rm gap}\equiv E_0(1/2)$; and $\mathbf{1}$ which is also continuous starting at $E_0(1)$. \textbf{Right:} Degeneracy for all states with $J=0$. These come from $\mathbf{0}$, the delta function at $E=0$, $\mathbf{1/2}$, starting at $E_{\rm gap}$, and $\mathbf{1}$, starting at $E_0(1)$. All other supermultiplets do not have a $J=0$ component.}
    \label{fig:my_label}
\end{figure}
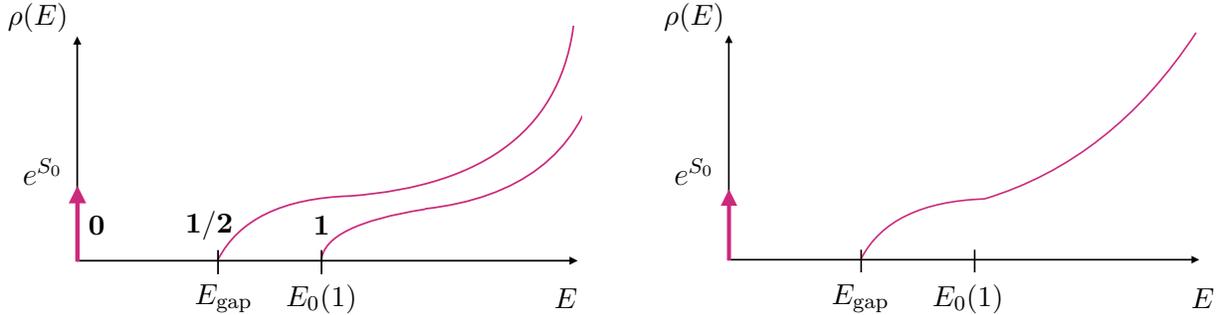

Using this result we can get a simple picture of the shape of the spectrum. First we have a number $e^{S_0}$ of states at exactly $E=0$ which are all in the supermultiplet $\mathbf{0}$, an $SU(2)$ singlet. These are the extremal BPS states of the black hole as we will see in the next section. For small energies there are no states until we reach the gap in the spectrum given by the threshold energy for the supermultiplet $\mathbf{\frac{1}{2}}=(\frac{1}{2})\oplus 2(0)$, given by 
\beq
E_{\rm gap} = \frac{1}{8 \Phi_r},
\eeq
and for $E>E_{\rm gap}$ we have a continuum of states. Something similar is true for higher multiplets $J > 1/2$, but now the continuum starts at a supermultiplet-dependent gap
\beq
E_{0}(J) = \frac{1}{2\Phi_r} J^2.
\eeq
It is perhaps not surprising that states with spin $J$ start at $E_{0}(J)$. The surprising feature is that there are no states with $J=0$ at energies $0<E<E_{\rm gap}$. 

In figure \ref{fig:my_label} we depict the shape of the spectrum of the $\cN=4$ super-Schwarzian theory. In the left panel we show the degeneracy of supermultiplets, with the supermultiplet-dependent gap mentioned above. In the right we show the degeneracy of states (not supermultiplets) with $J=0$ as an illustration. We see the delta function from $\mathbf{0}$ but also contributions from the continuum coming from $(0)\subset\mathbf{1/2}$ and $(0)\subset\mathbf{1}$.

For each supermultiplet, the density of states has a spectral edge $\rho_{\rm cont}(J,E) \sim (E-E_{0}(J))^{1/2}$ characteristic of a hermitian random matrix model, although we will not follow this direction in this paper.

We can make some comments regarding the calculation of the $\cN=4$ super-Schwarzian Witten index. First it is straightforward to show that the fermion number can be computed in the following way 
\beq
(-1)^F = e^{2 \pi i T^3}
\eeq
acting on a $SU(2)$ representation of spin $J$.\footnote{Here $T^3 = T^3_L + T^3_R$ corresponding to the left and right $SU(2)$ symmetries of the $\cN=4$ super-Schwarzian.} Using the density of states derived above it is easy to extract the index. First all states with $E=0$ have $J=0$ and, therefore, $(-1)^F=1$. States with $J>1/2$ contribute in the following way 
\beq
e^{2\pi i J} \left[\chi_J(\alpha) - 2 \chi_{J-1/2}(\alpha)+\chi_{J-1}(\alpha)\right].
\eeq
It is easy to see that when $\alpha=0$ this combination exactly vanishes since there is the same number of bosonic and fermionic states in the supermultiplet. The same is true for $\mathbf{1/2}$ which gives $\chi_{1/2}(\alpha)-2\chi_0(\alpha)$ and also vanishes for $\alpha=0$. Therefore, the Witten index of the $\cN=4$ super-Schwarzian theory is given by $e^{S_0}$ and counts the number of ground states.

As a final comment, there are two different definitions of the $\cN=2$ super-Schwarzian theory that differ on the presence of a 't Hooft anomaly, as we review in Appendix \ref{app:N2}. Since the gauge group of the $\cN=4$ super-Schwarzian is $SU(2)$ we think there cannot be such anomaly \cite{Kapec:2019ecr} and the theory is unique, but this deserves further investigation.

\subsection{Comparison with a pure bosonic theory} 
\label{sec:compbosth}
To finish this section we would like to compare this solution to a non-supersymmetric version of the theory, such as the one in \cite{Iliesiu:2020qvm}. Imagine we have a bosonic Schwarzian theory coupled to an $SU(2)$ mode. The action is 
\beq\label{ejksjw}
I =-\Phi_r \int d\tau \hspace{0.1cm} \Sch(f,\tau) + K \int d\tau  {\rm Tr} \left( g^{-1} \partial_\tau g\right)^2 ,
\eeq
where $K$ and $\Phi_r$ are independent parameters. This theory can be solved exactly \cite{Mertens:2019tcm}. The density of states as a function of energy and angular momentum $J$ is given by
\beq
\rho_{bos.}(J,E) = e^{S_0} \sinh \left( 2\pi \sqrt{2\Phi_r\left( E- E_{bos.}(J)\right)}\right) \Theta (E-E_{bos.}(J)),~~~E_{bos.}(J)\equiv\frac{J(J+1)}{2K}.
\eeq
The bosonic sector of the supersymmetric theory is special in two ways. First of all it necessarily has $K=\Phi_r$. Therefore, at least semiclassically one can compute the gap scale by measuring the following quantity 
\beq
\left( \frac{\partial J}{\partial \Omega} \right)_{T=0,\Omega=0}= K = \Phi_r,
\eeq
where $\Omega=i\alpha/\beta$ is the potential conjugated to $J$ (from 4D perspective, angular velocity). The second, and more important, feature is the fact that for the bosonic theory \eqref{ejksjw}, even if $K=\Phi_r$, there are states with $J=0$ for any energy $E>0$, namely $\rho(J=0,E>0)\neq 0$ and $\rho(J=0,E=0)=0$. The $\cN=4$ supersymmetric Schwarzian theory is completely different. We find a delta function at $E=0$ describing $e^{S_0}$ states. Moreover, even for $J=0$, there are no states in the range $0<E<\frac{1}{8\Phi_r}$. This is the surprising feature we are emphasizing in this paper.

\section{Near-BPS Black Holes in Flat Space}
\label{sec:higherD}

As discussed in the Introduction~\ref{sec:intro}, extremal Reissner-Nordstr\"om black holes are a basic object of study in supergravity and string theory because in many examples they preserve supersymmetry and are exact solutions of the theory. The BPS nature of such gravitational solutions allows one to identify them with a corresponding D-brane configuration; such brane models allow for a precise microstate counting \cite{Strominger:1996sh} which matches the result of the Bekenstein-Hawking entropy. While AdS$_2$ solutions of various supergravities are well studied, we focus on a minimalistic choice of black hole and AdS$_2 \times S^2$ solutions of pure, ungauged, 4D $\mathcal{N} = 2$ supergravity. As we will review, such a 4D theory already has the right ingredients for $\frac{1}{2}$-BPS extremal black holes. Additionally, the pure $\mathcal{N}=2$ supergravity theory will generically appear as a subsector of a more general matter coupled $\mathcal{N}=2$ supergravity~\cite{Freedman:2012zz}.

In this section, we will first recall some facts about 4D supergravity, then analyze the extremal AdS$_2 \times S^2$ black hole solution. This solution has ordinary isometries corresponding to the AdS and sphere factors, and additionally permits covariantly constant spinors which are solutions of the Killing spinor equations. As is known in the literature \cite{Kallosh:1997qw,Claus:1998yw,Boonstra:1998yu} the group of isometries for the near-horizon region enlarges, from the bosonic $SL(2, \mathbb R)$, to the supergroup $PSU(1,1|2)$ and we will review this in some detail. The appearance of this supergroup corresponds to the symmetries of the $\mathcal{N}=4$ JT and Schwarzian theories utilized in Section~\ref{sec:N=4superJT}.

In this section, we are really interested in black holes which parametrically approach their BPS limits (i.e.~that have temperature of the same order or smaller than the conjectured $E_\text{gap}$). In the effective AdS$_2$ near horizon dilaton supergravity, we allow fluctuations of the dilaton as well as Kaluza-Klein gauge fields and fermions. In the limit of a large black hole with small near BPS fluctuations, we will derive the effective 2D $\cN=4$ JT supergravity. This will allow us to match the microscopic and thermodynamic properties of 4D black holes with the output of the $\mathcal{N}=4$ Schwarzian from Section~\ref{sec:density-of-states}.

\subsection{4D $\mathcal{N}=2$ Supergravity}
\label{ssec:4dSugra}
We begin with some notions and notations for gravity in 4D. Some of these conventions will be different than those used in Section~\ref{sec:N=4superJT} since we will be working in four-dimensional notation. In the end we will explain how to translate the results to be compatible with the two-dimensional $BF$ theory.

We use a mostly plus Lorentzian metric, and will eventually Wick rotate to Euclidean signature in 2D. We denote 4D curved space indices $M,N = 0 \dots 3$ and the tangent space indices $A, B = 0, \dots 3$. The 4D metric $G_{MN} = \eta_{AB} E^A_M E^B_N$ is expressed in terms of the vierbein $E^A_M$. In 2D, the curved and tangent space indices are respectively $\mu, \nu = 0,1$ and $a,b = 0,1$, and we use the vielbein $e^a_\mu$ and dualized spin connection $\omega_\mu$. We will occasionally use indices for the internal $S^2$ space, which will take values $m,n = 2,3$ for the coordinate indices and $a',b' = 2,3$ for the frame indices. When we are working in the near-horizon region, with the AdS$_2$ $\times S^2$ product manifold, the vielbein will decompose as:
\begin{equation}
    E^A_M = e^a_\mu \, , \, \, e^{a'}_m \, .
\end{equation}
 We will generally suppress spinor indices when possible, but our spinor conventions are given in Appendix~(\ref{app:spinorssugra}).

Pure, ungauged $\mathcal{N}=2$ supergravity is a minimal theory containing only the graviton, $G_{MN}$; an $SU(2)_R$ doublet of gravitinos, $\Psi_M^I$, $I=1,2$; and a $U(1)$ gauge field $A_M$ under which a black hole solution is electrically or magnetically charged. Without the addition of extra vector or hypermultiplets, this gravity theory does not have a precise embedding in string theory.\footnote{As explained in \cite{Becker:2007zj}, compactification of either Type II theory on a Calabi-Yau threefold will always produce at least one $\mathcal{N}=2$ hypermultiplet along with the gravity multiplet discussed here. This follows from the reduction of the 10D graviton, dilaton, and 2-form.} Nevertheless, we argue that universal features of supersymmetric black holes are already present in this model, through the relationship with 2D Super-JT gravity.

We write the Lagrangian of pure 4D $\mathcal{N}=2$ supergravity, following the conventions of \cite{Freedman:2012zz}:
\begin{equation}
    E^{-1} \mathcal{L} = \kappa^{-2}\bigg(\frac{1}{2}R - \bar{\Psi}_{I M} \Gamma^{MNP}D_{N}\Psi^I_P - \frac{1}{4}F_{MN}F^{MN} + \frac{\varepsilon^{IJ}}{2\sqrt{2}}\bar{\Psi}^M_I(F_{MN} + i\, \star  F_{MN}\Gamma_5)\Psi_J^{N}  
    + \textrm{4 gravitino} \bigg)\, ,
    \label{Eq:4DCompLag}
\end{equation}
with the standard definitions $D_M = \partial_M + \frac{1}{4}\omega^{AB}_M \Gamma_{AB} $ and $ F= dA$. We also write the 4D Newton constant as $  8 \pi G_N = \kappa^2$. In the above action and throughout this subsection, we follow the convention that a spinor $\epsilon^I$ with an upper $SU(2)_R$ index has positive chirality, while a lower index has negative chirality:
\begin{equation}
    \Gamma_5 \epsilon^I = + \epsilon^I \, , \,\,\,\,\, \Gamma_5 \epsilon_I = - \epsilon_I \, .
    \label{Eq:chiralindex}
\end{equation}
In analyzing the supersymmetry and dimensional reduction later, we will often instead use Dirac spinors where the chirality will be implicit.

While we have explicitly displayed the Pauli terms in Eq.~\eqref{Eq:4DCompLag} proportional to the antisymmetric symbol $\varepsilon^{IJ}$, we have omitted the more complicated 4-gravitino term. The form of this term is known in general,\footnote{A general matter coupled $\mathcal{N}=2$ supergravity theory including additional fermion terms is found, for instance, in \cite{Andrianopoli:1996vr}.} but it is vanishing in the bosonic backgrounds corresponding to extremal black hole solutions. In the dimensional reduction to 2D, we will use other methods to fix the higher order fermion terms. With the understanding that some expressions are augmented by these terms higher order in fermions, Eq.~\eqref{Eq:4DCompLag} is invariant under the following local supersymmetry transformations:
\begin{align}
    \delta_{\epsilon} E^A_M &= \frac{1}{2}\bar{\epsilon}^I \Gamma^A \Psi_{MI} + \textrm{h.c.} \, , \\
    \delta_{\epsilon} A_M &= \frac{1}{\sqrt{2}} \varepsilon^{IJ} \bar{\epsilon}_I\Psi_{MJ} + \textrm{h.c.} \, , \\
    \delta_{\epsilon}\Psi^I_M &= (\partial_M + \frac{1}{4}\omega^{AB}_M \Gamma_{AB})\epsilon^I  - \frac{1}{4 \sqrt{2}}\Gamma^{AB}  F_{AB}\Gamma_M \varepsilon^{IJ}\epsilon_J \, . \label{eq:gravsusytransform}
\end{align}

The Lagrangian and supersymmetry transformations above make manifest the $SU(2)_R$ symmetry of $\mathcal{N}=2$ supergravity. In the application of this theory to nearly AdS$_2$ $\times$ S$^2$ black holes described in Section~\ref{sec:N=4superJT}, we considered a theory with $PSU(1,1|2)$ symmetry and bosonic subgroup $SL(2,\mathbb{R})\times SU(2)$. This latter $SU(2)$ is realized geometrically by the spatial rotations of the black hole, and is independent of the $SU(2)_R$ symmetry present in 4D. This feature is characteristic of the \emph{small} $\mathcal{N}=4$ superconformal algebra, which has the R-symmetry $SU(2) \times SU(2)_{outer}$. In the context of this paper, it is possible to show that the outer R-symmetry corresponds to the 4D $SU(2)$ global symmetry.\footnote{We thank E. Witten for emphasizing to us that the 4D $SU(2)$ global symmetry is only an approximate symmetry in supergravity; it should not be an exact symmetry of the near extremal black holes. Only the $SU(2)$ symmetry realized as rotations of the $S^2$ is a physical symmetry in a more complete model such as string theory.} Because the outer symmetry plays no role in our analysis, it will be convenient to use a formulation of $\mathcal{N}=2$ supergravity with Dirac rather than Majorana gravitinos. This will hide the outer $SU(2)$ symmetry, but simplify many formulas in what follows. This version of the theory is presented in \cite{Freedman:1976aw,Romans:1991nq}, and the passage to this formalism involves introduction of the Dirac spinors $\Psi_M = \Psi^1_M + i \Psi_M^2$.

\subsection{The Black Hole and AdS$_2 \times S^2$ Solutions}
\label{Sec:extremalBH}
 The extremal black hole of $\mathcal{N} =2$ supergravity is a standard solution of Einstein-Maxwell theory which preserves 4 real supercharges of the 8 total. This is in contrast to the near horizon limit of this solution, AdS$_2 \times S^2$, which preserves all supersymmetries. The supersymmetry enhancement~\cite{Kallosh:1997qw, Claus:1998yw} corresponds to the addition of conformal supercharges, and the AdS background should have a full superconformal set of isometries. In this section, we briefly review these two solutions, which are by now somewhat standard.

The bosonic part of our action using our normalization conventions from 
Eq.~\eqref{Eq:4DCompLag} is the typical Einstein-Maxwell action:
\begin{equation}
    S_{4D}^{\text{(bosonic)}} =  \frac{1}{\kappa^2} \int d^4x \sqrt{-G}\left (\frac{1}{2}R - \frac{1}{4}F_{MN}F^{MN} \right ) \, .
    \label{eq:4Dbosoniclag}
\end{equation}

Solving the equations of motion, one can obtain the metric and gauge field of the celebrated extremal Reissner–Nordstr\"om black hole solution, here given by: 
\begin{align}
    ds^2 &= -\left(\frac{r_0}{r} - 1\right)^{2}dt^2 + \left(\frac{r_0}{r} - 1\right)^{-2}dr^2 + r^2 d\Omega_2^2 \, , \label{eq:extremalRNmetric}\nn \\
    F_{tr} &= -\frac{\kappa Q}{\sqrt{4\pi} r^2} \, .
\end{align}
We also have the relationships between the extremal radius, mass, and charge, $r_0 = G_N M$ and $Q = \sqrt{r_0 M}$. For simplicity, we have chosen an electrically charged black hole and set the magnetic charge $P=0$, but the results of this section can be easily generalized to this case.
In terms of the charge which we are keeping fixed, we have chosen units such that the extremal mass and extremal Bekenstein-Hawking entropy are:\footnote{This choice is slightly non-standard, but will simplify some formulas later.}
\be 
\label{eq:mass-and-entropy-BH-SUGRA}
M_0 = Q/\sqrt{G_N}\,, \qquad \qquad S_0 = \frac{\pi r_0^2}{G_N}= \pi Q^2\,.
\ee

In addition to the familiar bosonic symmetries of the solution~\eqref{eq:extremalRNmetric} corresponding to translations and rotation, there are additional supersymmetries represented by covariantly constant spinors on the manifold. Since all fermions vanish on this background, we only need to analyze the gravitino supersymmetry transformation, Eq~\eqref{eq:gravsusytransform}, adapted to Dirac spinors. Preserved Killing spinors are spacetime dependent $\epsilon$ which are annihilated by the gravitino transformation rule:\footnote{A discussion of the relationship between supersymmetry preserving black branes, Killing spinors, and Killing vectors may be found in \cite{Gibbons:1982fy,Duff:1986hr,Lu:1998nu,Gauntlett:1998kc}.}
\begin{equation}
    \delta_{\epsilon}\Psi_M = \hat{D}_M \epsilon =  (\partial_M + \frac{1}{4}\omega^{AB}_M \Gamma_{AB})\epsilon  + \frac{i}{4 \sqrt{2}}\Gamma^{AB}  F_{AB}\Gamma_M \epsilon = 0\, .
    \label{eq:KillingSpinor4D}
\end{equation}
We will typically use Dirac spinors and do not make use of the chiral components.

The extremal black hole of Eq.~\eqref{eq:extremalRNmetric} can be thought of as a supergravity soliton which interpolates between two maximally supersymmetric vacuua; Minkowski space in the asymptotic far region, and AdS$_2 \times S^2$ in the near horizon region. Because we are interested in an effective AdS$_2$ theory approximated by supersymmetric JT gravity, we now turn our attention to the AdS$_2 \times S^2$ metric with the purely electric Bertotti-Robinson ansatz. 
Starting with Eq.~\eqref{eq:extremalRNmetric}, we perform standard manipulations to take the near horizon limit. We shift the location of the event horizon then rescale the radial coordinate with to get the Poincare patch metric:
\begin{align}
    ds^2 = \frac{r_0^2}{z^2}(-dt^2 + dz^2) + r_0^2 d \Omega_2^2 \, , \,\,\,\, F = \frac{1}{2} \frac{\kappa Q}{\sqrt{4\pi}}\frac{1}{z^2} dt \wedge dz\, .
    \label{eq:poinAdSmetric}
\end{align}

To study this background, including the Killing spinors, we make the further change of coordinates $z = r_0 e^{-r/r_0}$:
\begin{equation}
    ds^2 = -e^{2r/r_0}dt^2 + dr^2 + r_0^2 d \Omega_2^2 \, , \, \, \, \,
    F = -\frac{1}{2} \frac{\kappa Q}{\sqrt{4\pi} r_0^2}e^{r/r_0} dt \wedge dr\, .
\end{equation}
At the risk of notational clutter, we denote frame indices with a hat, and in frame coordinates, the bosonic fields are:
\begin{align}
    E^{\hat t} &= e^{r/r_0} dt \, , \,\,\,\,
    E^{\hat r} = dr \, , \,\,\,\,
    E^{\hat m} = (r_0 d \theta \, , \, \, r_0 \sin{\theta} d \phi) \, , \,\,\,\, 
    F_{\hat{t} \hat{r}} = - \frac{\kappa Q}{\sqrt{4\pi} r_0^2} = - \frac{\sqrt{2}}{r_0} \label{eq:ads2s2frames}\, . \\
        \omega^{\hat{t}\hat{r}} &= \frac{1}{r_0}E^{\hat t} = \frac{1}{r_0}e^{r/r_0} dt \, , \,\,\,\,
    \omega^{\hat{\phi} \hat{\theta}} = \omega^{\phi \theta}_{S^2} = \cos \theta d\phi \nn \, .
\end{align}
In the expression for the flux, we used the definitions for $\kappa$, $Q$, and $r_0$ in the extremal limit. The constant flux background is analogous to the Freund-Reubin ansatz~\cite{Freund:1980xh} in that the spin connection also has a $1/r_0$ behavior, and it is the cancellation between these two terms which will permit a maximally supersymmetric vacuum.

Before plugging the explicit background into Eq.~\eqref{eq:KillingSpinor4D}, it is useful to first examine the integrability condition for the supercovariant derivative with a general background acting on a Killing spinor. A necessary condition for the existence of Killing spinors is
\begin{equation}
    [\hat{D}_M,\hat{D}_{N}]\epsilon = 0 \, .
\end{equation}
Computing this in the case of pure ungauged $\mathcal{N}=2$ supergravity \cite{Romans:1991nq} gives
\begin{equation}
    \left ((R_{MN}^{\, \, AB}\Gamma_{AB} - \frac{1}{8} \slashed{F}\Gamma_{[M}\slashed{F}\Gamma_{N]}) +\frac{i}{\sqrt{2}}\Gamma_{AB}\Gamma_{[N}(\nabla_{M]}F^{AB}) \right )\epsilon = 0 \, .
\end{equation}
It is possible to further simplify the Gamma matrix contractions for a general flux background, but the case of AdS$_2 \times S^2$ has further simplifications as the third term vanishes identically. Allowing $a,b = 0,1$;  $a',b' = 2,3$ as frame indices and $\mu,\nu$;  $m,n$ as coordinate indices for AdS and the sphere respectively, we obtain the familiar conditions:
\begin{align}
     \big ( R_{\mu \nu}^{\,\, ab}\gamma_{ab} &+ \frac{1}{r_0^2}(e^a_\mu e^b_\nu - e^a_\nu e^b_\mu) \gamma_{ab}  \big )\epsilon = 0 \, , \nn \\
        \big ( R_{m n}^{\,\, a'b'}\gamma_{a'b'} &- \frac{1}{r_0^2}(e^{a'}_m e^{b'}_n - e^{a'}_n e^{b'}_m) \gamma_{a'b'}  \big ) \epsilon = 0 \, .
\end{align}
These are identically satisfied by the Bertotti-Robinson metric as the AdS and sphere are maximally symmetric. We do not obtain any new algebraic conditions on the Killing spinors. These conditions, as well as the form of the spinors themselves, must be determined from the first order equation, Eq.~\eqref{eq:KillingSpinor4D}. Inserting $F_{ab} = -\frac{\sqrt{2}}{r_0}\epsilon_{ab}$ into this Killing spinor equation, we find that it splits into a pair of equations for the AdS and sphere parts:
\begin{align}
(\partial_\mu + \frac{1}{4}\omega^{ab}_\mu \gamma_{ab}   - \frac{i}{2 r_0}\gamma_3 \gamma_\mu )&\epsilon = 0 \, , \label{eq:adsKSequation} \\ 
(\partial_m + \frac{1}{4}\omega^{a'b'}_m \gamma_{a'b'}  - \frac{i}{2 r_0}\gamma_m  )&\epsilon = 0 \, , \label{eq:sphereKSequation}
\end{align}
where $\gamma_3 = - \gamma_0 \gamma_1 = \sigma_3 \otimes \mathbf{1}$ is the pseudo-chirality operator on the AdS space. These can be obtained by applying the conventions of Appendix \ref{app:spinorssugra}. We observe that these equations are of the same form studied in \cite{Lu:1998nu}, and we will give an explicit construction of the Killing spinors.

To find expressions for the Killing spinors, we make a slight abuse of notation in writing $\epsilon_{4D} = \epsilon \otimes \eta$. We will see that the complex 2 component $\epsilon$ and $\eta$ are Killing spinors on the AdS$_2$ and S$^2$ factors. Inserting the expressions of Eq.~\eqref{eq:ads2s2frames}, we find the four Killing spinor equations,
\begin{align}
&(\partial_t - \frac{1}{2r_0}e^{r/r_0} \gamma_{3} (1+ i \gamma_0)) \epsilon = 0 \, ,\nn \\ 
&(\partial_r + \frac{i}{2 r_0} \gamma_0 ) \epsilon = 0 \, , \nn\\ 
&(\partial_\theta - \frac{i}{2}\sigma_1 )  \eta = 0 \, ,\nn\\
&(\partial_\phi - \frac{1}{2}\cos\theta \sigma_{13} - \frac{i}{2}\sin \theta \sigma_3 ) \eta = 0 \, .
\end{align}
To solve these, we first introduce constant anticommuting Dirac spinors $\epsilon^{0}_{\pm}$ and $\eta^\alpha$. Here the $\alpha$, $\beta$ $= 1,2$ spinor indices are those for the internal $SU(2)$ on the sphere; this contrasts with the notation of Section~\ref{sec:N=4superJT}, where these indices were denoted by $p,q$.

The $\epsilon^0_\pm$ are eigenstates under the supersymmetric projector:
\begin{equation}
\label{eq:bpsprojector}
        \epsilon^{0}_{\pm} = \mathcal{P}_\pm \epsilon^{0}_{\pm} = \frac{1}{2}(1 \pm  i \gamma_0 ) \epsilon^{0}_{\pm} \, .
\end{equation}
A convenient basis for the $\eta^\alpha$ is the one of definite chirality on the S$^2$. From Appendix~\ref{app:spinorssugra}, we see that we may choose $\eta^\alpha = (\eta^1, \eta^2)$ with $\sigma_2 \eta^1 =  \eta^1$ and $\sigma_2 \eta^2 = - \eta^2$. For clarity, this explicit basis choice amounts to the familiar eigenstates: $\epsilon^{0}_{+} = (1, i)$, $\epsilon^{0}_{-} = (1, -i)\,$, $\eta^1 = (1, i)\,$, $\eta^2 = (1, -i)$.

With these projections, we find the AdS$_2$ $\times$ S$^2$ solutions to the Killing spinor equations~\eqref{eq:KillingSpinor4D} are:
\begin{equation}
\epsilon^\alpha_- \equiv e^{\frac{r}{2r_0}} e^{\frac{i \theta}{2} \sigma_1} e^{\frac{\phi}{2}  \sigma_{13}} (\epsilon^0_{-} \otimes \eta^\alpha )\, .
\label{eq:KS-spinor}
\end{equation}
\begin{equation}
\epsilon^\alpha_+ \equiv  \left (1 + \frac{t e^{r/r_0}}{r_0} \gamma_3 \right ) e^{-\frac{r}{2r_0}} e^{\frac{i \theta}{2} \sigma_1} e^{\frac{\phi}{2}  \sigma_{13}} (\epsilon^0_{+} \otimes \eta^\alpha )\, .
\label{eq:KS+spinor}
\end{equation}
Similar expressions hold for the Dirac conjugate $\bar{\epsilon}^\alpha_\pm$, which we will need to form Killing vectors.\footnote{Some of our conventions for Killing spinors may be different than elsewhere in the literature. For more, see \cite{Fujii:1985bg,Lu:1998nu,Banerjee:2010qc,Banerjee:2011jp}.}

The super-isometry algebra of Killing vectors and spinors is typically determined by forming bilinears of spinors and checking that these correspond to the bosonic symmetries of the metric. Knowledge of the Killing spinors may be used to construct spacetime supercharges as integrals of the supercurrent~\cite{Gauntlett:1998kc}; for our purposes it is sufficient to check that all bosonic isometries are generated by Killing spinor bilinears. This allows us to loosely interpret $\epsilon^\alpha_- \sim G^\alpha_{-1/2}$ and $\epsilon^\alpha_+ \sim G^\alpha_{1/2}$, and we will compute the Killing spinor analog of $\{\bar{G}^\alpha_{\pm}, G^\beta_{\pm'}\}$ which were the anticommutators appearing in~\eqref{eq:smalln4gen4}. Therefore, we consider objects of the form,
\begin{equation}
    \bar{\epsilon}^\alpha_{\pm}\Gamma^M \epsilon^\beta_{\pm'} E_M \, ,
    \label{Eq:KillvecSpin}
\end{equation}
which transforms as a vector field. In fact, if we form bilinears from Killing spinors, this result will automatically be a Killing vector. A typical bilinear of the form Eq.~\eqref{Eq:KillvecSpin} may vanish depending on the commutation of  $\mathcal{P}_\pm$ with the various gamma matrices. 

To simplify the formulas, we will choose a unit normalization for the $\eta^\alpha$ and $\epsilon^0_\pm$ which solve Eq.~\eqref{eq:bpsprojector}, but in general one would find constant spinor contractions appearing on the right hand side of the formulas to follow. As shown in Appendix~\ref{ss:PSUblackhole}, the Killing vectors determined by \eqref{Eq:KillvecSpin},\eqref{eq:KS-spinor}, \eqref{eq:KS+spinor} are:
\begin{align}
    H &= \partial_t \, , \,\,\,\, D = t \partial_t + z\partial_z \, , \, \, \, \, \,K = (z^2 +t^2)\partial_t + 2 tz \partial_z \, , \\
    T_1 &= \sin \phi \partial_\theta + \cot \theta \cos \phi \partial_\phi \, , \,  T_2 =  \cos \phi \partial_\theta - \cot \theta \sin \phi \partial_\phi \, , \, \, T_3 =  \partial_\phi \, .
    \label{eq:4dS2Killing}
\end{align}
which form the $SL(2)\times SU(2)$ algebra, i.e.~the bosonic part of the supergroup $PSU(1,1|2)$. This motivates the conclusion that the near horizon theory should have this symmetry.

\subsection{Dimensional Reduction}
\label{ss:dimensionalreduction4d}

Our goal is to now demonstrate that the effective 2D gravity theory describing the near extremal perturbations of Eq.~\eqref{eq:poinAdSmetric} ultimately reduces to the supersymmetric completion of JT gravity described in Section~\ref{sec:N=4superJT}. We will start with a convenient Kaluza-Klein ansatz for 4D black holes with a dilaton $\chi$ capturing the fluctuating horizon area~\cite{Duff:1986hr,Gibbons:2003gp}. Without the inclusion of extra massless fields, this ansatz does not generally satisfy all the higher dimensional Einstein equations~\cite{Cvetic:2000dm}. However, as in the non-supersymmetric case~\cite{Iliesiu:2020qvm}, for sufficiently small masses and spins above extremality, the reduced effective 2D theory does capture the higher dimensional near extremal black holes.

In our discussion of the Kaluza-Klein ansatz, distorted 4D quantities will be hatted. We begin with the distorted metric, written in terms of coordinates $x^M = (x^\mu, y^m)$:
\begin{equation}
    d\hat{s}^2_{4D} = \frac{r_0}{\chi^{1/2}}g_{\mu \nu}dx^\mu dx^\nu + \chi \, h_{mn}(dy^m + T^m_{i} B^{i}_{\mu}dx^{\mu})(dy^n + T^n_{j} B^{j}_{\nu}dx^{\nu}) \, ,
\end{equation}
with an arbitrary 2D metric $g_{\mu \nu}$ and the unit $S^2$ metric:
\begin{equation}
    h_{mn}dy^mdy^n = d\theta^2 + \sin^2 \theta d\phi^2 \, .
\end{equation}
We parametrize the angular fluctuations with an $SU(2)$ gauge field $B^i_\mu$, which transforms as a $\mathbf{3}$ of $SO(3)$ and enters through the S$^2$ Killing vectors~\eqref{eq:4dS2Killing}.\footnote{In components, these obey the following identities in terms of the scalar harmonics $\mu_{i}$~\cite{Gibbons:2003gp}:
\begin{align}
    \mu_1 = \sin\theta \cos\phi \, , \, \, \, \mu_2 = \sin\theta \sin\phi   \, , \, \, \, \mu_3 = \cos\theta  \, , \, \, \, 
    T_{i}^{a'} = \epsilon^{a'b'}\partial_{b'} \mu_{i} \, , \,\,\,
    \delta_{a'b'} T^{a'}_{i} T^{b'}_{j}&= \delta_{i j} - \mu_{i} \mu_{j} \, .\nn
\end{align}
}

The vierbein associated with this decomposition is
\begin{equation}
\label{Eq:Vielbeinreduction}
    E_{M}^A dX^M \rightarrow \hat{e}^a = \frac{r_0^{\frac{1}2}}{\chi^{\frac{1}4}} e_{\mu}^a dx^{\mu} \textrm{, }\, \,  \hat{e}^{a'} = \chi^{\frac{1}2} e_{m}^{a'}(dy^m + T^{m}_{i} B^{i}_{\mu}dx^{\mu}) \, ,
\end{equation}
with the metrics for the arbitrary 2D theory and the explicit metric of S$^2$:
\begin{equation}
    g_{\mu \nu} = \eta_{ab} e_{\mu}^a e_{\nu}^b\, \textrm{, } \, \, \, \,  h_{mn}= \delta_{a'b'} e_{m}^{a'} e_{n}^{b'}
\end{equation}

The dimensional reduction of the Einstein-Hilbert term as well as the gravitino kinetic term both require the spin connection compatible with the vierbeins above. Even though the torsion tensor is not in general zero in supergravity due to fermions, for now we are free to use the torsion free first Cartan structure equations:\footnote{Here, we use a hat notation to distinguish the modified vielbeins and connections from those corresponding to their 2D symmetric space versions, which satisfy:
\begin{align}
d e^a + \omega^{ab}\wedge e_b = 0\, , \qquad
 d e^{a'} + \omega^{a'b'}\wedge e_{b'} =0 \, .\nn
\end{align}}
\begin{align}
    d \hat{E}^A &+ \hat{\omega}^{AB}\wedge \hat{E}_B = 0 \, .
\end{align}
We will ultimately pass to the first order formulation of gravity in the effective AdS$_2$ theory. This first order theory does not automatically enforce the torsion free constraint, so we will only use the 4D distorted spin connection as an intermediate step. The expressions for the distorted spin connection, Ricci tensors and Ricci scalar are given in Appendix~\ref{app:spinorssugra}.

We may now write the reduced gravitational action explicitly after dropping a total derivative on $\chi$:\footnote{Once we pass to the first order formalism in terms of the vielbein and spin connection, it is no longer generally true that integration by parts is valid due to the presence of torsion. In this dimensional reduction, all derivative terms on the dilaton $\chi$ are true total derivatives, independent of torsion.}
\begin{align} 
S_{2D,\text{ metric}} = \frac{2 \pi}{\kappa^2} \int \! \! d^2x \sqrt{-g} \left (\chi R + \frac{2r_0}{\sqrt{\chi}} -\frac{1}{6} \frac{\chi^{\frac{5}2}}{r_0} H^{i}_{ab}H_{i}^{ab} \right ) \, .
\end{align}

Our ansatz for $F$ in the distorted case is a term which generalizes the constant flux background by the inclusion of the dilaton; in local coordinates it is 
\begin{equation}
    \hat{F}_{ab} = - \sqrt{2} \frac{r_0}{\chi} \epsilon_{ab} \, .
\end{equation}
As we will explain shortly, fixing the field strength in the near-horizon region is well motivated and we do not need to be concerned with quantum fluctuations of the $U(1)$ gauge field around this saddle. Thus, such a mode will not contribute to the Euclidean path integral when studying black holes in the canonical ensemble.

Once we expand around the near extremal background, this will have the effect of introducing a cosmological constant term. Additionally, we will perform the Wick rotation to Euclidean time, which will give the gravitational path integral an interpretation as a statistical partition function. In total, the reduction of the 4D bosonic Euclidean action becomes:
\begin{align}
S_{2D}^{\text{(bosonic)}} = -\frac{2 \pi}{\kappa^2} \int \! \! d^2x \sqrt{g} \left (\chi R + \frac{2r_0}{\chi^{1/2}} -\frac{2r_0^3}{\chi^{3/2}} -\frac{1}{3} \frac{\chi^{5/2}}{r_0} \tr_{SU(2)} H_{ab}H^{ab} \right )  \, .
\end{align}
This is the bosonic part of a more complicated 2d dilaton supergravity, see for example \cite{Grumiller:2007ju}. This model is difficult to quantize directly, so we must make use of the near extremal condition in order to make contact with an extension of the JT model. To do this, we expand the dilaton to first order around the extremal squared radius, with a multiple of $G_N$ to obtain a convenient normalization:
\begin{equation}
   \frac{ \chi(x)}{G_N} = \frac{r_0^2}{G_N} + 2  \Phi(x) + \mathcal{O}\left(\frac{G_N \Phi^2}{r_0^2}\right)\,.
\end{equation}
To this order, the action becomes
\begin{align}
S_{2D}^{\text{(bosonic)}} = -\int \! \! d^2x \sqrt{g} \left[\frac{r_0^2}{4G_N} R -\frac{1}{12} \frac{r_0^4}{G_N} \tr_{SU(2)}H_{ab}H^{ab} + \frac{\Phi}2\left(R + \frac{2}{r_0^2} -\frac{5}{6} r_0^2\,\tr_{SU(2)}H_{ab}H^{ab} \right)\right]  \, .
\end{align}
The first term does not involve the dilaton, and is proportional to the Euler characteristic once we restore the Gibbons-Hawking-York boundary term containing the extrinsic curvature. This term thus sets the extremal entropy, and this will remain true independently of how we supersymmetrize the model. Thus, we find:
\begin{equation}
    S_0 = \frac{\pi r_0 ^2}{ G_N}=\pi Q^2 \, ,
\end{equation}
which is precisely the extremal entropy of Eq.~\eqref{eq:mass-and-entropy-BH-SUGRA}.

To make contact with the bosonic part of the BF description presented in section \ref{sec:PSU(1,1|2)-BF-thy}, we must introduce Lagrange multiplier fields for this gauge field. This is a scalar field $b_i$ transforming in the adjoint of $SU(2)$; we will additionally look for a form of the BF coupling $i b_i H^i$. In a large $r_0 $ expansion, this part of the action becomes:
\begin{equation}
    S_{2D}^{\text{(bosonic)}} \supset -i \int \,   \tr_{SU(2)}b H  + \int \! \! d^2x \sqrt{g}\, \frac{3G_N}{2r_0 ^4} \,\left (1 + \mathcal{O}\left(\frac{G_N \Phi}{r_0 ^2}\right) \right ) b_i b^i \, .
\end{equation}
We see the quadratic terms in $b_i$ are suppressed by large powers of the extremal radius, so in the large mass near extremal limit ($r_0 \gg 1/G_N^{1/2}$), the effective description is the unit normalized BF theory for $SU(2)$.\footnote{One may worry that in integrating out the gauge field, the dilaton coupling leads to a divergent 1-loop determinant. There are various ways of addressing this, we refer the reader to \cite{Iliesiu:2019lfc}.} Consequently, the dimensionally reduced bosonic part of the action is
\begin{align}
\label{eq:2D-bosonic-action}
S_{2D}^{\text{(bosonic)}} = S_0 - \frac{1}{2} \int \! \! d^2x \sqrt{g} \, \Phi \left (R + \frac{2}{r_0 ^2}\right ) -  i \int   \tr_{SU(2)}b H \, + \cO\left(\frac{G_N}{r_0^2}\right).
\end{align}
This has the form of the standard JT bulk term plus a decoupled BF term. 
  This matches the fact that the extremal AdS$_2$ $\times$ S$^2$ metric possesses $SL(2,\mathbb{R})$ $\times$ $SU(2)$ as bosonic symmetries. Of course, to the action \eqref{eq:2D-bosonic-action}, one in principle has to add the GHY term and the boundary term for the $SU(2)$ gauge field. Nevertheless, since the correct boundary term was derived in directly in the first order formalism in section \ref{sec:PSU(1,1|2)-BF-thy}, we will solely discuss the matching of bulk terms in this section.

  Our next goal is to restore the appropriate fermionic terms to the 2D action. 
  Kaluza-Klein supergravity reductions on spheres are important for the AdS/CFT correspondence, but performing this procedure for the full, consistent, nonlinear supergravity is somewhat more involved than what we will attempt here.\footnote{One typically needs extra massless fields in higher dimensions, transgression terms in the KK ansatz, and higher order terms in fermions. For a review of some of the developments on this front, see \cite{Duff:1986hr,deWit:1986oxb,Nastase:1999kf,Cvetic:2000dm}.} Rather than performing a full Kaluza-Klein reduction, we will instead derive only the terms that are linear (or less than linear) in the dilaton $\Phi$. Higher powers of $\Phi$ will come suppressed by factors of $G_N/r_0^2$ (i.e.~similar to the higher powers in $\Phi$ in \eqref{eq:2D-bosonic-action}) and can thus be neglected for sufficiently large black holes. 
. 

The basic fermion which must be included is the 2D gravitino. To ultimately match $PSU(1,1|2)$ BF theory, which naturally uses complex fermions, we will use the Dirac formulation of $\mathcal{N}=2$ supergravity as discussed above. After dimensional reduction on $S^2$, the Dirac gravitinos $\psi_\mu^{\alpha}$, $\bar{\psi}_{\mu \alpha}$ will transform in the $\mathbf{2}$ and $\bar{\mathbf{2}}$ representations of the $SU(2)$ symmetry, which in this section, is parametrized by the spinor index $\alpha=1,2$. The form of the gravitinos follows from the standard Kaluza Klein ansatz $\Psi_\mu \sim \psi_\mu^{\alpha}\eta_\alpha$, where $\eta_\alpha$ is a Dirac spinor which solves the angular equation \eqref{eq:sphereKSequation}, i.e.~just the $S^2$ Killing spinor equation. 

The angular solution is given by taking only the angular part of~\eqref{eq:KS-spinor} and \eqref{eq:KS+spinor} in terms of constant $\eta^\alpha$ spinors. A similar relation holds for the dilatino $\lambda^\alpha$, $\bar{\lambda}_\alpha$ which come from the $S^2$ vector components of the gravitino.\footnote{Some details of this are given for the asymptotically AdS$_4$ case in~\cite{Forste:2020xwx}. }

The standard gravitino kinetic term $\bar{\psi}_{\mu \alpha} \gamma^{\mu \nu \rho}D_\nu \psi_\rho^\alpha$ vanishes in two dimensions. This is related to the Einstein-Hilbert term $R$ being a topological invariant, and the supersymmetric variation of this is automatically zero. The basic term arising from dimensional reduction after choosing the correct normalization for fermions and dropping higher order terms in $r_0 $ and the dilaton fields is:
\begin{equation}
   S_{2D} \supset 2i \int d^2x \, \varepsilon^{\mu \nu} {\lambda}^\alpha \mathcal{D}_\mu \bar \psi_{\nu\alpha} \, , + \textrm{ h.c.}
\end{equation}
for $\mathcal{D}_\mu$ a suitable gauge supercovariant derivative whose form we will address directly in the first-order formalism.\footnote{Such a term is natural in the generalization of JT theory to include a supercurvature~\cite{Teitelboim:1983uy, Chamseddine:1991fg,Cardenas:2018krd}.} The next fermionic term, arising from the dimensional reduction, couples the dilaton $\Phi$ to the 2D gravitinos $\psi$ and $\bar \psi$. Such a term comes from the 4D terms $- \bar{\Psi}_{I M} \Gamma^{MNP}D_{N}\Psi^I_P + \frac{\varepsilon^{IJ}}{2\sqrt{2}}\bar{\Psi}^M_I(F_{MN} + i \star \! F_{MN}\Gamma_5)\Psi_J^{N}$ in the full supergravity Lagrangian \eqref{Eq:4DCompLag}. After passing to the Dirac spinors $\Psi_M = \Psi_M^1 + i \Psi_M^2$, the resulting term in the 2D action is 
\be 
 S_{2D} \supset \int  \frac{2}{r_0}\Phi \, (\bar \psi_\a  \wedge \psi^\alpha) \, , 
\ee
with all higher powers in $\Phi$ suppressed by factors of $G_N^{1/2}/r_0$. The final term that involves quadratic fermionic terms couples $\psi$ and $\bar \psi$ to the $SU(2)$ field strength $H$ arising from the KK reduction. Such a term arises from the dimensional reduction of  $- \bar{\Psi}_{I M} \Gamma^{MNP}D_{N}\Psi^I_P$ by using the dependence of the 4D spin connection on the field strength $H$ (see appendix \ref{app:spinorssugra}), again dropping higher order terms. After integrating-in the zero form field $b$, the 2D gravitinos then couple to $b$ instead of to the field strength $H$, resulting in the term
\be 
 S_{2D} \supset i  \int \frac{b^i}{r_0} \bar \psi_\a (\gamma_3') (\sigma^i)^\a{}_\b \wedge \psi^\b 
\ee
In principle integrating-out $b$ generates a four gravitino term, which is suppressed in $r_0$ and can be absorbed in a shift of the four gravitino term which we have only schematically written in the action \eqref{Eq:4DCompLag}. Because such higher power fermionic terms are suppressed by factors of $G_N^{1/2}/r_0$ we will not be concerned with them when computing the linearized gravitational action.

We can now put  together the bosonic and fermionic terms discussed above. In passing to the first order formulation, $\omega^{ab}_\mu$ will be promoted to a dynamical variable. Additionally, we will introduce standard 1-form notation for all fields, $e^a = e^a_\mu dx^\mu$, etc. We can now also now move to working in Euclidean signature, for which we will use the appropriate gamma matrices $\gamma'^a$, as outlined in Appendix~\ref{app:spinorssugra}. In terms of these, the first order bulk action (which will soon need to be modified) is: 
\be
\label{eq:reducedbadaction}
S_{2D} &\supset - \int \Phi \left ( d\omega + \frac{1}{2r_0 ^2} \epsilon_{ab}e^a \wedge e^b - \frac{2}{r_0} \bar \psi_\a  \wedge \psi^\a \right )+\frac{i}{2} b_{i}\left(H^{i} - \frac{2}r_0 \bar \psi_\a (\gamma_3')(\sigma^i)^\a_\b \wedge \psi^\b \right) \nn \\ &- 2i {\lambda} \mathcal{D}  \bar\psi  + \textrm{ h.c.} \,,
\ee
up to orders of $\cO(G_N/r_0^2)$ which we have neglected. Here, the supercovariant gauge exterior derivative is
\begin{equation}
\mathcal{D} \psi^\alpha=  d \psi^\alpha + \frac{1}{2}\omega \gamma_3' \wedge \psi^\alpha  - \frac{1}{2 r_0 } \gamma_3' \gamma_a' e^a \wedge \psi^\alpha + i B^i (\sigma_i)^\alpha{}_\beta \wedge \psi^\beta \, .
\label{eq:2dsuperexterior}
\end{equation}
The problem with the action~\eqref{eq:reducedbadaction} is that there is nothing to ensure the vanishing of the torsion tensor which has bosonic part $\tau^a = de^a + \varepsilon^{ab}\omega \wedge e_b $. In deriving Eq.~\eqref{eq:4DRicci}, this was implicitly assumed. Additionally, in supergravity, it is only the supertorsion which vanishes. This will be remedied by introducing extra Lagrange multipliers to fix the supertorsion constraint explicitly. Therefore, we can add to the action an additional term $\int \phi_a \tau^a$, where $\phi_a$ are new scalars. 

The various 1-forms and scalars can now be collected into super-multiplets as:
\begin{equation}
\label{eq:2d-SUGRA-supermultiplets}
    (\omega \, , \, e^a  \, , \, B^i \, , \, \psi^{\alpha} \, , \bar{\psi}_{\alpha})\, \, \, , \, \, \,  (\Phi \, , \, \phi^a \, , \, b^i \, , \, \lambda^\alpha \, , \, \bar{\lambda}_\alpha) \, ,
\end{equation}
which we recognize as the matter content of equations~\eqref{eq:gauge-field-ansatz} and~\eqref{eq:Lagr-multiplier}, respectively. 

Thus, after adding the super-torsion terms, the total linearized (in the dilaton) action arising from the dimensional reduction is:\footnote{Alternatively, one can obtain \eqref{eq:final-action-from-dim-red-nhr-2d} by restoring the $\cN=4$ supersymmetry for the 2D bosonic action obtained in \eqref{eq:2D-bosonic-action}. }
\begin{align}
\label{eq:final-action-from-dim-red-nhr-2d}
S_{2D} = - \int &\Phi \left ( d\omega + \frac{1}{2r_0 ^2} \varepsilon_{ab} e^a \wedge e^b -\frac{2}{r_0 }\bar{\psi}_\alpha  \wedge \psi^\a \right ) \nonumber \\ &+i \frac{\phi_a}{r_0} \left (d e^a + \varepsilon^{ab} \omega \wedge e_b - 2\,\bar{\psi}_\alpha \gamma'^a \wedge  \psi^{\a} \right )\nonumber \\ & +\frac{i}{2} b_{i}\left(H^{i} -\frac{2}{r_0}\bar{\psi}_\alpha (\gamma_3') (\sigma^i)^{\alpha}{}_\beta \wedge \psi^\beta \right ) - \left[2i \lambda_\a \mathcal{D} \bar \psi^\a  + \textrm{ h.c.}\right] \,
\end{align}
where the index $\alpha$ is an $SU(2)$ index and the fermionic indices are contracted with the Euclidean gamma matrices $\gamma_a'$ defined in Appendix~ \ref{app:spinorssugra}. 

Comparing  \eqref{eq:final-action-from-dim-red-nhr-2d} to the $\cN=4$ JT gravity action \eqref{eq:JT-N=4-action} with $\Lambda = 2/r_0^2$, one finds an exact match after some simple field redefinitions having to do with the reality of dilaton and the choice of basis for the gamma matrices.\footnote{In order to impose that the $PSU(1, 1|2)$ connection is flat the field $\Phi$ should be integrated along an imaginary contour, thus identifying $\Phi= i\phi_0$. 
} 
Before proceeding to obtain the full partition function of such black holes by using the results from section \ref{sec:N=4-super-Schw}, we first address two subtleties about the dimensional reduction: how to obtain the boundary conditions presented in section \ref{sec:PSU(1,1|2)-BF-thy}, and why other KK modes (besides the massless KK modes we have included thus far) do not contribute to the partition function in a significant way.

 \subsection{Subtleties about the dimensional reduction}
  \label{sec:subtleties}
  
 Before presenting the final results regarding the near-extremal spectrum of black holes in these supergravity theories, we will mention some subtleties and clarifications about the dimensional reduction. 
 
 \subsubsection*{Boundary conditions}
 
  This paper aims to perform the path integral in the 4D geometry by separately analyzing the near-horizon contribution and the far away (asymptotically flat) region. The far away contribution is trivial, but importantly propagating the boundary conditions from infinity to the edge of the throat, we can derive the boundary conditions we should impose on the fields living in $AdS_2$ \cite{Nayak:2018qej, Moitra:2019bub}. We will argue that the boundary conditions derived in this way are the same as the one used in section \ref{sec:N=4superJT} to analyze $\cN=4$ super-JT.
 
 This exercise has been done for the metric appearing in JT gravity and for the dilaton. We will follow the presentation of \cite{Iliesiu:2020qvm}. The throat and far-away regions are separated by an arbitrary curve with fixed dilaton value $\chi_b$, fixed intrinsic boundary metric $h_{\tau\tau}=1/\varepsilon^2$ and proper length $\ell = \int \sqrt{h}= \beta L_2 / \varepsilon$, where $L_2$ is the AdS$_2$ radius. Curves of constant dilaton are fixed at some radial distance $r_0 + \delta r_{\rm bdy}$, with $r_{\rm bdy}=L_2^2/\varepsilon$. Looking at the extremal solution in the far-away region we can derive the boundary condition for the dilaton $\Phi_b = \Phi_r/\varepsilon$ with $$\Phi_r = \frac{r_0 L_2^2}{G_N}.$$ In our case the AdS$_2$ radius is $L_2=r_0$. The definition of $\Phi_r$ through the dilaton boundary condition agrees with the one used in section \ref{sec:N=4superJT}, and this calculation relates it to 4D quantities.

Next, we can analyze the $SU(2)$ gauge field boundary condition. We will derive this indirectly in the following way, ignoring for now its coupling to other fields. In \cite{Iliesiu:2020qvm} we integrated out the 2D gauge field everywhere. We can see from the exact solution which boundary condition should be chosen in the throat in order to reproduce the angular-momentum dependence of the partition function. This gives the mixed boundary condition between the zero-form $b^i$ and the field strength $B^i$: $\delta(2i\Phi_r B_\tau - b)\big|_{\rm NHR bdy} = 0 $. To prove this, we can consider the classical solution for the $SU(2)$ gauge field, in the region far-away from the horizon. We fix the holonomy of the gauge field $h = \cP \exp \left(\oint_{r\to\infty} B\right) = \exp(i 2\pi \alpha\sigma^3) $ at the asymptotic boundary, with $\alpha \sim \alpha +1$.   Following appendix A in \cite{Iliesiu:2020qvm}, we will work in a gauge where $B_r = 0$, to obtain the general solution for the zero-form field $b$ and the gauge field $B$:
\be 
B= i\frac{2\pi \alpha T^3}{\b} \left(1+\frac{\mC}{r^3}\right) d\tau\,, \qquad H_{r\tau} =-i\frac{6\pi \alpha T^3}{\b} \frac{\mC}{r^4}\,, \qquad b = \frac{4\pi \mC \,\alpha\, T^3}{\b}\,,
\ee
where $\mC$ is an undetermined constant. Solving for $\mC$, at the boundary that separates the near-horizon region from the asymptotic region we find that $B_\tau$ and $b$ are related as
\be
\label{eq:SU(2)-gauge-field-sol}
B_\tau  =\frac{ 2\pi  i\alpha T^3}{\b} + \frac{i b \,\lpl^2\,}{r^3}= \frac{ 2\pi  i \alpha T^3}{\b} + \frac{i b }{2\Phi_r}
\ee
Thus, the boundary condition which we want to choose for the  $SU(2)$ field is
\be
\label{eq:bc-SU(2)-gauge-field}
\delta\left(2\Phi_r i  B_\tau - b \right)= 0\,,
\ee
which is consistent with the boundary condition in $\cN=4$ super-JT gravity discussed in section \ref{sec:super-JT-bdy-cond}, given that $\tau \sim \tau+\beta$ in both the diffeomorphism gauges of \eqref{eq:FG-gauge-metric} and for the Euclidean black hole metric. From a 4D perspective, the boundary holonomy of the $SU(2)$ gauge field is proportional to the boundary angular velocity $ \Omega \sim i \alpha/ \beta$.

Next, we mention why we could fix the flux of the $U(1)$ gauge field in the near-horizon region when working in the canonical ensemble. In such a case, the field strength (and not the gauge field itself) is fixed as $r\to\infty$. Once again, we can study how this happens by solving the equations of motion in the far-away region. The general solution takes the same form as \eqref{eq:SU(2)-gauge-field-sol} for non-abelian $SU(2)$ gauge fields; however, since we now fix the field strength at $r \to \infty$ the constant $\mC$ in  \eqref{eq:SU(2)-gauge-field-sol}  is also fixed. Therefore, the field strength is completely fixed at $r = r|_{\rm NHR bdy}$, instead of fixing the linear combination \eqref{eq:bc-SU(2)-gauge-field} as in the case of the $SU(2)$ gauge field. Since no massless field in the gravity, sector is charged under the $U(1)$ gauge field, one can integrate-out the 2D gauge field exactly and work in a sector of fixed charge. Since there is no temperature-dependent one-loop determinant when integrating out this field, the computation simply amounts to replacing the field strength in the near-horizon region with its classical values. Thus, this motivates our dimensional reduction in section  \eqref{ss:dimensionalreduction4d}, where the $U(1)$ gauge field was absent from all supermultiplets.

Finally, we will choose boundary conditions in the asymptotically flat space region for the fermions to vanish. Then it is reasonable to impose the same boundary condition in the boundary of the throat. This is also consistent with the choice made in section \ref{sec:super-JT-bdy-cond}.

  \subsubsection*{Massive KK modes}

Performing the reduction in the previous sections, we have identified the massless Kaluza-Klein modes in the spectrum when reducing the supergravity sector of the 4D theory to 2D. We have ignored so far the contribution from towers of massive KK modes or other matter fields present in 4D. The reason is that they do not affect the temperature dependence of the partition function to leading order \cite{Iliesiu:2020qvm}. 

The first step to see this is to realize that in the throat, all interactions are suppressed by factors of $S_0$. This means that in $AdS_2$ besides having the JT mode, we have a tower of free fields to integrate out. Let's begin with the case of massless fields in 2D, which arise from an s-wave reduction of a massless field in 4D. We can couple it to JT and integrate it out. The answer has the following form
\beq\label{eq:intoutout}
\log Z_{m=0} = \delta E_0 \beta + \delta S_0 + \mathcal{O}(\varepsilon, \beta^{-2}),
\eeq
where $\delta E_0 \sim 1/\varepsilon$ and $\delta S_0 \sim \log L_2$. The shift in energy is divergent and can be removed by a counterterm. The parameter $\delta S_0$ simply shifts the overall prefactor of the partition function $e^{S_0}\to e^{S_0 + \delta S_0}$. Since $L_2 \sim r_0$ this correction is logarithmic in the extremal area. The terms of order $\varepsilon$ include couplings between matter and gravity, through the Schwarzian mode \cite{Yang:2018gdb}. They are multiplied by a factor of the cut-off $\varepsilon$ and therefore suppressed. Any other correction is further suppressed at low temperatures. The same conclusions are true for massive fields, and their contributions have the same structure as \eqref{eq:intoutout}, producing only a shift of $S_0$. Finally, we can ask what happens when we integrate out a tower of massive KK modes coming from dimensionally reducing an individual field in 4D. This has been studied extensively by Sen and collaborators \cite{Banerjee:2010qc,Banerjee:2011jp,Sen:2011ba,Sen:2012cj}. The answer always has the form \eqref{eq:intoutout} and, in general, $\delta S_0 \sim \log S_0$ with a prefactor depending on the 4D matter spectrum.    

The only exception to this rule are dimensional reductions of extra gauge fields in 4D. They reduce to gauge fields in 2D that can affect the temperature dependence in general. Nevertheless, one can focus on boundary condition of fixed charge for all these fields. This choice makes their contribution trivial, and temperature independent.\footnote{An explanation of this is given in \cite{Iliesiu:2020qvm}). }

 \subsection{The black hole spectrum}
 \label{sec:4D-BH-spectrum}
 
We can now put all results together. We argued the temperature dependence of the partition function of the near-extremal black hole is captured by $\cN=4$ super-JT gravity, and in the previous sections, we precisely solved this theory. By matching boundary conditions, we have found the parameters of the JT theory in terms of their 4D origin 
\beq
S_0 = \pi Q^2 ,~~~~~\Phi_r = \sqrt{G_N} \hspace{0.1cm} Q^3.
\eeq
We are looking at the spectrum at fixed $U(1)$ charge and, therefore, these are fixed parameters. To extract the near-extremal black hole spectrum, we can use this identification in equations \eqref{sksks} and \eqref{sksks2}. In these expressions, the parameter $J$ is interpreted as the black hole angular momentum in 4D.  

Since this spectrum was already analyzed in the introduction, so we will not repeat it here. The main features we want to point out are the following. If we look at states with zero angular momentum, we find a large extremal degeneracy and the presence of a mass gap in the black hole spectrum, given by
\beq
\text{Degeneracy of extremal BH}=e^{\pi Q^2},~~~~E_{\rm gap}=\frac{1}{8 \sqrt{G_N}\hspace{0.1cm} Q^3}.
\eeq
Moreover, our analysis in section \ref{sec:density-of-states} shows that all extremal states are bosonic, and therefore a calculation of the index would match with the black hole degeneracy.
 \begin{figure}
     \centering
     \includegraphics[scale=0.65]{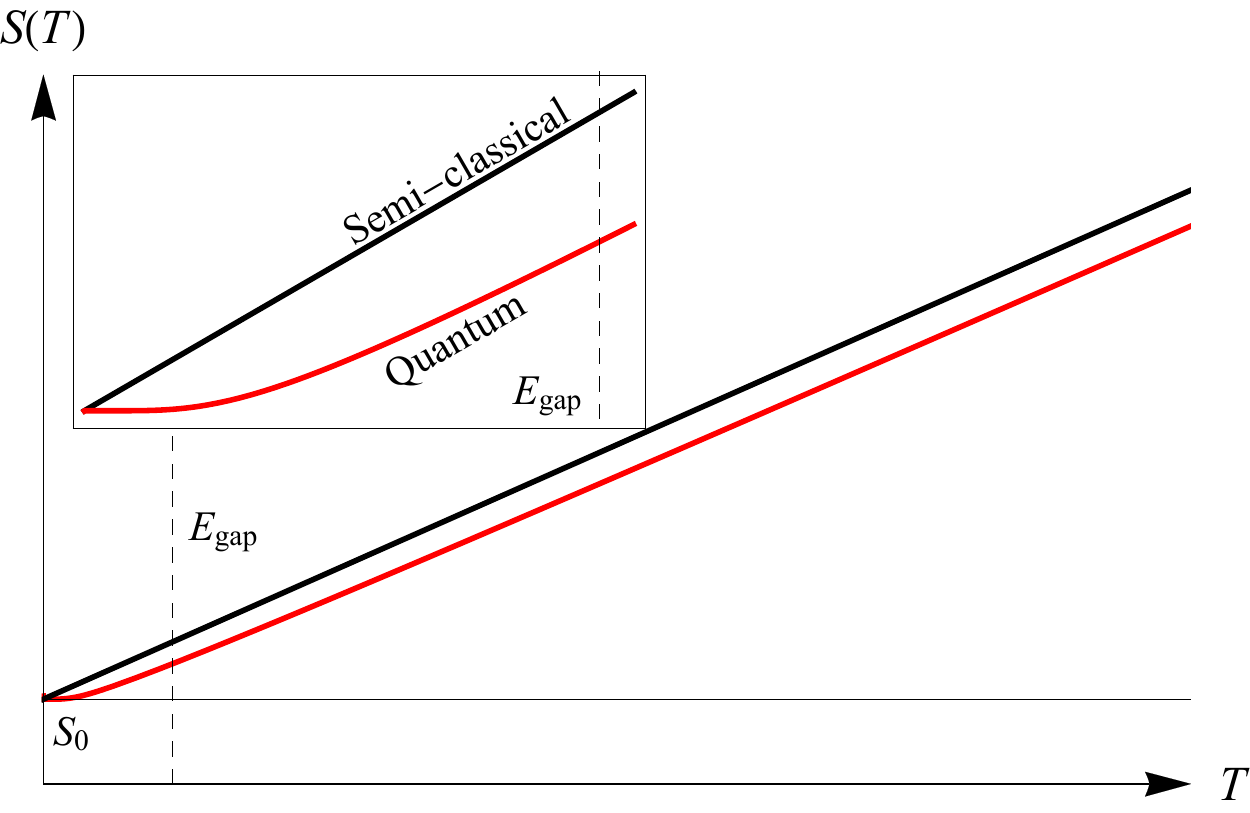}\hspace{0.15cm}
     \includegraphics[scale=0.65]{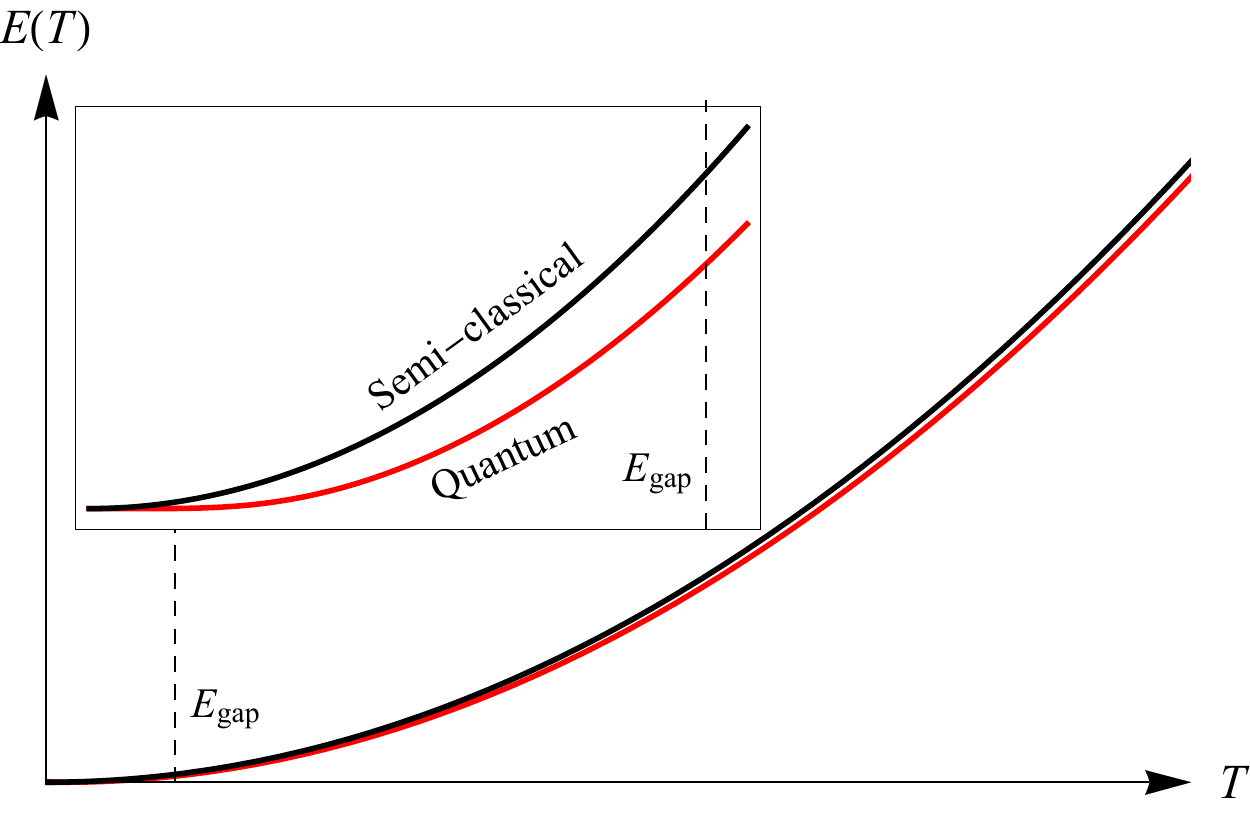}
     \caption{Plot of the entropy and energy dependence on temperature for near-BPS black holes in $\cN=2$ ungauged supergravity. The black curves represent results obtained from the naive semiclassical computation while the red curves account for quantum corrections. The inset figures zoom in on the temperature range smaller than $E_\text{gap}$ where quantum corrections become important. Both the entropy and energy approach zero temperature as $\# + \# e^{-E_\text{gap}/T}$ while in the semiclassical analysis, they approach zero temperature as $\# T$ and $\# T^2$, respectively.  }
     \label{fig:SPlot}
 \end{figure}
 
We can also look at other angular momentum sectors. The correction to the extremal energy we find when we turn on rotation is given by $$E_{\rm ext}(J)=\frac{J^2}{2\Phi_r}= \frac{J^2}{2\sqrt{G_N}\hspace{0.1cm}Q^3} \, ,$$ which is measured with respect to the $J=0$ extremal mass $M_0 = Q/\sqrt{G_N}$. This expression matches the expectation from 4D gravity: if we start with the Kerr-Newman metric, take the extremal limit with $Q\neq 0$ $J\neq 0$, we obtain precisely this correction when expanded to leading order in small $J$. Interestingly, we find the delta function giving the large extremal degeneracy at $J=0$ disappears at $J\neq0$. This makes sense since extremal black holes with $J\neq0$ do not preserve any supersymmetry, and they resemble the results in non-supersymmetry gravity in \cite{Iliesiu:2020qvm}.

Another interesting aspect of the theory is that the supersymmetry in the throat relates the heat capacity $\Phi_r$ of the black hole to a very different quantity $\left(\frac{\partial J}{\partial \Omega} \right)_{T=0,\Omega=0}=\Phi_r$, where $\Omega$ is the black hole angular velocity. In non-supersymmetric theories, these are independent quantities. Moreover, by measuring the dependence of the angular momentum with the angular velocity at zero temperature, we can also determine the gap in the spectrum (since it is controlled by the heat capacity $\Phi_r$). We comment on this in section \ref{sec:compbosth} as well.

Finally, if we decide to fix the metric in the asymptotically flat region and fix the angular velocity to vanish, it is equivalent to fixing the $SU(2)$ chemical potential to zero. With this choice we can use the exact results \eqref{sksks} and \eqref{sksks2} for the black hole spectrum to analyze how quantum corrections affect $S(T)$ and $E(T)$ when $\alpha=0$. The results are shown in figure \ref{fig:SPlot}. At large temperatures, the results above match the semiclassical result. At low temperatures, both the energy and the entropy behave exponentially as $E(T) \sim e^{-1/(8T  \Phi_r)}$ and $S(T) \sim S_0 + e^{-1/(8 T  \Phi_r)}$. This behavior is dictated by the fact that there is an actual gap in the spectrum.

\section{Near-BPS Black Holes in AdS$_3$}\label{sec:ads3}

In this section, we will describe near-extremal near-BPS black holes in theories of supergravity in $AdS_3$. We show, in some cases, the theory again reduces to $\cN=4$ super-JT near the horizon. In those cases, gravity predicts a large extremal entropy and a gap. We will mention concrete examples in string theory where this is relevant, and both the extremal states and the gap can be accounted for using D-brane techniques. 

\subsection{$(4,4)$ Supergravity in $AdS_3$}
We will summarize features of supergravity in $AdS_3$ and compute the partition function with a torus boundary. We will focus on the case with $\cN=(4,4)$ supersymmetry. These cases appear in string theory constructions that we will describe in the next section. We will also mention some features of theories with $\cN=(4,0)$ in the end.

When these theories are obtained from a dimensional reduction of six-dimensional supergravity on $AdS_3 \times S^3$ we have a specific spectrum of matter and gauge fields coupled to gravity. In the near extremal limit, we already saw that as long as we pick boundary conditions with fixed charges for gauge fields, both them and the matter fields do not affect the partition function's temperature dependence \cite{Iliesiu:2020qvm}. Therefore, in this section, we will focus on the pure supergravity sector on $AdS_3$. We will also give an argument for this fact near the end combining $AdS_3/CFT_2$ with the modular bootstrap (which is a supersymmetric version of \cite{Ghosh:2019rcj}).

Theories of supergravity in $AdS_3$ appearing in dimensional reductions of string theory can be written as a difference between two Chern-Simons theories \cite{Achucarro:1987vz}. We will follow the presentation in \cite{deBoer:1998kjm}. Generally for a set of left and right moving symmetry groups $G_L \times G_R $ we can define a gravity theory through the following action
\beq
I_{G_L \times G_R}= \frac{k}{4\pi}\int \left( \mathcal{A} \wedge d\mathcal{A} + \frac{2}{3} \mathcal{A} \wedge \mathcal{A} \wedge \mathcal{A} \right)-\frac{k}{4\pi} \int  \left(\bar{\mathcal{A}} \wedge d\bar{\mathcal{A}} + \frac{2}{3} \bar{\mathcal{A}} \wedge \bar{\mathcal{A}} \wedge \bar{\mathcal{A}} \right)
\eeq
where $\mathcal{A}$ is a connection for $G_L$ and $\bar{\mathcal{A}}$ a connection for $G_R$. We will begin studying the $\cN=(4,4)$ case, which corresponds to $G_L \times G_R= PSU(1,1|2)_L \times PSU(1,1|2)_R$. Expanding the gauge connection essentially involves two copies of the expansion performed in section \ref{sec:PSU(1,1|2)-BF-thy} for BF theory. 

The field content is a three-dimensional metric, a set of gravitini, and a $SU(2)_L \times SU(2)_R$ Chern-Simons gauge field with level $k$. The action in terms of these fields can be found adapting the analysis of \cite{Nishimura:1998ud}, for example. The metric part of the action in second-order formalism can be written as 
\beq
I_{G_L\times G_R} \supset \frac{1}{16 \pi G_3}\int d^3 x \sqrt{g}\left( R+\frac{2}{\ell^2}\right),
\eeq
where $\ell$ is the $AdS_3$ radius and $G_3$ the three-dimensional Newton constant. The Chern-Simons level is given by $k=\ell/(4G_3)$. The fact that this is also a level of $SU(2)$ implies $k$ is quantized.

To study this theory in asymptotically $AdS_3$ we consider a supersymmetric generalization of the Brown-Henneaux boundary conditions \cite{Brown:1986nw}. The asymptotic symmetries in the $(4,4)$ case are given by two copies, left and right moving, of the small $\cN=4$ Virasoro algebra. The generators of the bosonic Virasoro algebra are $L_n$, with integer $n$. The generators of the fermionic symmetries are $G_r^a$ where $a=1,2$ transforms in a double of $SU(2)$ and $r$ can be an integer (R-sector) or half-integer (NS-sector). Finally there is an $SU(2)$ current algebra at level $k$ generated by $T_n^i$, with integer $n$ and $i=1,2,3$ in the adjoint of $SU(2)$. The commutation relations were already written down above in equations \eqref{eq:smalln4gen1}\eqref{eq:smalln4gen2}\eqref{eq:smalln4gen3}\eqref{eq:smalln4gen4} and we will not repeat them here. The central charge of the algebra is $c$, and it is related through supersymmetry to the level $k$ of the $SU(2)$ algebra by $c=6k$. 

We will be interested in computing the partition function on three dimensional surfaces with torus boundaries
\beq
ds^2 = d\tau^2 + d\varphi^2, ~~~~(\tau,\varphi) \sim(\tau,\varphi+2\pi)\sim (\tau+\beta,\varphi+\theta). 
\eeq
We take the moduli of the torus to be $q=e^{2\pi i \tau}$ and $\bar{q}=e^{-2\pi i \bar{\tau}}$, where
\beq
\tau=\frac{\theta+i\beta}{2\pi}= \frac{i \beta_L}{2\pi}~~~~{\rm and}~~~~\bar{\tau}=\frac{\theta-i\beta}{2\pi}= -\frac{i \beta_R}{2\pi},
\eeq 
and it will be useful to think in terms of left and right moving temperatures. We also consider fixing the boundary chemical potential for the $SU(2)$ Chern-Simons field. We denote its fugacity by $z$ associated to the left-moving charge $T_0^3$ and $\bar{z}$ for the right-moving charge $\bar{T}_0^3$. Finally, we can pick the left- and right-moving fermions to be in the NS or R sector. In terms of $AdS/CFT$, the partition function with these boundary conditions can be interpreted as a trace over a $CFT_2$ Hilbert space as
\beq
Z_{NS/R-NS/R} = {\rm Tr}_{NS/R-NS/R}\left[ q^{L_0-\frac{c}{24}}\bar{q}^{\bar{L}_0-\frac{c}{24}} z^{T_0^3} \bar{z}^{\bar{T}^3_0} \right],
\eeq
and we have four possible partition functions.

\paragraph{Pure $\mathbf{AdS_3}$:}  To begin, we start with considering the pure $AdS_3$ solution of this theory, and the fermions in the NS sector. In the bosonic case, the path integral of pure gravity including quantum fluctuations around this geometry can be exactly computed and it is one-loop exact \cite{Maloney:2007ud}. An analogous argument applies for the supersymmetric generalizations considered here. The answer for the path integral around $AdS_3$ is given by the product of left- and right-moving vacuum super-Virasoro characters
\beq\label{skekwe}
Z_{NS-NS} = \chi_{{\rm id},NS}^{\cN=4} (q,z) \chi_{{\rm id},NS}^{\cN=4} (\bar{q},\bar{z}),
\eeq
These characters were computed by Eguchi and Taormina \cite{Eguchi:1987wf} and are given by the following expression in the NS sector
\beq
\chi_{{\rm id},NS}^{\cN=4} (q,z) =  q^{-\frac{k-1}{4}} \frac{i \theta_3(q,-z) \theta_3(q,-z^{-1})}{\eta(q)^3 \theta_1(q,z^2)} \left[ \mu(z,q) - \mu(z^{-1},q) \right],
\eeq
where $\theta_3(q,z)$ is the Jacobi theta function and we defined the function 
\beq
\mu(z,q) \equiv \sum_{n\in\mathbb{Z}} \frac{q^{(k+1)n^2+n}z^{2(k+1)n + 1}}{(1- z q^{n+\frac{1}{2}})(1- z q^{n+\frac{1}{2}})}.
\eeq
The characters in the Ramond sector can be obtained by a simple replacement $z\to z q^{1/2}$. A similar expression applies for the right-mover sector.

\paragraph{BTZ:} Now we can consider quantum fluctuations around the BTZ solution. Classically, the metric is parametrized by an energy $E$ and angular momentum $P$ (for simplicity we turn off the $SU(2)$ fields now whose generators are denoted above by $T^i$). The metric is given by 
\beq
ds^2 =  f d\tau^2 + \frac{\ell^2 dr^2}{f} + r^2 \left( d\varphi - i\frac{r_+r_-}{r^2 } d\tau\right)^2,~~~f=\frac{(r^2-r_+^2)(r^2-r_-^2)}{r^2},
\eeq
where some useful thermodynamics quantities given in terms of the inner and outer horizon radii $r_pm$ are given by 
\beq
E = \frac{r_+^2 + r_-^2}{8 G_3 \ell},~~~P=\frac{r_+ r_-}{4 G_3 \ell},~~~T =\frac{r_+^2-r_-^2}{2\pi \ell r_+},~~~S = \frac{2 \pi r_+}{4 G_3}.
\eeq
The chemical potential for the momentum along the circle is $\theta=i \frac{1}{T}\frac{r_-}{r_+}$. For simplicity we will work with a boundary torus with all chemical potentials fixed (both for rotation and $SU(2)$ charges). The partition function including quantum effects around the BTZ solution with these boundary conditions can be obtained from the pure AdS result \eqref{skekwe} by a modular transformation. We will be interested in the R-R sector since the fermion becomes periodic in spatial $S^1$ and supersymmetry is preserved in the throat. The final answer is 
\beq
Z_{R-R} = e^{2\pi i k \frac{z^2}{\tau}-2\pi i k \frac{\bar{z}^2}{\bar{\tau}}}~ \chi_{{\rm id},NS}^{\cN=4} (q',z') ~\chi_{{\rm id},NS}^{\cN=4} (\bar{q}',\bar{z}'),
\eeq
where we take into account that the modular transformation exchanges sectors and 
\beq
q'=e^{-2 \pi i/\tau},~~z'=e^{2\pi i \alpha/\tau},~~~\bar{q}'=e^{2 \pi i/\bar{\tau}},~~\bar{z}'=e^{2\pi i \bar{\alpha}/\bar{\tau}},
\eeq
where $\alpha$ is the chemical potential for the $SU(2)$. In the next section we will use these expressions in the near extremal near-BPS limit. The origin of the first two factors is explained in section 5 of \cite{Kraus:2006nb}. We can rewrite this answer as follows in a more transparent way
\beq\label{eq:RRsumsaddle}
Z_{R-R} = \sum_{n,\bar{n}\in\mathbb{Z}}
  Z_{1-loop}^{n,\bar{n}} e^{\frac{i \pi k}{2\tau}(1-4(\alpha+n)^2)-\frac{i \pi k}{2\bar{\tau}}(1-4(\bar{\alpha}+\bar{n})^2)}
\eeq
where the one-loop determinant is 
\beq\label{eq:RRoneloop}
Z_{1-loop}^{n,\bar{n}} = \left|\frac{(1-q')}{\eta(q')} \frac{z'{}^{2n}(q'{}^{(\frac{1}{2}+n)^2} z' - q'{}^{(\frac{1}{2}-n)^2}z'{}^{-1})}{\theta_{1}(q,-z'{}^2)} \frac{\theta_{3}^2(q',-z')}{\eta(q')^2 (1-z' q'{}^{n+\frac{1}{2}})^2(1-z'{}^{-1} q'{}^{-n+\frac{1}{2}})^2}\right|^2.
\eeq
This way of rewritten the product of vacuum characters shows that the answer is given by a sum over saddles labeled by integers $n$ and $\bar{n}$, with each contribution having a classical action written in \eqref{eq:RRsumsaddle} and a one-loop determinant \eqref{eq:RRoneloop} given by the product of the graviton, $SU(2)$ Chern-Simons and gravitino contributions.

\paragraph{Preserving only $\mathbf{(4,0)}$ SUSY:}  This theory is also given by a combination of Chern-Simons actions with group $G_L \times G_R= PSU(1,1|2)_L \times \left( SL(2,\mathbb{R})\times SU(2)\right)_R$. In the $(4,0)$ case, we have as asymptotic a small $\cN=4$ Virasoro algebra for left-movers and a bosonic Virasoro and an $SU(2)$ Kac-Moody algebra for the right-mover. The expressions for the partition functions look similar to the cases above, but the right-moving character has to be replaced by the bosonic ones.

\subsection{Near-BPS near-extremal limit}
In this section, we will consider a near-extremal rotating black hole state, for the theory considered in the previous section, at fixed $SU(2)$ charges.  

We will take the limit directly for the exact pure gravity result. If matter is present, we need to be more careful, divide the geometry into the throat and far-away region, and dimensionally reduce in the throat to $\cN=4$ super-JT. To simplify, we also assume we pick fixed charge boundary conditions for any other gauge field that is not in the gravity sector. Then for the same reasons as in 4D \cite{Iliesiu:2020qvm} the temperature dependence of the partition function is given by the $\cN=4$ super-Schwarzian. 

Using AdS$_3$/CFT$_2$, we can give another perspective of this universality using the modular bootstrap: in the near-extremal limit, the vacuum block in the modular-dual channel dominates up to corrections suppressed by the twist gap \cite{Ghosh:2019rcj}.\footnote{If extra currents are present, they are easy to take into account and in the near-extremal limit looking at states where their charge is fixed, their fluctuations are frozen (this can be derived using results in \cite{Mertens:2018fds})} This is a special case of the extended Cardy regime studied for example in \cite{Benjamin:2019stq}.

We will be interested in the following limit. First, we take large $\beta_R\to0$. The ensemble is then dominated by black holes with a very large spin $P>0$.\footnote{For the opposite orientation, we can take $\beta_L\to0$ then we have large $P<0$.} Second, we take large $k\gg 1$ so that the gravity description in the bulk is accurate. Finally, we take $\beta_L \sim k \gg 1$, which implies that we are looking at very low temperatures or states with $E \sim P$. Since the state without left-moving excitations preserves supersymmetry, this is also a near-BPS limit \cite{Coussaert:1993jp}. 

We will follow the calculation first in the fixed $\beta_L,\beta_R$ ensemble. As explained in \cite{Ghosh:2019rcj} when taking this near-extremal limit we can inverse Fourier transform to obtain fixed $P$ ensemble by basically replacing $\beta_R \to 2\pi \sqrt{c/(24 P)}$ and $\beta_L \to 2 \beta$ at the end of the calculation. We also keep the left-moving $SU(2)$ chemical potential $\alpha$ fixed in this limit, and consider zero right-moving charge. Either taking the limit of the character or doing the reduction, the near-extremal near-BPS limit of the partition function is  
\beq\label{eq:RReqnschwarzian}
Z_{R-R}(\beta,\alpha) \sim e^{2\pi \sqrt{k P}} \sum_{n\in \mathbb{Z}} \frac{\beta}{k} \frac{\cot(\pi \alpha)(\alpha+n)}{(1-4(n+\alpha)^2)^2} e^{\frac{ \pi^2 k}{2\beta}(1-4(n+\alpha)^2)}
\eeq
We will not repeat the calculation here since it is completely analogous to section \ref{sec:meth2cq}. The first term $e^{2\pi \sqrt{k P}}$ comes from the right-moving identity character which is basically evaluated in the usual Cardy limit, since $\bar{q}' \to 0$. The rest comes from the evaluation of the left-moving identity character, in the limit $q'\to 1$. 

The parameters describing the near-extremal near-BPS effective theory analyzed in section \ref{sec:N=4-super-Schw} are given in terms of the level $k$ and the angular momentum $P$, by 
\begin{equation}
    S_0 = 2 \pi \sqrt{k P} ,~~~~\Phi_r = \frac{k}{4}.
\end{equation}
Taking the inverse Laplace transform of this, we obtain the same density of states as the $\cN=4$ super-Schwarzian theory with these parameters. In particular, we find a large degeneracy of BPS states given by $e^{2\pi \sqrt{ k P}}$, we find a gap to the first excited black hole state $E_{gap}=1/(2k)$, and this predicts the index matches with the black hole degeneracy. 

This can be easily generalized to cases with non-zero $SU(2)_R$ charge $\bar{T}_0^3=J_R$. In this cases states with $T_0^3=0$ are still BPS and the contribution from the right-moving sector replaces $S_0 \to 2\pi \sqrt{k P - J_R^2}$. With this modification, the spectrum as a function of temperature and $SU(2)_L$ chemical potential is still given by \eqref{eq:RReqnschwarzian}, controlled by the $\mathcal{N}=4$ super-Schwarzian.

\paragraph{Alternative construction  -- preserving $\mathbf{(4,0)}$ SUSY:} In this case we obtain a similar conclusion for $\beta_R \to 0$ and large $\beta_L$. The difference now is that we can take instead $\beta_L \to 0 $ and $\beta_R$ large. This extremal limit breaks supersymmetry since the right-moving sector of the theory is purely bosonic. Therefore, we expect that in this case, the black hole spectrum has, to leading order, no extremal states and no gap, similar to \cite{Ghosh:2019rcj} (or \cite{Iliesiu:2020qvm}).

\subsection{Comparison to string theory constructions}
In this section, we will very briefly mention some string theory constructions using D-branes that can be dimensionally reduced to the $AdS_3$ supergravity theories studied above.

For example, take type IIB string theory compactified on $M^4$, being either $T^4$ or $K3$. We consider the D1-D5 system, which at low energies can be described by either a $(4,4)$ 2D superconformal field theory or supergravity on $AdS_3 \times S^3$. Compactifying down to $AdS_3$ gives the $(4,4)$ supergravity theory we studied above. The bosonic part of the spectrum is a 3D metric on $AdS_3$, and a gauge symmetry coming from the $S^3$ factor in the metric $SO(4)\sim SU(2)_L \times SU(2)_R$ separated into left- and right-movers. For the reasons explained in the previous section, we expect the presence of other fields to leave the conclusions below unchanged. For this theory, we can derive the level of the $SU(2)$ current algebra by matching the chiral anomaly
\begin{equation}
    k = Q_1 Q_5 ~~{\rm for}~T^4,~~~~k=Q_1Q_5+1~~{\rm for}~K3.
\end{equation}
For concreteness, we look at the $T^4$ case below. 

So far we have vacuum $AdS_3$. We can add some momentum $P$ along the D1-string direction, which is identified in the BTZ gravitational description with the angular momentum $P$ defined above \cite{Cvetic:1998xh}. The BPS extremal states correspond to no left-moving excitations of the string. Looking at low temperatures we have a near-extremal near-BPS black hole string with $AdS_2 \times S^1 \times S^3$ horizon. The parameters of the effective low-energy $AdS_2$ theory from the microscopic model is
\begin{equation}
    S_0 = 2 \pi \sqrt{Q_1 Q_5 P},~~~~~\Phi_r = \frac{Q_1Q_5}{4}
\end{equation}
Using our solution we see we have $e^{S_0}$ states at extremality, consistent with \cite{Strominger:1996sh}. Our analysis also explains why the index matches with the black hole degeneracy. This can be easily generalized to near BPS states with non-zero $SU(2)_R$ charges corresponding at zero temperature to BPS black holes with angular momentum in $S^3$ \cite{Breckenridge:1996is}.

From our gravitational analysis giving the low-temperature dependence of the partition function, we have also derived the gap to the first excited black hole, and it is $E_{gap}=1/(8\Phi_r)$. In terms of the microscopic model parameters, it is  
\begin{equation}
    E_{gap} = \frac{1}{2 Q_1 Q_5}.
\end{equation}
This answer matches with the string theory approach from \cite{Maldacena:1996ds}. From this perspective, the extremal black hole states come from counting string configurations at the brane system with only left movers. The lowest energy excitation comes from the first excitation of long strings wound $Q_1Q_5$ times along the branes.

\paragraph{Alternative construction  -- Black holes in type I string theory:} We can also analyze a similar model in type I string theory instead of type II. We consider a D1-D5 brane system but now the supergravity theory emerging in $AdS_3$ has $(4,0)$ supersymmetry \cite{Johnson:1998vd, Oz:1999it}. When we have an extremal black hole made out of right-movers, we expect the spectrum near-extremality to be analogous to the $\cN=4$ super-Schwarzian. On the other hand, when the extremal black hole is made out of left-movers, supersymmetry is broken, and the near-extremal spectrum will look like the non-supersymmetric cases studied in \cite{Ghosh:2019rcj} or \cite{Iliesiu:2020qvm}.

Finally, there are other interesting compactifications which we do not analyze in this paper, but whose role we briefly mention in the discussion section.

\section{Discussion}
\label{sec:conclusion}

In this paper, we have defined and solved $\cN=4$ super-JT gravity. We show it reduces to a $\cN=4$ generalization of the Schwarzian theory, which can be exactly solved. We argue that this theory captures the temperature-dependence of the gravitational path integral evaluated around near-extremal black holes in higher dimensions. We showed that both $\cN=2$ ungauged supergravity in 4D flat space and $(4,4)$ supergravity in $AdS_3$ reduce to $\cN=4$ super-JT in the near-horizon region of near-extremal black hole backgrounds. We found a gravitational explanation of the large extremal black hole degeneracy and for the presence of a gap in the spectrum. Thus our work addresses the strong tension between the non-supersymmetric results of \cite{Iliesiu:2020qvm} and past micro-state countings in string theory \cite{Strominger:1996sh}.

We finish here with some open questions and future directions:

\subsection*{Generalization to other black holes in AdS}

While in this paper, we have focused on near-BPS black holes in 4D $\cN=2$ supergravity in flatspace or in $(4,4)$ supergravity in $AdS_3$, there are numerous other near-extremal black hole solutions which are of interest in the AdS/CFT correspondence.\footnote{Similar considerations are also useful in computing quantum corrections to the Hartle-Hawking wavefunction in dS \cite{Maldacena:2019cbz}.}

The first near-BPS solutions which we have not fully analyzed are those on $AdS_3 \times S^3 \times S^3 \times S^1$ \cite{Elitzur:1998mm}. The special feature about this compactification is that the extremal solution exhibits a large $\cN=4$ symmetry, with $SU(2)_k\times SU(2)_{k'}\times U(1)$ current algebra. It would be interesting to analyze the 2D theory emerging in the throat for near-extremal near-BPS black holes. In this case, the symmetry of the boundary mode is now $D(2,1,\alpha)$. We do not yet know how to study this version of the $\cN=4$ Schwarzian theory, and we hope to address such a construction in future work.

We would also like to briefly mention the existence of near-BPS solutions in higher dimensional AdS. Extremal black holes in such theories typically preserve a smaller amount of supersymmetry; for instance, in AdS$_4$, such black holes exhibit an $OSp(2|2)$ isometry in the near-horizon region. The effective theory capturing the breaking of $OSp(2|2)$ was found to be $\cN=2$ super-JT gravity \cite{Forste:2020xwx} and the boundary dynamics is analogously given by the $\cN=2$ super-Schwarzian (whose properties we have reviewed in appendix \ref{app:N2}). As we explain in appendix \ref{app:N2}, the $\cN=2$ super-Schwarzian has a gap depending on the value of $\hat q$ (which gives the periodicity of the identification of the $U(1)$ field $\sigma$) and on whether or not the theory exhibits an anomaly (related to how we weigh the different saddles in the path integral). Thus, to conclude whether near-BPS black holes in such a theory exhibit a mass gap, we need to perform a rigorous analysis to account for all possible massless Kaluza-Klein modes that can appear in the near-horizon region, determine the analog of $\hat q$ in supergravity and understand the situations in which the action of the boundary mode can exhibit an anomaly.

One purpose for studying the partition function of near-BPS black holes from the bulk perspective is to understand the gap in scaling dimensions between BPS and the near-BPS states in the dual CFT. If we find that the effective theory which captures the near-horizon dynamics exhibits a gap (as it did for the black holes in flatspace studied in this paper), then this translates to a scaling dimension gap, $\Delta_{\text{gap}}\sim 1/N^2$. It would be interesting to understand whether this gap in scaling dimensions is consistent with predictions from the large charge bootstrap \cite{Jafferis:2017zna} in SCFTs. Finally, in comparing the partition function on the CFT side to the black hole partition function within a fixed large charge sector, there may be a mismatch coming from configurations with multiple black holes. Thus, it would be interesting to understand such corrections coming from multi-centered black hole solutions \cite{Denef:2007vg, Denef:2000nb, Bates:2003vx, Sen:2007pg}.\footnote{We thank G.~Moore for pointing out past works on this issue.}

\subsection*{On a possible $\cN=4$ SYK model}

Another interesting possibility is whether there is a UV completion of the $\mathcal{N}=4$ Super-Schwarzian theory in some quantum mechanical models, along the lines of \cite{Fu:2016vas} for $\mathcal{N}=1,2.$ To be more specific, we would like a random quantum mechanical model, with a stable unitary nearly conformal fixed point at low energies, with unbroken $\mathcal{N}=4$ supersymmetry. A model involving dynamical bosons typically causes instability and exhibits supersymmetry breaking in the infrared (shown either as an operator with complex scaling dimension in the spectrum as in \cite{Giombi:2017dtl,Klebanov:2018fzb}, or the absence of supersymmetric Dyson-Schwinger solutions as in \cite{Anninos:2016szt,Chang:2018sve} ), and is rather inconvenient to study at finite $N$. In fact, in such a theory with a supermultiplet with $(b, \psi, \dots),$ and in a supersymmetric nearly conformal fixed point, $\Delta_{\psi}=\Delta_b+\frac{1}{2}.$ The Dyson-Schwinger equation of the dominant interaction in the infrared would constrain the dimensions of various fields so that the sum of scaling dimensions of the fields in the interaction is one, i.e.~$ n \Delta_b+ 2m \Delta_{\psi}+ \dots=1$, with $n,m\in \mathbb{Z}_{\geq 0}$.  For the nearly conformal fixed point to be unitary, we require $\Delta_{b},\Delta_{\psi}\geq 0.$ Together with the supersymmetric constraint, we conclude the only non-trivial solutions possible is that 
\begin{equation}\label{bosondim}
    \Delta_b=0, \Delta_{\psi}=\frac{1}{2}\,.
\end{equation}
However, such a solution indicates that the two-point function of $b$ must be logarithmic and typically causes a divergence in the Dyson-Schwinger equations (or \eqref{bosondim} ceases to be a solution as in \cite{Popov:2019nja}). 

On the other hand, we may consider a theory with a fermionic super-multiplet. This scenario brings about yet another complication. Unlike the case of $\mathcal{N}=1,2$, there is no relevant deformation of the $\mathcal{N}=4$ that exists in the UV free theory. To see this, we can work in the $\mathcal{N}=4$ superspace,\footnote{Here, we note that this argument does not rule out possible theories without any kind of superspace realization. In particular, in one dimension one can consider a first derivative action in bosons as in \cite{Tikhanovskaya:2020elb}, and this modifies the supersymmetry constraint to $\Delta_b=\Delta_{\psi}$. Thus it allows a greater number of relevant interactions. } and the allowed action is 
\begin{equation} 
    \mathcal{L} \sim \int d^2\theta W(\Psi)+ h.c. + \int d^4\theta K(\Psi, \bar{\Psi}),
\end{equation}
where schematically 
\begin{equation}
    \Psi= \psi+ \theta b+ \dots,
\end{equation}
 where $\Psi$ can sit in any representation of $SU(2)$ with half-integer $J$, and $\psi$ is the lowest component. As a result, any local interaction must have dimension at least one due to $\int d^2\theta \dots $, which implies that it cannot be relevant. This contrasts with the constructions of $\mathcal{N}=1,2$ SYK-like models, where $\int d\theta W(\Psi)$ is allowed and can produce relevant interactions. We can still ask if it's possible to have a marginally relevant deformation. Even if this were the case, in the infrared, a similar argument to (\ref{bosondim}) would suggest 
 \begin{equation}
     \Delta_{\psi}=0, \Delta_b=\frac{1}{2}, \dots,
 \end{equation}
which coincides with the dimensions of the free theory. This analysis suggests that constructing an interacting IR fixed point is difficult when starting from a UV theory with the same amount of supersymmetry. Therefore, one might be tempted to consider a scenario in which the $\cN=4$ supersymmetry solely emerges in the IR and is not present in the UV. We leave a more thorough investigation into these issues for future work.

\subsection*{Higher genus corrections to super-JT}

Motivated by the existence of the gap in the leading density of states for the $\cN=2$ and $\cN=4$ super-Schwarzian, it would be interesting to understand whether the gap survives when accounting for corrections coming from higher genus geometries contributing to the 2D theory. Relatedly, due to the existence of the gap, it is interesting to note that the contribution from disk topologies to the spectral form factor $\< Z(\beta-i t) Z(\beta + it)\>$ dominates even at very late times. This result contrasts with non-supersymmetric or $\cN=1$ JT gravity, where at late times, the cylindrical topology starts dominating, leading to a ``ramp'' in the spectral form factor, followed by a plateau at even later times. It would also be interesting to understand whether the genus expansion of the $\cN=2$ and $\cN=4$ super-JT gravity has an interpretation in terms of a matrix integral; this interpretation needs to go beyond the three Dyson
ensembles \cite{Dyson:1962es} and the seven Altland-Zirnbauer ensembles \cite{altland1997nonstandard}, whose gravitational interpretation was studied in \cite{Stanford:2019vob}. 

These non-perturbative corrections are relevant from the perspective of solving 2D $\cN=4$ gravity exactly. It is not clear whether these corrections could be reliable from the higher dimensional picture, but it would be interesting if the presence of supersymmetry could help better understand issues of factorization in the D1/D5 system, as one example. We leave this for future work.

\subsection*{Acknowledgements}

We thank R. Campos Delgado, A.~Castro, G.~Horowitz, I.~Klebanov, J.~Maldacena, S.~Pufu, D.~Stanford, H.~Verlinde and E.~Witten for valuable discussions and comments on the draft. MTH is supported in part by Department of Energy Grants DE-SC0007968, DE-SC0009988, and the Princeton Gravity Initiative. LVI was supported in part by the Simons Collaboration on the Conformal Bootstrap, a Simons Foundation Grant with No. 488653, and by the Simons Collaboration on Ultra-Quantum Matter, a Simons Foundation Grant with No. 651440. GJT is supported by a Fundamental Physics Fellowship. WZ was supported in part by the US NSF under Grants No. PHY-1620059 and PHY-1914860.

\appendix

\section{$\cN=2$ super-Schwarzian spectrum}\label{app:N2}
 In this appendix, we analyze the $\cN=2$ super-Schwarzian theory \cite{Fu:2016vas}. We review the solution given by \cite{Stanford:2017thb} using localization and the exact spectrum derived in \cite{Mertens:2017mtv}. Even though we will not work out the reduction in detail, we expect these results to be relevant to the spectrum of near-extremal near-BPS black holes in gauged supergravity in higher dimensional AdS.
 
\subsection{The action} 

We follow the definition of the $\cN=2$ super-Schwarzian theory given in \cite{Fu:2016vas}. The theory can be described by superspace coordinates $(\tau, \theta, \bar{\theta})$. We will consider super-reparametrization $(\tau , \theta, \bar{\theta}) \to (\tau', \theta' , \bar{\theta}')$, which satisfy the constrains $D \bar{\theta}'=\bar{D} \theta' =0$, $D \tau' = \bar{\theta}' D \theta'$ and $\bar{D} \tau' = \theta' \bar{D} \bar{\theta}'$. Here we used the super-derivative $D \equiv \partial_\theta + \bar{\theta} \partial_\tau$ and $\bar{D} \equiv \partial_{\bar{\theta}} + \theta \partial_\tau$. We will parametrize the solutions by 
\bea
\tau' &=& f(\tau) + \ldots,\\
\theta' &=& e^{i \sigma(\tau)} \sqrt{f'(\tau)} \theta + \eta(\tau) + \ldots \\
\bar{\theta}' &=& e^{-i \sigma(\tau)} \sqrt{f'(\tau)} \bar{\theta} + \bar{\eta}(\tau)+ \ldots,
\eea
where the dots denote higher order terms necessary to solve the super-reparametrization constrains. We introduce with this the time-dependent fields $f(\tau)$, $e^{i \sigma(\tau)}\in U(1)$ (or more precisely the loop group) and fermions $\eta(\tau)$ and $\bar{\eta}(\tau)$. These will become the degrees of freedom of the Schwarzian theory. The Schwarzian derivative is defined as 
\beq
S(f,\sigma, \eta,\bar{\eta}) = \frac{\partial_\tau \bar{D} \bar{\theta}'}{\bar{D}\bar{\theta}'} - \frac{\partial_\tau D \theta'}{D \theta'} - 2 \frac{\partial_\tau \theta' \partial_\tau \bar{\theta}'}{(\bar{D} \bar{\theta}' )(D\theta')}= \ldots + \theta \bar{\theta} S_b(f,\sigma,\eta,\bar{\eta})
\eeq
Finally, the $\cN=2$ super-Schwarzian action is given by $I_{\cN=2}=-\Phi_r \int d\tau d\theta d\bar{\theta} S = - \Phi_r \int d\tau S_b$. The explicit expression is complicated but the bosonic part is naturally given by
\beq
I_{\cN=2}= \Phi_r \int \Sch(f,\tau) + 2 (\partial_\tau \sigma)^2 + ({\rm fermions}),
\eeq
and the terms we omit involve the fermions $\eta$ and $\bar{\eta}$. Finally the action is invariant under a global $SU(1,1|1)$ acting on the fields $f,\sigma, \eta,\bar{\eta}$ which can be found explicitly in \cite{Fu:2016vas}. In this section, we will review the calculation of 
\beq
Z(\beta,\alpha) = \int \frac{\mathcal{D}f \mathcal{D}\sigma \mathcal{D}\eta \mathcal{D}\bar{\eta}}{SU(1,1|1)} e^{\Phi_r \int d\tau S_b(f,\sigma,\eta,\bar{\eta})}.
\eeq
The partition function depends on the inverse temperature $\beta$ and the $U(1)$ chemical potential $\alpha$. They appear in the path integral as boundary conditions for the fields $f(\tau+\beta)=f(\tau)$, $e^{i\sigma(\tau+\beta)}=e^{2 \pi i \alpha} e^{i \sigma(\tau)}$, $\eta(\tau+\beta)=-e^{-2\pi i \alpha}\eta(\tau)$ and similarly for $\bar{\eta}$. The partition also depends on the parameter $\hat{q}$, which is the periodicity of the $U(1)$ field $\sigma \sim \sigma + 2 \pi \hat{q}$.

\subsection{The partition function} 
The partition function can be computed either from localization \cite{Stanford:2017thb} or quantizing the (symplectic) integration space \cite{Mertens:2017mtv}. The answer depends on whether there is an anomaly or not. 
\beq
Z(\beta,\alpha) =e^{S_0} \sum_{n\in \mathbb{Z}} (-1)^{n\nu} \frac{2\hat{q}\cos\left( \pi  (\alpha+ \hat{q} n)\right)}{\pi\left(1-4 (\alpha+ \hat{q} n)^2\right)} e^{\frac{2\pi^2 \Phi_r}{\beta}(1-4 (\alpha+ \hat{q} n)^2)}
\eeq
From a localization perspective this is a sum over saddles labeled by an integer $n$ and the difference between both theories is whether they are weighted by $1$ or $(-1)^n$. Therefore, we introduced the parameter $\nu$ which take the values 
\bea
&&\nu=0:~~~{\rm no~anomaly},\\
&&\nu=1:~~~{\rm anomaly}.
\ea
This will simplify some expressions below. For the theory $\nu=1$ we will see the spectrum of charges is half-integer and the partition function is not periodic under $\alpha\to\alpha+\hat{q}$.

\subsection{The spectrum}
Now we will extract the spectrum from the partition function above. This was done in section 6.2 of \cite{Mertens:2017mtv}. The goal is to rewrite the partition function as a sum over supermultiplets,\footnote{We thank E. Witten for this suggestion.} which for $\mathcal{N}=2$ are $(Q)\oplus (Q-1)$ for states with $E\neq 0$ and just $Q$ for states with $E=0$. Then we aim at an expansion
\beq
Z(\beta,\alpha) =\sum_{Q} e^{2\pi i \alpha Q} \rho_{\rm ext}(Q)+ \sum_{Q}\int dE e^{-\beta E} \left( e^{i 2\pi \alpha Q} + e^{i 2\pi \alpha (Q-1)} \right) \rho_{\rm cont}(Q,E).
\eeq
We will write down the range of $Q$ below determined from the exact partition function. We will also find states at exact $E=0$ which are in shorter representations.

We can start by doing an inverse Laplace transform of the partition function with respect to inverse temperature 
\beq
Z=\int dE e^{-\beta E} \left( D_{\rm ext}(\alpha) \delta(E) + D_{\rm cont}(\alpha,E)\right)
\eeq 
The continuous can be given by two different expressions depending on the theory
\beq
D_{\rm cont} (E,\alpha) =e^{S_0}\sum_{n\in \mathbb{Z}} (-1)^{n\nu} \frac{2\hat{q}\sqrt{2 \Phi_r} \cos \left(\pi  (\alpha+n\hat{q})\right)}{\sqrt{E(1-4(\alpha+n\hat{q})^2)}} I_1\left(2\sqrt{2\pi^2 \Phi_r E(1-4(\alpha+n\hat{q})^2)}\right)
\eeq
To get an efficient description of the spectrum we can rewrite the density of states derived in \cite{Mertens:2017mtv} in the following way
\beq
D_{\rm cont}(E,\alpha) =e^{S_0} \sum_{Q\in \frac{\mathbb{Z}}{\hat{q}} + \frac{\nu}{2\hat{q}}} \left( e^{2\pi i \alpha Q } + e^{2\pi i \alpha (Q-1)}\right)\frac{\sinh{\left(2\pi \sqrt{2\Phi_r(E-E_0(Q))}\right)}}{2\pi E} \Theta(E-E_0(Q)).
\eeq
 We define the function $E_0(Q)$ giving the edge of the spectrum of each supermultiplet
\beq
E_0(Q)= \frac{1}{8\Phi_r}\left(Q-\frac{1}{2}\right)^2.
\eeq
From the expressions above a very simple picture emerges. The continuous component of the $Q$ supermultiplet density of states as a function of $Q$ and $E$ is given by 
\beq
\rho_{\rm cont}(Q,E) = e^{S_0}\frac{\sinh{\left(2\pi \sqrt{2\Phi_r(E-E_0(Q))}\right)}}{2\pi E} \Theta(E-E_0(Q)).
\eeq
First of all this is completely independent on $\hat{q}$ which only appears through the unit of charge. Second, this expression is valid for both type of theories with or without anomaly, the only difference being whether $Q\in \frac{1}{\hat{q}} \cdot \mathbb{Z}$ (no anomaly) or $Q\in \frac{1}{\hat{q}} \cdot \mathbb{Z} + \frac{1}{2\hat{q}}$ (anomaly).

 The presence of a gap in the spectrum depends on both whether $\hat{q}$ is even or odd and whether we have an anomaly or not. If $\hat{q}$ is odd, then the non-anomalous theory has a gap controlled by the supermultiplets labeled by $Q_0=\frac{1}{2}\pm \frac{1}{2\hat{q}}$ with $E_{\rm gap} \equiv E_0(Q_0) = 1/(32 \Phi_r \hat{q}^2)$. This is the actual gap of the theory since any other supermultiplet $Q$ has $E_0(Q)>E_{\rm gap}$. If the theory with odd $\hat{q}$ is anomalous then the spectrum of $Q$ is half-integers and there is no gap since for $Q=1/2$, $E_0(1/2)=0$. Similarly, for the case with $\hat{q}$ even, the non-anomalous theory has no gap, while this time the anomalous theory has a gap of the same scale as before.

We can check the same with the `extremal' states with support at $E=0$. From the partition function we get  
\beq
D_{\rm ext}(\alpha)\delta(E)= e^{S_0}\delta(E)\sum_{n\in \mathbb{Z}} (-1)^{n\nu} \frac{2\hat{q}\cos\left(\pi (\alpha+n\hat{q})\right)}{\pi\left(1-4(\alpha+n\hat{q})^2 \right)},
\eeq
This expression can be rewritten as a sum over charges in the following way 
\beq
D_{\rm ext}(\alpha) = \sum_{Q\in \frac{1}{\hat{q}}\mathbb{Z}+\frac{\nu}{2\hat{q}}, |Q|<\frac{1}{2} } e^{2 \pi i \alpha Q} ~e^{S_0} \cos \left(\pi Q\right)
\eeq
which was found in \cite{Mertens:2017mtv} using the modular transform of the character. Finally, the density of states of extremal states per charge is given by
\beq
\rho_{\rm ext}(Q,E) = e^{S_0} \cos \left(\pi Q \right) ~\Theta\left(2|Q|-1\right) .
\eeq
This is always positive since the cosine is always evaluated between $(-\pi/2,\pi/2)$. This formula, like before, is valid for either theory as long as $Q$ is taken over the correct set.

\begin{figure}
    \centering
\begin{tikzpicture}[scale=0.65]
 \pgftext{\includegraphics[scale=0.5]{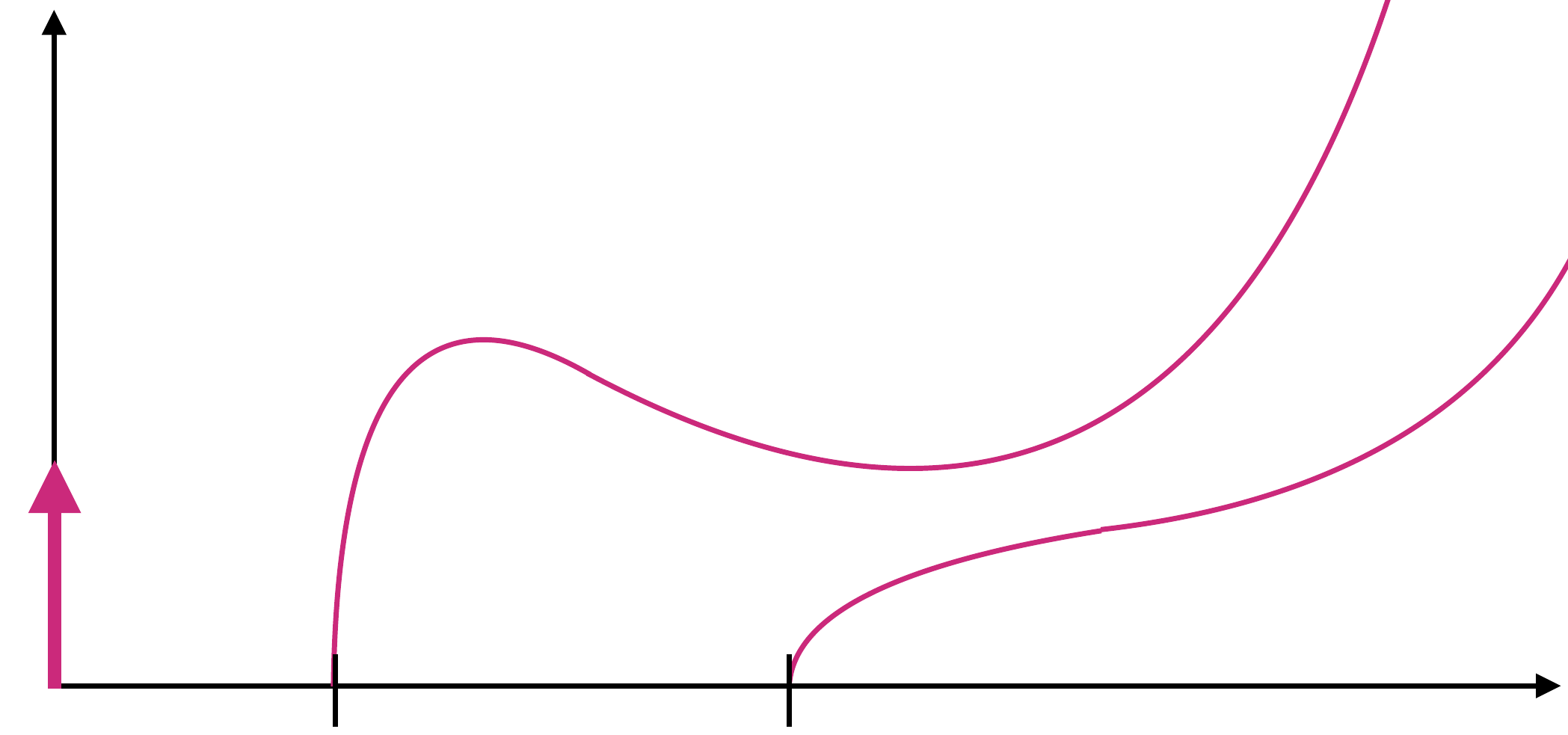}} at (0,0);
 \draw (-5.7,-0.4) node  {$e^{S_0}$};
   \draw (-5.8,2.75) node  {\small $\rho(E)$};
  \draw (-2.6,-3) node  {\small $E_{\rm gap}$};
  \draw (3.4,2.2) node  {\small $Q_0$};
   \draw (4.6,0.5) node  {\small $Q$};
  \draw (-0,-3) node  {\small $E_0(Q)$};
   \draw (5,-3) node  {\small $E$};
     \draw (-0.5,-4.3) node  {\small No anomaly};
  \end{tikzpicture}
  \hspace{0.7cm}
     \begin{tikzpicture}[scale=0.65]
 \pgftext{\includegraphics[scale=0.5]{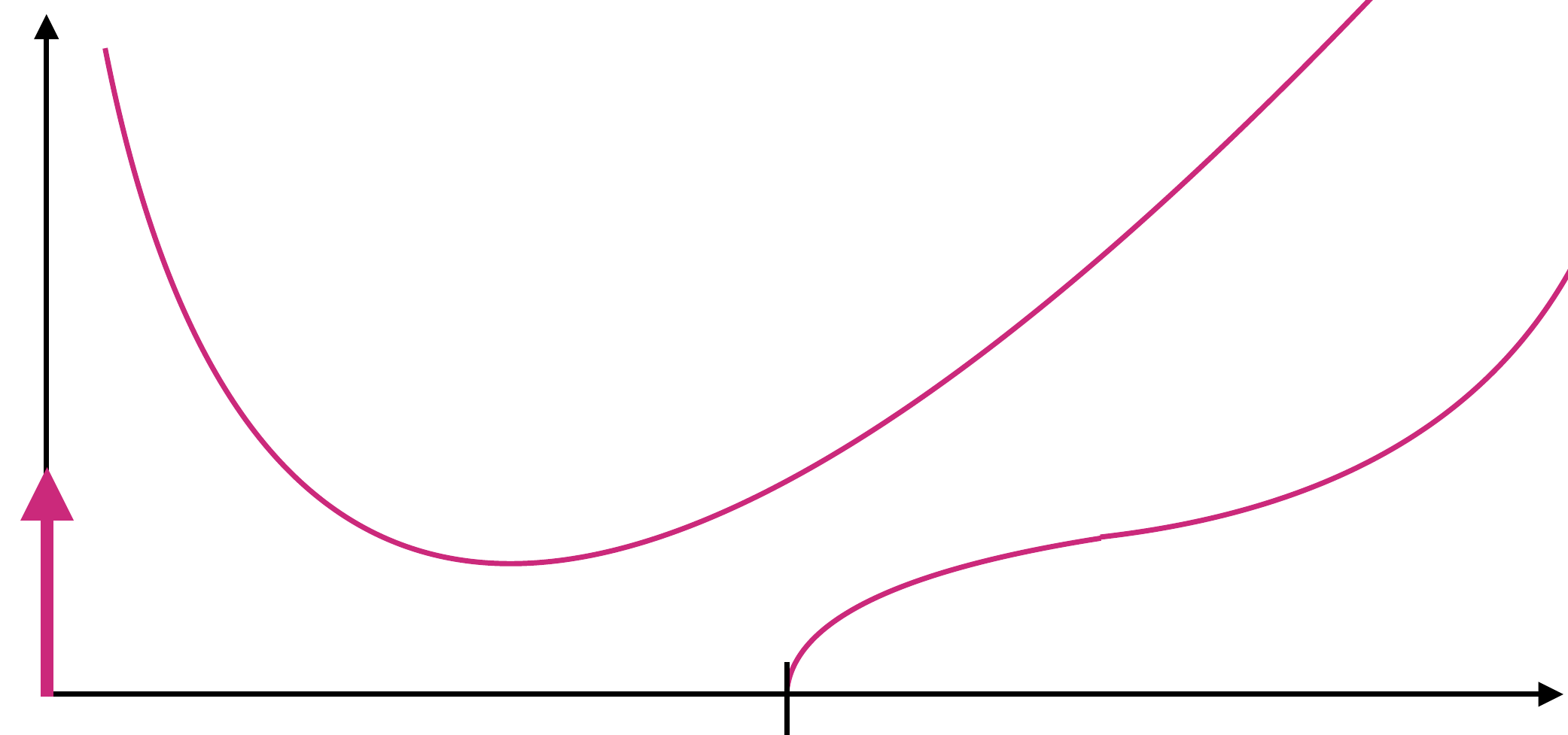}} at (0,0);
 \draw (-5.7,-0.4) node  {$e^{S_0}$};
   \draw (-5.8,2.75) node  {\small $\rho(E)$};
   \draw (-0,-3) node  {\small $E_0(Q)$};
   \draw (4.6,0.5) node  {\small $Q$};
   \draw (2,2) node  {\small $Q=1/2$};
   \draw (5,-3) node  {\small $E$};
     \draw (-0.5,-4.3) node  {\small Anomaly};
  \end{tikzpicture}
    \caption{ Density of supermultiplets as a function of energy $E$ and charge $Q$. \textbf{Left:} Odd $\hat{q}$ and no anomaly. The delta function at $E=0$ involves charges in the range $|Q|<1/2$. The supermultiplet with the lowest gap has $Q_0=1/2\pm 1/(2\hat{q})$ with $E_{\rm gap}=E_0(Q_0)$. Other supermultiplets labeled by $Q$ start at higher energies as shown. \textbf{Right:} Odd $\hat{q}$ and anomaly. The delta function at $E=0$ involves charges in the range $|Q|<1/2$. The supermultiplet with $Q=1/2$ has no gap. Other supermultiplets have a gap, as shown.}
    \label{fig:N2spexctrumappB}
\end{figure}
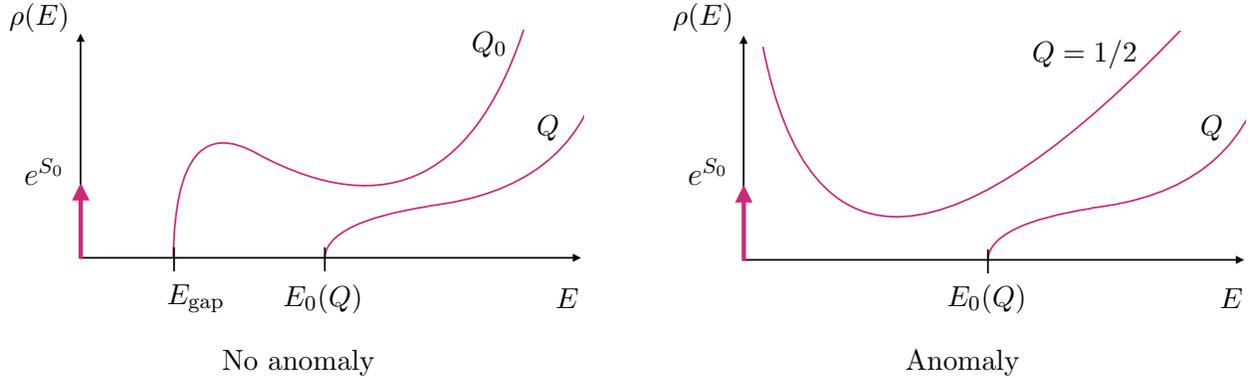

As a summary, we show the density of supermultiplets for the anomalous and non-anomalous cases in figure \ref{fig:N2spexctrumappB}. These results for the spectrum of the $\cN=2$ super-Schwarzian theory have been also obtained analytically from double-scaled $\cN=2$ SYK in \cite{Berkooz:2020xne}.

\section{Conventions and useful relations in Supergravity}
\label{app:spinorssugra}
In this appendix, we briefly summarize our conventions for supergravity. In two dimensions, we will use the flat space gamma matrices $\gamma^a$, which may be written in terms of the standard Pauli matrices $\sigma_i$ as:
\begin{align}
    \gamma^0 = i\sigma_2 = 
    \begin{pmatrix}
    0 & 1 \\
    -1 & 0
    \end{pmatrix} \, , \, \, \, \, \gamma^1 = \sigma_1 =
        \begin{pmatrix}
    0 & 1 \\
    1 & 0
    \end{pmatrix} \, , \, \, \, \, \gamma_3 = -\gamma_0 \gamma_1 =\sigma_3 =
        \begin{pmatrix}
    1 & 0 \\
    0 & -1
    \end{pmatrix} \, .
\end{align}
Our convention for the 4D gamma matrices $\Gamma^A$ is:
\begin{align}
    \Gamma^0 = i \sigma_2 \otimes \mathbf{1} \, , \, \, \, \,
    \Gamma^1 = \sigma_1 \otimes \mathbf{1} \, , \, \, \, \,
    \Gamma^2 = \sigma_3 \otimes \sigma_1 \, , \, \, \, \,
    \Gamma^3 = \sigma_3 \otimes \sigma_3 \, .
\end{align}
Note that these choices mean the gamma matrices are purely real. The curved space Dirac matrices satisfy $\Gamma_M = E^A_M \Gamma_A$, where $\Gamma_A$ satisfy $\{\Gamma_A, \Gamma_B\} = 2 \eta_{AB}$. 
Additionally, we will sometimes make use of the 4D parity Clifford element:
\begin{align}
\Gamma_5 = i \Gamma_0 \Gamma_1 \Gamma_2 \Gamma_3  = \gamma_3 \otimes (-\sigma_2) \, 
\end{align}
Our choice is convenient for the product manifold AdS$_2$ $\times$ S$^2$ because the 4D gamma matrices may be written as
\begin{equation}
    (\Gamma^a, \, \, \Gamma^{bc}) = (\gamma^a, \, \, \varepsilon^{bc} \gamma_3) \otimes \mathbf{1} \, , \, \, \, \, (\Gamma^{a'}, \, \, -\Gamma_5) = \gamma_3 \otimes \sigma_{i} \, .
\end{equation}
where we introduced the $i=1,2,3$ label for the $SU(2)$ Pauli matrices of the sphere. This gamma matrix decomposition manifests the $SU(1,1)\otimes SU(2)$ subgroup of $PSU(1,1|2)$ in terms of the higher dimensional gamma matrices.

In the gravitational theories with 2 or 4 component fermions, we also define the various conjugate fermions. The Dirac complex conjugate is
\begin{equation}
    \bar{\psi} = i \psi^\dagger \Gamma^0 \, .
\end{equation}
The Majorana conjugate is
\begin{equation}
    \bar{\psi} = \psi^T C \, .
\end{equation}
The necessary charge conjugation matrix is 
\begin{equation}
    C = - \sigma_2 \otimes \mathbf{1} = i \Gamma^0\, .
\end{equation}

For the two-dimensional action of Section~\ref{sec:higherD}, we Wick rotate to Euclidean signature. In addition to $t\rightarrow - i t$, we must also change the gamma matrices to
\begin{align}
    \gamma'^0 = \sigma_2 = 
    \begin{pmatrix}
    0 & -i \\
    i & 0
    \end{pmatrix} \, , \, \, \, \, \gamma'^1 = \sigma_1 =
        \begin{pmatrix}
    0 & 1 \\
    1 & 0
    \end{pmatrix} \, , \, \, \, \, \gamma'_3 = -\gamma'_0 \gamma'_1 =
        \begin{pmatrix}
    -i & 0 \\
    0 & i
    \end{pmatrix} \, .
\end{align}
These are also Pauli matrices for $SU(1,1)$, and various other expressions follow by continuation $\gamma_0' = -i \gamma_0$. In comparing to the BF formulation, we will also use the different set of Euclidean gamma-matrices $\bar{\gamma}_1=\sigma_1,\,\, \bar{\gamma}_2=-\sigma_3$, and $\bar \gamma_3 = \bar \gamma_1 \bar \gamma_2 = i\sigma_2$.

In the Kaluza-Klein dimensional reduction of Section~(\ref{ss:dimensionalreduction4d}), we made use of a number of results from \cite{Gibbons:2003gp}. To simplify calculations, we introduce $e^{2 \rho} = r_0 \chi^{-1/2}$.  In terms of the metric ansatz, the torsion free spin connection is found to be:
\begin{align}
    \hat{\omega}^{ab} &= \omega^{ab}+ (e^a \partial^b - e^b \partial^a )\rho -\frac{1}{2}r_0^2e^{-6 \rho} e_{im}T^m_{i} H^{iab} (e^{a'} + e^{a'}_n T^{n}_{j} B^{j}) \, , \label{eq:distortedspin} \\
        \hat{\omega}^{a a'} &= -\frac{1}{2}r_0 e^{-3 \rho} e^{a'}_m T^m_{i} H^{ia}_{b} e^b +2 r_0e^{-3 \rho}\partial^a\rho(e^{a'} + e^{a'}_n T^{n}_{j} B^{j})\, , \\
            \hat{\omega}^{a'b'} &= \omega^{a'b
            } -\nabla^{a'} e^{b'}_mT^m_{i} B^{i}\, ,
\end{align}
where $H^{i} = dB^{i} + \epsilon^{i}_{j k} B^{j} \wedge B^{k}$ is the field strength corresponding to the $SO(3)$ Kaluza-Klein gauge field. From the spin connections, one finds the components of the Ricci tensors to be
\begin{align}
    \hat{R}^{ab} &= e^{-2 \rho}\left( R^{ab} - \eta^{ab}\square \rho  + 4(\nabla^a \nabla^b \rho + \eta^{ab}(\nabla \rho)^2 ) -16 \nabla^a \rho \nabla^b \rho\right ) - \frac{1}{2}e^{-8 \rho} r_0^2 H^{ia}_{\, c}H^{jbc}T^{a'}_{i} T_{ja'} \, , \label{Eq:RicciAdS}\\
    \hat{R}^{aa'} &= \frac{1}{2}e^{-5\rho} \left (D^b H^{ia}_{\, \, b} T_{i}^{a'} -10H^{i a}_b T_{i}^{a'} \nabla^b \rho \right ) \, , \label{Eq:RicciMix} \\
    \hat{R}^{a'b'} & = \frac{e^{4 \rho}}{r_0^2}R^{a'b'} +2 e^{-2 \rho}\delta^{a'b'}\left (\square \rho -4 (\nabla \rho)^2 \right ) + \frac{1}{4}e^{-8 \rho}r_0^2 H^{i}_{ab}H^{jab}T^{a'}_{i} T_{j}^{b'} \label{Eq:RicciS2} \, .
\end{align}
These expressions are symmetric, and we recall that the unhatted $R^{a'b'}$ is the ordinary Ricci tensor for the unit $S^2$, in our case $R^{a'b'} = \delta^{a'b'}$. Finally, using the expressions above
\begin{align}
\label{eq:4DRicci}
    \hat{R} = e^{-2 \rho} R +2 r_0^2 e^{4\rho}  + 6 e^{-2 \rho}(\square \rho  -4 (\nabla \rho)^2 ) - \frac{1}{4}e^{-8 \rho}r_0^2 H^{i}_{ab}H^{jab}(\delta_{i j} - \mu_{i} \mu_{j})
\end{align}
We recognize the dilaton coupled to the 2D Ricci scalar, a dilaton potential, a total derivative, and a term quadratic in the field strengths $H^{i} = dB^{i} + \epsilon^{i}_{j k} B^{j} \wedge B^{k}$. We may use this expression for the Ricci scalar directly in the 4D action~\eqref{eq:4Dbosoniclag}; the integrals over the S$^2$ require the identity:
\be 
     \int \! \! d\theta d\phi \sin\theta \, (\delta_{i j} - \mu_{i} \mu_{j}) H^{i}_{ab}H^{jab} = \frac{8 \pi}{3} H^{i}_{ab}H_{i}^{ab} \,\nn .
\ee

\section{$PSU(1,1|2)$ Symmetry of the Extremal Black Hole}
\label{ss:PSUblackhole}

Here we will further comment on the $PSU(1,1|2)$ symmetry appering in the near-horizon region of extremal black holes. The main goal in this appendix is to check that the Killing vectors and Killing spinors discussed in section \ref{ssec:4dSugra} indeed satisfy the $\mathfrak{psu}(1,1|2)$ superalgebra.

Using the explicit expressions~\eqref{eq:KS-spinor},\eqref{eq:KS+spinor} in which we have picked a unit normalization for the constant spinors, the only non-vanishing bilinear built from $\epsilon^\alpha_-$ is
\begin{align}
        (\bar{\epsilon}^\alpha_-\Gamma^M \epsilon^\beta_-) E_M = (\bar{\epsilon}^\alpha_-\Gamma^t \epsilon^\beta_-) E_t = \delta^{\alpha \beta}\partial_t \, ,
\end{align}
which we recognize as proportional to $H = \partial_t$, the generator of time translations of the metric Eq.~\eqref{eq:poinAdSmetric}~\cite{Lu:1998nu}. The bilinear built from a pair of $\epsilon^\alpha_+$ is
\begin{align}
       (\bar{\epsilon}^\alpha_+\Gamma^M \epsilon^\beta_+) E_M  = \delta^{\alpha \beta}\left ((z^2 + t^2) \partial_t + 2tz \partial_z \right) \, ,
\end{align}
which is identified with $K = (z^2 +t^2)\partial_t + 2 tz \partial_z$, the generator of special conformal isometries of the AdS$_2$ Poincare patch. The final set of bilinears involves both $\epsilon^\alpha_+$ and $\epsilon^\alpha_-$, and is interpreted as the $\{Q, S\}$ commutator in a superconformal algebra; the nonvanishing terms are:
\begin{align}
        (\bar{\epsilon}^\alpha_-\Gamma^t \epsilon^\beta_+) E_t &= \delta^{\alpha \beta} t\partial_t \, , \\
        (\bar{\epsilon}^\alpha_-\Gamma^r \epsilon^\beta_+ ) E_r &= \delta^{\alpha \beta} z\partial_z \, , \\
                (\bar{\epsilon}^\alpha_-\Gamma^\theta \epsilon^\beta_+ ) E_\theta&  = (\sigma_2)^{\alpha \beta} \cos\phi \partial_\theta + (\sigma_1)^{\alpha \beta} \sin\phi \partial_\theta\, , \\
                (\bar{\epsilon}^\alpha_-\Gamma^\phi \epsilon^\beta_+ ) E_\phi& = (\sigma_2)^{\alpha \beta}\cot \theta \sin\phi\partial_\phi - (\sigma_1)^{\alpha \beta}\cot \theta \cos\phi\partial_\phi + (\sigma_3)^{\alpha \beta}\partial_\phi\, .
\end{align}
In the first two lines, we identify the generator of AdS$_2$ scale transformations, $D = t \partial_t + z\partial_z$. The second two lines may be combined and simplified if we introduce
\begin{align}
    T_1 = \sin \phi \partial_\theta + \cot \theta \cos \phi \partial_\phi \, , \,  T_2 =  \cos \phi \partial_\theta - \cot \theta \sin \phi \partial_\phi \, , \, \, T_3 =  \partial_\phi \, .
\end{align}
These satisfy the $SU(2)$ algebra $[T_i, T_j] = \epsilon_{ijk}T_k$. This leads to the relatively simple expression:
\begin{align}
        (\bar{\epsilon}^\alpha_-\Gamma^M \epsilon^\beta_+) E_M &= \delta^{\alpha \beta} D + (\sigma_i)^{\alpha \beta}  T^i\, .
\end{align}
In total, we have then 
\begin{align}
    [D,H] = H \, , \, \, \, [D,K] = -K \, , \, \, \, [H,K] = D \, , \, \, \, [T_i, T_j] = \epsilon_{ijk}T_k \, ,
\end{align}
which is isomorphic to the bosonic part, $SL(2)\times SU(2)$ of the supergroup $PSU(1,1|2)$ with superalgebra \eqref{eq:psu(1,1|2)-superalgebra}.

\bibliographystyle{utphys2}
{\small \bibliography{Biblio}{}}

\end{document}